\documentclass[aps,floats]{revtex4}
\usepackage{amsmath,amssymb}
\usepackage{graphicx,epsfig}
\usepackage[greek,english]{babel}
\usepackage{bbold}

\begin{document}
\bibliographystyle {plain}
\pdfoutput=1
\def\oppropto{\mathop{\propto}} 
\def\opsimeq{\mathop{\simeq}}
\def\opoverderline{\mathop{\overline}}
\def\operarrow{\mathop{\longrightarrow}}
\def\opsim{\mathop{\sim}}

\def\fig#1#2{\includegraphics[height=#1]{#2}}
\def\figx#1#2{\includegraphics[width=#1]{#2}}


\title{ Large deviations for Markov processes with stochastic resetting :  \\
analysis via the empirical density and flows or via excursions between resets    } 


\author{ C\'ecile Monthus }
 \affiliation{Institut de Physique Th\'{e}orique, 
Universit\'e Paris Saclay, CNRS, CEA,
91191 Gif-sur-Yvette, France}

\begin{abstract}

Markov processes with stochastic resetting towards the origin generically converge towards non-equilibrium steady-states. Long dynamical trajectories can be thus analyzed via the large deviations at Level 2.5 for the joint probability of the empirical density and the empirical flows, or via the large deviations of semi-Markov processes for the empirical density of excursions between consecutive resets. The large deviations properties of general time-additive observables involving the position and the increments of the dynamical trajectory are then analyzed in terms of the appropriate Markov tilted processes and of the corresponding conditioned processes obtained via the generalization of Doob's h-transform. This general formalism is described in detail for the three possible frameworks, namely discrete-time/discrete-space Markov chains, continuous-time/discrete-space Markov jump processes and continuous-time/continuous-space diffusion processes, and is illustrated with explicit results for the Sisyphus Random Walk and its variants, when the reset probabilities or reset rates are space-dependent.

\end{abstract}

\maketitle


\section{ Introduction }

The recent interest into stochastic resetting (see the review \cite{review_reset} and references therein)
can be explained via its many applications : 
within the field of intermittent search strategies (see the review \cite{review_search} and references therein),
the resetting procedure is clearly the simplest one; 
jump-diffusions processes have been also much studied in mathematical finance \cite{book_finance,review_finance}), in biology for integrate-and-fire neuronal models \cite{dumont,miles} and in ecology to describe fires \cite{daly_fire} or soil moisture \cite{daly_rain,daly_rainbis}; finally, for open quantum systems,
the unravelling of the Lindblad dynamics in terms of quantum trajectories involve 
quantum jumps that are analogous to resetting procedures \cite{quantum2.5doob},
and it is thus interesting to better understand the similarities and differences with resetting
in classical stochastic models  
\cite{quantum2.5doob,previousquantum2.5doob,garrahan_reset}.

Besides all the possible applications, another motivation
 to study stochastic resetting is that the resetting procedure towards the origin
will generically produce a non-equilibrium steady state localized around the origin, 
even if the process without resetting does not 
converge towards a stationary state \cite{review_reset,manette}. 
In the present paper, our goal will be thus to characterize Markov processes with stochastic 
resetting via the recent progresses made in the field of 
large deviations for non-equilibrium steady-states
(see the reviews with different scopes \cite{derrida-lecture,harris_Schu,searles,harris,mft,sollich_review,lazarescu_companion,lazarescu_generic,jack_review}, the PhD Theses \cite{fortelle_thesis,vivien_thesis,chetrite_thesis,wynants_thesis} 
 and the HDR Thesis \cite{chetrite_HDR}).
The large deviations at Level 2.5 has emerged in order to characterize the non-equilibrium steady-states
via the joint distribution of the empirical measure and of the empirical flows. While the theory of large deviations has a long history 
(see the reviews \cite{oono,ellis,review_touchette} and references therein), 
the explicit forms of rate functions at Level 2.5 have been obtained more recently 
 for discrete-time/discrete-space Markov chains \cite{fortelle_thesis,fortelle_chain,c_largedevdisorder,review_touchette},
for continuous-time/discrete-space Markov jump processes 
\cite{fortelle_thesis,fortelle_jump,maes_canonical,maes_onandbeyond,wynants_thesis,chetrite_formal,BFG1,BFG2,chetrite_HDR,c_ring,c_interactions,c_open,barato_periodic,chetrite_periodic}
and for continuous-time/continuous-space diffusion processes 
\cite{wynants_thesis,maes_diffusion,chetrite_formal,engel,chetrite_HDR,c_lyapunov},
while the generalization to the Lindblad dynamics 
\cite{previousquantum2.5doob,quantum2.5doob} offers new possibilities to analyze open quantum systems.
This Level 2.5 formulation plays an essential role, because any time-additive observable
of the dynamical trajectory can be reconstructed via its decomposition in terms of the empirical density and of the empirical flows. 
It is thus closely related to the studies focusing on the generating functions of time-additive observables 
via deformed Markov operators, that have attracted a lot of interest
in various models  \cite{derrida-lecture,sollich_review,lazarescu_companion,lazarescu_generic,jack_review,vivien_thesis,lecomte_chaotic,lecomte_thermo,lecomte_formalism,lecomte_glass,kristina1,kristina2,jack_ensemble,simon1,simon2,simon3,Gunter1,Gunter2,Gunter3,Gunter4,chetrite_canonical,chetrite_conditioned,chetrite_optimal,chetrite_HDR,touchette_circle,touchette_langevin,touchette_occ,touchette_occupation,derrida-conditioned,derrida-ring,bertin-conditioned,touchette-reflected,c_lyapunov,previousquantum2.5doob,quantum2.5doob},
 with the formulation of the corresponding 'conditioned' process via the generalization of Doob's h-transform.

In the field of stochastic resetting, large deviations have been already considered
 for time-additive observables
of various processes \cite{review_reset,touchette2015,harris_reset,pal,maj2019,coghi_reset}. 
In particular, for the special case of time-additive observables involving only the position
in one-dimensional diffusion models with a uniform resetting rate,
the studies  \cite{touchette2015,maj2019} have used the decomposition of the dynamical trajectory into excursions between resets in order to relate the generating functions with and without resetting.
As a consequence, besides the analysis of large deviations at Level 2.5 for 
the empirical density and of the empirical flows mentioned above, 
we will also consider the large deviations for 
the empirical excursions between two consecutive resets.
These two points of view will be then used to analyze the large deviations of the most general time-additive observables involving both the position and the increments of the dynamical trajectory,
 when the reset probabilities or reset rates are not uniform anymore but space-dependent,
 since this is relevant in many applications \cite{miles,daly_fire,daly_rainbis,sisyphus,optimal,pinsky,path}. 
 
Let us now describe in more detail the organization of the paper and mention the equations numbers 
where the main results can be found.

\subsection{Discrete-time/discrete-space Markov chains in sections \ref{sec_chain} and \ref{sec_chain1d} and in Appendix
\ref{sec_chaintree}} 

In section \ref{sec_chain}, Markov Chains in discrete time and discrete space 
with space-dependent resetting probabilities (Eq. \ref{markovchainreset})
are analyzed via the large deviation at Level 2.5 for the empirical density and flows (Eq. \ref{proba2.5chain})
and via the large deviations for the empirical excursions between two consecutive resets (Eq. \ref{probaempifin}).
The generating function of general time-additive observables (Eq. \ref{genek})
is analyzed via the tilted dynamics (Eq. \ref{Wktilt}), 
via the generator of the conditioned process obtained as the generalization of Doob's h-transform (Eq. \ref{Wsoludoubletildedoob}) and via the probability of excursions between resets of this conditioned process
(Eq. \ref{Psoludoubletilde}).

In section \ref{sec_chain1d}, this general formalism is applied 
to the example of the Sisyphus Random Walk \cite{sisyphus} on the semi-infinite lattice $x=0,1,2,..$, 
where Sisyphus at position $x$ at time $t$  (Eq \ref{markovchainreset1d})
can either return to the origin with probability $R_x$ or
move forward to the next position $(x+1)$ with the complementary probability $(1-R_x)$.
 Explicit expressions are given for the existence of a non-equilibrium steady-state (Eq. \ref{cvsteady1d}),
 for the large deviations at Level 2.5 (Eq. \ref{proba2.5chain1dsansj}) and at Level 2 
 for the empirical density alone (Eq. \ref{proba2.5chaintreeonlyrhosimpli1d}), as well as for the large deviations of excursions (Eq. \ref{probaexctree1d}).
 The generating function of general time-additive observables (Eq. \ref{genekchainsij})
 is analyzed via the tilted dynamics and via the tilted excursions to obtain 
 the same explicit result for the scaled cumulant generating function (Eqs \ref{Wktiltrightmu}
 and \ref{constraintsttsi}).

 Appendix \ref{sec_chaintree} contains another application of the general formalism of section \ref{sec_chain}
to the Sisyphus Random Walk on the Cayley Tree of branching ratio $b$, that generalizes
the results of the one-dimensional case $b=1$ described in section  \ref{sec_chain1d}.

\subsection{Continuous-time/discrete-space Markov jump processes in sections \ref{sec_jump} and \ref{sec_jump1d} } 

In section \ref{sec_jump}, Markov Jump processes in continuous time and discrete space 
with space-dependent resetting rates (Eq. \ref{mastereqexpli})
are analyzed via the large deviation at Level 2.5 for the empirical density and flows (Eq. \ref{level2.5master})
and via the large deviations of empirical excursions between two consecutive resets (Eq. \ref{probaempiexcjump}).
The generating function of general time-additive observables (Eq. \ref{genekj})
is analyzed via the tilted dynamics (Eqs \ref{wtiltedoff} and \ref{wtiltediag}),
via the generator of the conditioned process obtained as the generalization of Doob's h-transform (Eq. \ref{Wdoubletildedoobj}) 
and via the tilted excursions (Eq. \ref{pexctiltedj}) with the corresponding conditioned excursion probabilities
(Eq \ref{Psoludoubletildej}).

In section \ref{sec_jump1d}, this general formalism is applied 
to the example of the Sisyphus Jump process on the semi-infinite lattice $x=0,1,2,..$,
where Sisyphus at position $x$ at time $t$  (Eq \ref{treejumpdyn})
can return to the origin with rate $r_x$,
move forward to the next position $(x+1)$ with rate $w$ or remain at $x$.
Explicit expressions are given for the existence of a non-equilibrium steady-state (Eq. \ref{cvsteady1dj}),
 for the large deviations at Level 2.5 (Eq. \ref{level2.5mastertree})
 and for the large deviations of excursions (Eq. \ref{probaempiexcjumps}).
 The generating function of general time-additive observables (Eq. \ref{genekj1d})
 is analyzed via the tilted dynamics and via the tilted excursions to obtain 
 the same explicit result for the scaled cumulant generating function (Eqs \ref{tilteqmuk}
 and \ref{constraintstt1si}).

\subsection{ Continuous-time/continuous-space diffusion processes in sections \ref{sec_diff} and \ref{sec_neuronal} } 

In section \ref{sec_diff}, diffusion processes in a force field in dimension $d$
with space-dependent resetting rates (Eq. \ref{fokkerplanckreset})
are analyzed via the large deviation at Level 2.5 for the empirical density and currents (Eq. \ref{ld2.5diff})
and via the large deviations of empirical excursions between two consecutive resets (Eq. \ref{probaempifinjump}).
The generating function of general time-additive observables (Eq. \ref{genekdiff})
is analyzed via the tilted dynamics (Eq \ref{fokkerplancktilt}),
via the generator of the conditioned process obtained as the generalization
 of Doob's h-transform (Eqs \ref{fokkerdoob} and \ref{fokkerplancktt}) 
and via the tilted excursions (Eq. \ref{pexctiltediff}) with the corresponding conditioned excursion probabilities
(Eq \ref{Psoludoubletildediff}).

In section \ref{sec_neuronal},
 this general formalism is applied 
to the example of the continuous-time Sisyphus process  (Eq \ref{neuronal}) on the semi-infinite line $x \in [0,+\infty[$.
 Explicit expressions are given for the existence of a non-equilibrium steady-state (Eq. \ref{neuronalstcv}),
 for the large deviations at Level 2.5 (Eq. \ref{ld2.5diffsi}) and at Level 2 
 for the empirical density alone (Eq. \ref{ld2.5diffrho}), as well as
 for the large deviations of excursions (Eq. \ref{largedevnsi}).
 The generating function of general time-additive observables (Eq. \ref{genekdiff1dred})
 is analyzed via the tilted dynamics and via the tilted excursions to obtain 
 the same explicit result for the scaled cumulant generating function (Eqs \ref{eqmukdiff}
 and \ref{constraintstt1dsi}).



\section{ Markov Chain in discrete time and discrete space with resetting  }

\label{sec_chain}

\subsection{ Models and notations }

In this section, we focus 
on the Markov Chain dynamics for the probability $P_y(t)  $ to be at position $y$ at time $t$
\begin{eqnarray}
P_x(t+1) =  \sum_y W^{reset}_{x,y}  P_y(t)
\label{markovchain}
\end{eqnarray}
where the matrix 
\begin{eqnarray}
W^{reset}_{x,y} \equiv  W_{x,y} (1-R_y)      +  \delta_{x,0} R_y  
\label{markovchainreset}
\end{eqnarray}
has the following physical meaning.
When the particle is at position $y$ at time $t$, two types of moves are possible to obtain the new position $x$ at time $t+1$ :

(i) with probability $R_y \in [0,1]$,
there is a reset to the origin $x=0$, i.e. the particle makes a non-local spatial move from $y$ to $0$.

(ii)  with the complementary probability $(1-R_y)$, 
the particle follows the spatially-local Markov Chain described by the matrix $W_{x,y} $,
i.e. it jumps towards the some neighboring position $x$ of $y$ with probability $W_{x,y}$,
with the normalization
\begin{eqnarray}
  \sum_x W_{x,y} && =1
\label{markovnorma}
\end{eqnarray}

Although the case of uniform resetting probabilities $R_y=R$ is the most studied in the literature \cite{review_reset},
it is also interesting to consider the case of space-dependent resetting probabilities, where $R_y$
depends on the position $y$ either deterministically or randomly \cite{miles,daly_fire,daly_rainbis,sisyphus,optimal,pinsky,path}
since this is very relevant in many applications \cite{miles}. 
In the whole paper, we will thus consider that the resetting probability is space-dependent.

As stressed in the Introduction, the stochastic resetting towards the origin
will generically produce a non-equilibrium steady state localized around the origin  
even if the process without resetting does not 
converge towards a stationary state \cite{review_reset,manette}. 
In this section, we will thus assume that the steady-state solution $P^*_x$ of Eq. \ref{markovchain}
\begin{eqnarray}
P^*_x =  \sum_y W^{reset}_{x,y} P^*_y
\label{markovchainst}
\end{eqnarray}
exists in order to apply the large deviations at Level 2.5 for non-equilibrium steady-states.


\subsection{ Large deviations at level 2.5 for the empirical density and the empirical flows  }

The empirical 2-point density characterizes the flows between two consecutive positions in a very long trajectory $x(0 \leq t \leq T)$
\begin{eqnarray}
 \rho^{(2)}_{x,y} && \equiv \frac{1}{T} \sum_{t=1}^T \delta_{x(t),x}  \delta_{x(t-1),y} 
\label{rho2pt}
\end{eqnarray}
It contains the information on the empirical 1-point density
that can be obtained via the sum over the first or the second position 
 (up to a boundary term of order $1/T$ that is negligible for large duration $T \to +\infty$)
\begin{eqnarray}
 \rho_{x} && \equiv \frac{1}{T} \sum_{t=1}^T \delta_{x(t),x}  = \sum_y \rho^{(2)}_{x,y} =
 \sum_y \rho^{(2)}_{y,x}
\label{rho1pt}
\end{eqnarray}
with the normalization
\begin{eqnarray}
\sum_x \rho_{x} && = 1
\label{rho1ptnorma}
\end{eqnarray}

For this discrete-time/discrete-space Markov chain framework,
the joint probability to see the empirical 2-point and 1-point densities 
follows the large deviation form at Level 2.5
\cite{fortelle_thesis,fortelle_chain,c_largedevdisorder,review_touchette}
\begin{eqnarray}
P_T ( \rho^{(2)}_{.,.} ; \rho_. )
 && \opsimeq_{T \to +\infty}  C ( \rho^{(2)}_{.,.} ; \rho_.)e^{ - T I( \rho^{(2)}_{.,.} ; \rho_.)  } 
\label{proba2.5chain}
\end{eqnarray}
with the constraints discussed in Eqs \ref{rho1pt} and \ref{rho1ptnorma}
\begin{eqnarray}
&& C ( \rho^{(2)}_{.,.} ; \rho_.)
  = \delta \left( \sum_x \rho_{x} - 1 \right) 
 \prod_x \left[  \delta \left(   \sum_y \rho^{(2)}_{x,y} -\rho_{x} \right)
\delta \left(   \sum_y \rho^{(2)}_{y,x} -\rho_{x} \right) \right]
\label{constraints2.5chain}
\end{eqnarray}
while the rate function involves the kernel
 $W^{reset}_{x,y} $ of Eq. \ref{markovchainreset}
\begin{eqnarray}
  I( \rho^{(2)}_{.,.} ; \rho_. ) 
  =  
\sum_x \sum_y \rho^{(2)}_{x,y} \ln \left( \frac{\rho^{(2)}_{x,y}}{ W^{reset}_{x,y}  \rho_{y}}  \right) 
   =  
\sum_x \sum_y \rho^{(2)}_{x,y} \ln \left( \frac{\rho^{(2)}_{x,y}}{ \left[ W_{x,y} (1-R_y)      +  \delta_{x,0} R_y  \right]  \rho_{y}}  \right)
\label{rate2.5chain}
\end{eqnarray}
Then one needs to discuss the possible empirical flows $\rho^{(2)}_{x,y} $ due to the Markov chain $W_{x,y} $ and to the resetting procedure in order to obtain more explicit expressions,
as explained on the specific examples of section \ref{sec_chain1d}
and Appendix \ref{sec_chaintree}.


\subsection{ Large deviations for the empirical excursions between two consecutive resets }

\subsubsection{ Empirical density of excursions between two consecutive resets }

A very long dynamical trajectory $x(0 \leq t \leq T)$ starting at $x_0=0$
can also be analyzed via its decomposition into excursions between resets.
Let us note $t_l$ with $l=0,1,..,N-1$ the times where a reset towards the origin occurs $x(t_l)=0$,
with $t_0=0$.
It is convenient to introduce the durations of these excursions
\begin{eqnarray}
\tau_l \equiv t_{l+1}-t_l 
\label{tauk}
\end{eqnarray}
and the corresponding internal trajectory for $s=1,..,\tau_l-1$
\begin{eqnarray}
y^{(l)}(s) \equiv x(t_l+s)
\label{yks}
\end{eqnarray}
 between $y^{(l)}(s=0) = x(t_l)=0 $ and $y^{(l)}(\tau_l)= x(t_l+\tau_l)=x(t_{l+1})=0$.

For large time $T$, we assume that the density $n=\frac{N}{T}$ of resets will be finite,
so the total number $N=nT$ of excursions will be large,
and we may neglect the fact that the last excursion does not finish exactly at time $T$,
in order to analyze the empirical density of excursions of duration $\tau$ and internal trajectory $y( 1 \leq s \leq \tau-1) $
\begin{eqnarray}
n[\tau ; y(.) ]\equiv n[\tau ; y( 1 \leq s \leq \tau-1) ] && \equiv  \frac{1}{T} \sum_{l=0}^{N-1}    \delta_{\tau_{l},\tau} 
\left( \prod_{s=1}^{\tau-1} \delta_{y^{(l)}(s),y(s)} \right)
\label{nexcur}
\end{eqnarray}
The total density of excursions is then
\begin{eqnarray}
n \equiv \frac{N}{T} = \sum_{\tau=1}^{+\infty} \sum_{y( .)} n[\tau ; y(.) ] 
\label{nexcurtot}
\end{eqnarray}
while the normalization is given by the total duration of the excursions
\begin{eqnarray}
 \sum_{\tau=1}^{+\infty} \tau \sum_{y(.) } n[\tau ; y(.) ] 
 = \frac{1}{T} \sum_{l=0}^{N-1} \tau_l =1
\label{normaexcur}
\end{eqnarray}
Note that if we had not neglected the fact that the last excursion does not finish exactly at time $T$,
there would be a correction of order $1/T$ in Eq. \ref{normaexcur}
that would be the analog of the boundary correction of order $1/T$ that one neglects in Eq. \ref{rho1pt}.


\subsubsection{ Large deviations for the empirical density of excursions between two consecutive resets  }

For the dynamics of Eq. \ref{markovchainreset},
the probability to have an excursion of duration $\tau$ and of internal trajectory $y( 1 \leq s \leq \tau-1) $
reads
\begin{eqnarray}
P^{exc}[ \tau ; y(.)  ] \equiv P^{exc}[ \tau ; y( 1 \leq s \leq \tau-1)  ] 
&& = R_{y(\tau-1)} W_{y(\tau-1),y(\tau-2) } (1-R_{y(\tau-2)}) ... W_{y(1),0 } (1-R_{0})
 \nonumber \\ &&
 =  R_{y(\tau-1)}
 \left[    \prod_{s=1}^{\tau-1} W_{y(s),y(s-1)}  (1-R_{y(s-1)} )\right]
\label{pinter}
\end{eqnarray}
For $\tau=1$ corresponding to a consecutive reset towards the origin, 
there is no-internal trajectory $y( 1 \leq s \leq \tau-1) $
and the probability reduces to
\begin{eqnarray}
P^{exc}[ \tau=1]  = R_{0} 
\label{pinter0}
\end{eqnarray}

The framework of semi-Markov processes
\cite{fortelle_thesis,c_largedevdisorder,gaspard,maes_semi,zambotti,faggionato}
allows to write the large deviations properties of empirical intervals as follows.
The probability to see 
 the empirical density $n[\tau ; y( 1 \leq s \leq \tau-1) ]  $ of excursions between resets
and the total density $n$ follows the large deviation form 
\begin{eqnarray}
P_T ( n[.;..], n ) && \opsimeq_{T \to +\infty}  C ( n[.;..] ,  n) e^{ - T I( n[.;..], n )  } 
\label{probaempifin}
\end{eqnarray}
with the constraints of Eqs \ref{nexcur} and \ref{normaexcur}
\begin{eqnarray}
&& C ( n[.;..] ,  n)
  = \delta \left[ \sum_{\tau=1}^{+\infty} \sum_{y(.)} n[\tau ; y(.) ] 
- n \right]
 \delta \left[ \sum_{\tau=1}^{+\infty} \tau \sum_{y(.) } n[\tau ; y(.) ] -1 \right]
\label{constraints}
\end{eqnarray}
and the rate function that involves the probability of excursions (Eq. \ref{pinter})
\begin{eqnarray}
&&  I( n[.;..], n )  =  
 \sum_{\tau=1}^{+\infty} \sum_{y( .)} 
  n[\tau ; y(.)  ]
  \ln \left( \frac{n[\tau ; y( .)  ]  }{ P^{exc}[ \tau ; y(.)  ]  n  }  \right) 
\label{rateexc}
\end{eqnarray}


\subsubsection{ Simplifications for uniform reset probability $R_y=R$}

When the reset probability is uniform $R_y=R$, the probability of excursions of Eq. \ref{pinter}
becomes factorized 
\begin{eqnarray}
P^{exc}_{uniform}[ \tau ; y(.)  ] 
 = p^{geo}(\tau) p^{config}_{\tau} [ y(.)  ] 
\label{pinterfactor}
\end{eqnarray}
into the geometric distribution that describes the probability of the duration $\tau=1,2,..$ of the excursion
independently of the internal trajectory $y( 1 \leq s \leq \tau-1) $
\begin{eqnarray}
p^{geo}(\tau) =  R (1-R)^{\tau-1} 
\label{pgeo}
\end{eqnarray}
while
\begin{eqnarray}
 p^{config}_{\tau} [ y(.)  ] =    \prod_{s=1}^{\tau-1} W_{y(s),y(s-1)}  
\label{pconfig}
\end{eqnarray}
represents the probability of the internal trajectory $y( 1 \leq s \leq \tau-1) $ once the duration $\tau$ is fixed,
and coincides with the probability of the unperturbed Markov Chain of kernel $W_{x,y}$ starting at $y(0)=0$.
This decoupling is at the origin of most results obtained for the case of uniform resetting.
In particular, the statistical properties of the durations of the intervals between resets is completely independent from the configurational part.
For instance, the probability of the total density $n=\frac{N}{T}$ of excursions simply follows the binomial distribution
with its corresponding large deviation form
\begin{eqnarray}
P^{Binomial}_{T}(n) = \frac{ T! }{ (nT)! ((1-n)T)! }  R^{nT} (1-R)^{(1-n)T} 
\oppropto_{T \to +\infty} e^{- \displaystyle  T \left[  n \ln \left(  \frac{ n }{ R }\right)
+(1-n)  \ln \left(  \frac{ 1-n }{  1-R }\right) \right] }
\label{binom}
\end{eqnarray}

On the contrary, whenever the reset probability $R_y$ depends on the position $y$, 
the duration $\tau$ and the internal trajectory $y( 1 \leq s \leq \tau-1)  $ are coupled via Eq. \ref{pinter}.


\subsection{ Large deviations for time-additive observables via three points of view }

\label{sec_additivechain}

In this section, we focus on time-additive observables of the trajectory $x(0 \leq t \leq T)$
that involves some function $\alpha_x$ 
\begin{eqnarray}
 A_T  && \equiv   \frac{1}{T} \sum_{t=1}^T \alpha_{x(t)}  
\label{additiveA}
\end{eqnarray}
or that involves some function $\beta_{x,y}$ 
\begin{eqnarray}
 B_T  && \equiv   \frac{1}{T} \sum_{t=1}^T \beta_{x(t),x(t-1)} 
\label{additiveB}
\end{eqnarray}
In the following, we consider the observable $(A_T+B_T )$ and
 analyze its large deviations properties via the scaled cumulant generating function $\mu(k)$
appearing in the asymptotic behavior
\begin{eqnarray}
Z_T(k) \equiv    < e^{ T k \left( A_T + B_T \right) } >  \opsimeq_{T \to +\infty} e^{T \mu(k) }
\label{genek}
\end{eqnarray}

\subsubsection{ Analysis via the tilted dynamics }

As recalled in the Introduction, the standard method to analyze
the generating function of time-additive observables 
is based on the spectral analysis of deformed Markov operators  \cite{derrida-lecture,sollich_review,lazarescu_companion,lazarescu_generic,jack_review,vivien_thesis,lecomte_chaotic,lecomte_thermo,lecomte_formalism,lecomte_glass,kristina1,kristina2,jack_ensemble,simon1,simon2,simon3,Gunter1,Gunter2,Gunter3,Gunter4,chetrite_canonical,chetrite_conditioned,chetrite_optimal,chetrite_HDR,touchette_circle,touchette_langevin,touchette_occ,touchette_occupation,derrida-conditioned,derrida-ring,bertin-conditioned,touchette-reflected,c_lyapunov,previousquantum2.5doob,quantum2.5doob}.
Let us recall how this method works within the present Markov Chain framework.
The probabilities of the whole trajectories $x(0 \leq t \leq T)$
\begin{eqnarray}
{\cal P}[x(T),x(T-1),...,x(1),x(0)=0]  = \delta_{x(0),0 } \prod_{t=1}^T W^{reset}_{x(t) ,x(t-1)} 
\label{pwtraj}
\end{eqnarray}
allow to compute the generating function of Eq. \ref{genek} 
\begin{eqnarray}
Z_T(k) =   < e^{ \displaystyle k \sum_{t=1}^T\left(\alpha_{x(t)} + \beta_{x(t),x(t-1)}  \right) } >
&& =\delta_{x(0),0 } \sum_{x(1 \leq t \leq T)} 
 \prod_{t=1}^T \left[ W^{reset}_{x(t) ,x(t-1)} e^{ k \left(\alpha_{x(t)} + \beta_{x(t),x(t-1)}  \right) } \right]
 \nonumber \\
 && \equiv  \delta_{x(0),0 } \sum_{x(1 \leq t \leq T)} 
 \prod_{t=1}^T \left[ {\tilde W}^{[k]}_{x(t) ,x(t-1)}  \right]
\label{genektraj}
\end{eqnarray}
via the tilted matrix
\begin{eqnarray}
 {\tilde W}^{[k]}_{x ,y} \equiv W^{reset}_{x ,y} e^{ k \left(\alpha_{x} + \beta_{x,y}  \right) }
 = \left[ W_{x,y} (1-R_y)      +  \delta_{x,0} R_y  \right] e^{ k \left(\alpha_{x} + \beta_{x,y}  \right) }
\label{Wktilt}
\end{eqnarray}
The scaled cumulant generating function $\mu(k)$ of Eq. \ref{genek}
will then correspond to the logarithm of the highest eigenvalue $e^{\mu(k)}$
of the tilted matrix of Eq. \ref{Wktilt} that will dominate the propagator
\begin{eqnarray}
< x \vert  \left(W^{[k]}\right)^T \vert y>  \opsimeq_{T \to +\infty}  e^{T \mu(k) } {\tilde r}^{[k]}_x {\tilde l}^{[k]}_y 
\label{Wktiltprop}
\end{eqnarray}
where $ {\tilde r}^{[k]}_x$ is the corresponding positive right eigenvector
\begin{eqnarray}
e^{ \mu(k) }  {\tilde r}^{[k]}_x = \sum_y  {\tilde W}^{[k]}_{x ,y}  {\tilde r}^{[k]}_y
=  \sum_y   W_{x,y} (1-R_y)   e^{ k \left(\alpha_{x} + \beta_{x,y}  \right) }  {\tilde r}^{[k]}_y
+  \delta_{x,0} \sum_y    R_y  e^{ k \left(\alpha_{0} + \beta_{0,y}  \right) }  {\tilde r}^{[k]}_y
 \label{Wktiltright}
\end{eqnarray}
and where ${\tilde l}^{[k]}_y $ is the corresponding positive left eigenvector
\begin{eqnarray}
e^{ \mu(k) }  {\tilde l}^{[k]}_y = \sum_x  {\tilde l}^{[k]}_x {\tilde W}^{[k]}_{x ,y}  
= \sum_x  {\tilde l}^{[k]}_x  W_{x,y} (1-R_y)   e^{ k \left(\alpha_{x} + \beta_{x,y}  \right) } 
+  {\tilde l}^{[k]}_0   R_y  e^{ k \left(\alpha_{0} + \beta_{0,y}  \right) } 
 \label{Wktiltleft}
\end{eqnarray}
with the normalization
\begin{eqnarray}
 \sum_x  {\tilde l}^{[k]}_x   {\tilde r}^{[k]}_x=1
 \label{Wktiltnorma}
\end{eqnarray}


\subsubsection{ Analysis via the empirical density and empirical flows }

The empirical 1-point density $\rho_x$ of Eq. \ref{rho1pt}
allows to reconstruct any time-additive observable $A_T$ of the position (Eq. \ref{additiveA})
\begin{eqnarray}
 A_T  &&  = \sum_x \alpha_x \rho_x
\label{additiveArho}
\end{eqnarray}
while the empirical 2-point density $\rho^{(2)}_{x,y} $ of Eq. \ref{rho2pt}
allows to reconstruct any time-additive observable $B_T$ of the increments (Eq. \ref{additiveB})
\begin{eqnarray}
 B_T  &&  = \sum_x \sum_y \beta_{x,y} \rho^{(2)}_{x,y}
\label{additiveBrho2}
\end{eqnarray}
As a consequence, the generating function of Eq. \ref{genek} can be evaluated from
the joint probability of Eqs \ref{proba2.5chain} \ref{constraints2.5chain} \ref{rate2.5chain}
\begin{eqnarray}
Z_T(k) && =  \int {\cal D} \rho^{(2)}_{.,.} \int {\cal D} \rho_{.}
P_T ( \rho^{(2)}_{.,.} ; \rho_. )   e^{ \displaystyle T k \left(  \sum_x \alpha_x \rho_x +  \sum_x \sum_y \beta_{x,y} \rho^{(2)}_{x,y} \right) }  
\nonumber \\
&& \opsimeq_{T \to +\infty}   \int {\cal D} \rho^{(2)}_{.,.} \int {\cal D} \rho_{.}
 \delta \left( \sum_x \rho_{x} - 1 \right) 
 \prod_x \left[  \delta \left(   \sum_y \rho^{(2)}_{x,y} -\rho_{x} \right)
\delta \left(   \sum_x \rho^{(2)}_{x,y} -\rho_{y} \right) \right]
\nonumber \\
&&
e^{ \displaystyle  T \left[ - \sum_x \sum_y \rho^{(2)}_{x,y} \ln \left( \frac{\rho^{(2)}_{x,y}}{ W^{reset}_{x,y}  \rho_{y}}  \right) 
+  k  \sum_x \alpha_x \rho_x  +k  \sum_x \sum_y \beta_{x,y} \rho^{(2)}_{x,y} \right] } 
\label{genekrho}
\end{eqnarray}
Using the constraints, one can rewrite the functional in the exponential in a more compact form
involving the tilted generator of Eq. \ref{Wktilt}
\begin{eqnarray}
Z_T(k)   \opsimeq_{T \to +\infty}   \int {\cal D} \rho^{(2)}_{.,.} \int {\cal D} \rho_{.}
 \delta \left( \sum_x \rho_{x} - 1 \right) 
 \prod_x \left[  \delta \left(   \sum_y \rho^{(2)}_{x,y} -\rho_{x} \right)
\delta \left(   \sum_x \rho^{(2)}_{x,y} -\rho_{y} \right) \right]
e^{ \displaystyle  - T   \sum_x \sum_y \rho^{(2)}_{x,y} \ln \left( \frac{\rho^{(2)}_{x,y}}{  {\tilde W}^{[k]}_{x ,y}  \rho_{y}}  \right)
 } 
\label{genekrhot}
\end{eqnarray}
It is convenient to make the change of variables from the 2-point density $\rho^{(2)}_{x,y}  $
to the effective Markov-Chain kernel that would make $\rho^{(2)}_{.,.} $ and $\rho_. $ typical
\begin{eqnarray}
{\tilde  {\tilde W}}_{x ,y} \equiv \frac{\rho^{(2)}_{x,y}}{ \rho_{y}} 
\label{Wktdoob}
\end{eqnarray}
to obtain
\begin{eqnarray}
Z_T(k)   && \opsimeq_{T \to +\infty}   \int {\cal D} {\tilde  {\tilde W}}_{.,.} \int {\cal D} \rho_{.}
 \delta \left( \sum_x \rho_{x} - 1 \right) 
 \prod_x \left[  \delta \left(   \sum_y {\tilde  {\tilde W}}_{x ,y}  \rho_{y} -\rho_{x} \right)
\delta \left(   \sum_x {\tilde  {\tilde W}}_{x,y}  - 1 \right) \right]
 \nonumber \\ &&
e^{ \displaystyle  - T   \sum_y \rho_{y} \sum_x {\tilde  {\tilde W}}_{x ,y}  
       \ln \left( \frac{ {\tilde  {\tilde W}}_{x ,y}}{  {\tilde W}^{[k]}_{x ,y}  }  \right)
 } 
\label{genekrhott}
\end{eqnarray}
The two last constraints mean that ${\tilde  {\tilde r}}_x=\rho_x$
and ${\tilde  {\tilde l}}_x=1$ are the right and the left eigenvectors of the matrix $ {\tilde  {\tilde W}}$
associated to the eigenvalue unity.
In order to optimize the functional in the exponential in the presence of the constraints,
it is convenient to introduce the following Lagrangian
with the Lagrange multipliers 
$(\omega^{[k]},\nu^{[k]}_x, \lambda^{[k]}_y)$
\begin{eqnarray}
 {\cal L}  ( {\tilde  {\tilde W}}_{.,.} ; \rho_.) 
\!  && = 
\!
   -  \sum_y  \rho_{y}  \sum_x {\tilde  {\tilde W}}_{x ,y} 
       \ln \left( \frac{ {\tilde  {\tilde W}}_{x ,y}}{  {\tilde W}^{[k]}_{x ,y}  }  \right)
\!  - \omega^{[k]} \left( \sum_x \rho_{x} - 1 \right) 
\! +\sum_x  \nu^{[k]}_x \left(   \sum_y {\tilde  {\tilde W}}_{x ,y}  \rho_{y} -\rho_{x} \right)
\! + \sum_y  \lambda^{[k]}_y \left(   \sum_x {\tilde  {\tilde W}}_{x ,y}  - 1 \right) 
\nonumber \\
&& = \omega^{[k]}
  +  \sum_y  \rho_{y} \left[ -  \sum_x {\tilde  {\tilde W}}_{x ,y} 
       \ln \left( \frac{ {\tilde  {\tilde W}}_{x ,y}}{  {\tilde W}^{[k]}_{x ,y}  } \right)
         - \omega^{[k]} 
   + \sum_x  \nu^{[k]}_x {\tilde  {\tilde W}}_{x ,y}  
    -     \nu^{[k]}_y  
         \right]
\! + \sum_y  \lambda^{[k]}_y \left(   \sum_x {\tilde  {\tilde W}}_{x ,y}  - 1 \right)  
\label{lagrangian2.5chaintiltedt}
\end{eqnarray}
Again it is useful to use the constraints to obtain the more compact form
\begin{eqnarray}
{\cal L}  ( {\tilde  {\tilde W}}_{.,.} ; \rho_.) 
\!  = 
 \omega^{[k]}
  +  \sum_y  \rho_{y} \left[  -   \sum_x  {\tilde  {\tilde W}}_{x ,y}  
       \ln \left( \frac{ {\tilde  {\tilde W}}_{x ,y}}{  {\tilde W}^{[k]}_{x ,y}  e^{ - \omega^{[k]}+  \nu^{[k]}_x- \nu^{[k]}_y } }  \right) 
         \right]
\! + \sum_y  \lambda^{[k]}_y \left(   \sum_x {\tilde  {\tilde W}}_{x ,y}  - 1 \right) 
\label{lagrangian2.5chaintiltedcompact}
\end{eqnarray}
The optimization with respect to the density $\rho_y$ 
\begin{eqnarray}
0 = \frac{ \partial {\cal L}  ( {\tilde  {\tilde W}}_{.,.} ; \rho_.) }{\partial \rho_y} 
=  -  
    \sum_x  {\tilde  {\tilde W}}_{x ,y}  
       \ln \left( \frac{ {\tilde  {\tilde W}}_{x ,y}}{  {\tilde W}^{[k]}_{x ,y}  e^{ - \omega^{[k]}+  \nu^{[k]}_x- \nu^{[k]}_y } }  \right)
\label{lagrangian2.5chaintiltedcompactderirho}
\end{eqnarray}
yields directly that the optimal value of the Lagrangian will simply be given by the Lagrange multiplier $ \omega^{[k]}$
that should thus coincide with the scaled cumulant generating function $\mu(k)$ of Eq. \ref{genek}
\begin{eqnarray}
 {\cal L}^{opt} = \omega^{[k]}  =  \mu(k)  
 \label{mukoptiomegak}
\end{eqnarray}
In addition, 
the solution of Eq. \ref{lagrangian2.5chaintiltedcompactderirho} reads using Eq. \ref{mukoptiomegak}
\begin{eqnarray}
 {\tilde  {\tilde W}}_{x ,y} =  {\tilde W}^{[k]}_{x ,y}  e^{ - \mu(k)  +  \nu^{[k]}_x- \nu^{[k]}_y } 
 \label{Wsoludoubletilde}
\end{eqnarray}
The two last constraints of Eq. \ref{genekrhot} 
\begin{eqnarray}
\rho_{x} && =   \sum_y {\tilde  {\tilde W}}_{x ,y}  \rho_{y} 
= e^{ - \mu(k)+  \nu^{[k]}_x } 
\sum_y {\tilde W}^{[k]}_{x ,y}  e^{ - \nu^{[k]}_y }   \rho_{y} 
\nonumber \\
1 && =   \sum_x {\tilde  {\tilde W}}_{x ,y}  =  
e^{ -\mu(k)- \nu^{[k]}_y }
\sum_x  e^{   \nu^{[k]}_x } {\tilde W}^{[k]}_{x ,y} 
\label{2constr}
\end{eqnarray}
can be rewritten as eigenvalues equations for the tilted operator $ {\tilde W}^{[k]}_{x ,y} $
\begin{eqnarray}
e^{ \mu(k) } \left( e^{-  \nu^{[k]}_x } \rho_{x} \right) && 
= \sum_y {\tilde W}^{[k]}_{x ,y}  \left( e^{ - \nu^{[k]}_y }   \rho_{y} \right)
\nonumber \\
 e^{ \mu(k) } e^{ \nu^{[k]}_y } && =
\sum_x  e^{   \nu^{[k]}_x } {\tilde W}^{[k]}_{x ,y} 
\label{2constreigen}
\end{eqnarray}
where $ e^{ \mu(k) }$ is the eigenvalue associated to the left eigenvector
\begin{eqnarray}
{\tilde l}^{[k]}_y && = e^{ \nu^{[k]}_y } 
\label{lefttdk}
\end{eqnarray}
and to the right eigenvector
\begin{eqnarray}
{\tilde r}^{[k]}_x && = e^{-  \nu^{[k]}_x } \rho_{x}  = \frac{\rho_{x}}{{\tilde l}^{[k]}_x} 
\label{righttdk}
\end{eqnarray}
while the remaining constraint of Eq \ref{genekrhot}
concerning the normalization of the density $\rho_x$ reads
\begin{eqnarray}
1 = \sum_x \rho_{x} = \sum_x {\tilde l}^{[k]}_x{\tilde r}^{[k]}_x 
\label{normatdk}
\end{eqnarray}
So the equivalence with Eqs \ref{Wktiltright} \ref{Wktiltleft} \ref{Wktiltnorma} is complete,
and the matrix of Eq. \ref{Wsoludoubletilde}
\begin{eqnarray}
 {\tilde  {\tilde W}}_{x ,y} = e^{ - \mu(k) } {\tilde l}^{[k]}_x {\tilde W}^{[k]}_{x ,y} \frac{1}{{\tilde l}^{[k]}_y } 
  \label{Wsoludoubletildedoob}
\end{eqnarray}
corresponds to the generator of the conditioned process obtained via the generalization of Doob's h-transform.


\subsubsection{ Analysis via the empirical density of excursions between two consecutive resets }

Any time-additive observable of the position (Eq. \ref{additiveA})
can be rewritten in terms of the empirical density of excursions of Eq. \ref{nexcur}
as
\begin{eqnarray}
 A_T  &&  
=  \frac{1}{T} \sum_{l=0}^{N-1} \sum_{t=t_l+1}^{t=t_{l+1}}  \alpha_{x(t)}  
= \frac{1}{T} \sum_{l=0}^{N-1} \sum_{s=1}^{\tau_l}  \alpha_{y^{(l)}(s) }  
 = \sum_{\tau=1}^{+\infty} \sum_{y( .)} n[\tau ; y(.) ] 
\left[ \sum_{s=1}^{\tau-1}  \alpha_{y(s)} +\alpha_0 \right]  
\label{additiveAexc}
\end{eqnarray}
Similarly, any time-additive observable of the increments (Eq. \ref{additiveB})
can be rewritten as
\begin{eqnarray}
 B_T  &&  
=  \frac{1}{T} \sum_{l=0}^{N-1} \left( \sum_{t=t_l+1}^{t=t_{l+1}-1}  \beta_{x(t),x(t-1)}  + \beta_{x(t_{l+1}),x(t_{l+1}-1)} \right)
 = \frac{1}{T}  \sum_{l=0}^{N-1} \left(\sum_{s=1}^{\tau_l-1}  \beta_{y^{(l)}(s), y^{(l)}(s-1)}  
 + \beta_{0,y^{(l)}(\tau_l-1)}
 \right)
 \nonumber \\ &&
 = \sum_{\tau=1}^{+\infty} \sum_{y(.)} n[\tau ; y(.) ] 
\left[ \sum_{s=1}^{\tau-1}  \beta_{y(s),y(s-1)} + \beta_{0,y(\tau-1) } \right]
\label{additiveBexc}
\end{eqnarray}
So the generating function of Eq. \ref{genek} can be evaluated from
the joint probability of Eqs \ref{probaempifin} \ref{constraints} \ref{rateexc}
as 
\small
\begin{eqnarray}
&& Z_T(k)  = \int dn \int {\cal D} n[.;..] 
P_T ( n[.;..], n )
 e^{ \displaystyle T k \sum_{\tau=1}^{+\infty} \sum_{y(.)} n[\tau ; y(.) ] 
\left[ \sum_{s=1}^{\tau-1} ( \alpha_{y(s)} + \beta_{y(s),y(s-1)} )+\alpha_0+ \beta_{0,y(\tau-1) }\right]  }  
\label{genekexc}
 \\
&& \opsimeq_{T \to +\infty}  
\int dn \int {\cal D} n[.;..] 
\delta \left[ \sum_{\tau=1}^{+\infty} \sum_{y(.)} n[\tau ; y(.) ] - n \right]
 \delta \left[ \sum_{\tau=1}^{+\infty} \tau \sum_{y(.) } n[\tau ; y(.) ] -1 \right]
 \nonumber \\ &&
  e^{ \displaystyle T \left[ -  \sum_{\tau=1}^{+\infty} \sum_{y( .)}   n[\tau ; y(.)  ]  \ln \left( \frac{n[\tau ; y( .)  ]  }{ P^{exc}[ \tau ; y(.)  ]  n  }  \right)  
+  k \sum_{\tau=1}^{+\infty} \sum_{y( .)} n[\tau ; y(.) ] 
\left(  \sum_{s=1}^{\tau-1} ( \alpha_{y(s)} + \beta_{y(s),y(s-1)} )+\alpha_0+ \beta_{0,y(\tau-1) }  \right)
\right]  }
\nonumber
\end{eqnarray}
\normalsize

One can now proceed exactly as in the previous subsection.
In terms of the tilted non-conserved quantity
\begin{eqnarray}
{ \tilde P}^{exc}_k[ \tau ; y(.)  ] \equiv P^{exc}[ \tau ; y(.)  ] 
e^{\displaystyle k \left[ \sum_{s=1}^{\tau-1} ( \alpha_{y(s)} + \beta_{y(s),y(s-1)} )+\alpha_0+ \beta_{0,y(\tau-1) }\right] } 
\label{pexctilted}
\end{eqnarray}
the functional in the exponential of Eq. \ref{genekexc} can be written in the more compact form
\begin{eqnarray}
 Z_T(k)   && \opsimeq_{T \to +\infty}  
\int dn \int {\cal D} n[.;..] 
\delta \left[ \sum_{\tau=1}^{+\infty} \sum_{y(.)} n[\tau ; y(.) ] - n \right]
 \delta \left[ \sum_{\tau=1}^{+\infty} \tau \sum_{y(.) } n[\tau ; y(.) ] -1 \right]
 \nonumber \\ &&
  e^{ \displaystyle - T  \sum_{\tau=1}^{+\infty} \sum_{y(.)} 
   n[\tau ; y(.)  ]
  \ln \left( \frac{n[\tau ; y( .)  ]  }{ { \tilde P}^{exc}_k[ \tau ; y(.)  ]  n  }  \right) 
  }
\label{genet}
\end{eqnarray}

One can then make the change of variables from the excursion density $n[\tau ; y( .)  ]  $
to the excursion probability that would make $ n[\tau ; y( .)  ]$ and $n$ typical
\begin{eqnarray}
{\tilde  {\tilde P}}^{exc} [\tau ; y( .)  ] \equiv \frac{n[\tau ; y( .)  ]}{ n} 
\label{Pdoubletilde}
\end{eqnarray}
to obtain the new expression
\begin{eqnarray}
 Z_T(k)   && \opsimeq_{T \to +\infty}  
\int dn \int {\cal D} {\tilde  {\tilde P}}^{exc}[.;..] 
\delta \left[ \sum_{\tau=1}^{+\infty} \sum_{y(.)} {\tilde  {\tilde P}}^{exc}[\tau ; y(.) ] - 1 \right]
 \delta \left[ n \sum_{\tau=1}^{+\infty} \tau \sum_{y(.) } {\tilde  {\tilde P}}^{exc}[\tau ; y(.) ] -1 \right]
 \nonumber \\ &&
  e^{ \displaystyle - T   n \sum_{\tau=1}^{+\infty} \sum_{y(.)} 
    {\tilde  {\tilde P}}^{exc}[\tau ; y(.)  ]
  \ln \left( \frac{ {\tilde  {\tilde P}}^{exc}[\tau ; y( .)  ]  }{ { \tilde P}^{exc}[ \tau ; y(.)  ]    }  \right) 
  }
\label{genett}
\end{eqnarray}
The physical meaning of the two constraints is that $ {\tilde  {\tilde P}}^{exc}[\tau ; y( .) ]$
should be normalized, while $n$ should correspond to the inverse of the first moment of the duration $\tau$ of excursions.

In order to optimize the functional in the exponential in the presence of the constraints,
it is convenient to introduce the following Lagrangian
with the Lagrange multipliers $(\theta^{[k]},\zeta^{[k]})$
\begin{eqnarray}
{\cal L}  ( {\tilde  {\tilde P}}^{exc}[.;..] ,  n) 
&&  = 
   -  n \sum_{\tau=1}^{+\infty} \sum_{y(.)} 
    {\tilde  {\tilde P}}^{exc}[\tau ; y(.)  ]
  \ln \left( \frac{ {\tilde  {\tilde P}}^{exc}[\tau ; y( .)  ]  }{ { \tilde P}^{exc}[ \tau ; y(.)  ]    }  \right) 
 \nonumber \\  &&
  -  \theta^{[k]}  \left[ \sum_{\tau=1}^{+\infty} \sum_{y( .)}  {\tilde  {\tilde P}}^{exc}[\tau ; y( .) ] -  1 \right]
 -\zeta^{[k]} \left[ n \sum_{\tau=1}^{+\infty} \tau \sum_{y(.) }  {\tilde  {\tilde P}}^{exc}[\tau ; y(.) ] -1 \right]
 \nonumber \\
 &&  = \zeta^{[k]}
   -  n \left[ \sum_{\tau=1}^{+\infty} \sum_{y(.)} 
    {\tilde  {\tilde P}}^{exc}[\tau ; y(.)  ]
  \ln \left( \frac{ {\tilde  {\tilde P}}^{exc}[\tau ; y( .)  ]  }{ { \tilde P}^{exc}[ \tau ; y(.)  ]  e^{ -\tau  \zeta^{[k]}}   }  \right)   \right]
-  \theta^{[k]}  \left( \sum_{\tau=1}^{+\infty} \sum_{y( .)}  {\tilde  {\tilde P}}^{exc}[\tau ; y( .) ] -  1 \right)
\label{lagrangianexcchaintt}
\end{eqnarray}

The optimization with respect to the total density $n$ of excursions
\begin{eqnarray}
0 = - \frac{ \partial {\cal L}  (  {\tilde  {\tilde P}}^{exc}[.;..] ,  n) }{\partial n} 
= \sum_{\tau=1}^{+\infty} \sum_{y(.)} 
    {\tilde  {\tilde P}}^{exc}[\tau ; y(.)  ]
  \ln \left( \frac{ {\tilde  {\tilde P}}^{exc}[\tau ; y( .)  ]  }{ { \tilde P}^{exc}[ \tau ; y(.)  ]  e^{ -\tau  \zeta^{[k]}}   }  \right) 
\label{lagrangianexcchainttcompactderi}
\end{eqnarray}
yields directly that the optimal value of the Lagrangian will simply be given by the Lagrange multiplier $ \zeta^{[k]}$
that should thus coincide with the scaled cumulant generating function $\mu(k)$ of Eq. \ref{genek}
\begin{eqnarray}
 {\cal L}^{opt} = \zeta^{[k]}  =  \mu(k)  
 \label{mukzeta}
\end{eqnarray}
In addition, 
the solution of Eq. \ref{lagrangianexcchainttcompactderi} reads using Eq. \ref{mukzeta}
\begin{eqnarray}
 {\tilde  {\tilde P}}^{exc}[\tau ; y( .)  ] = { \tilde P}^{exc}[ \tau ; y(.)  ]  e^{ -\tau  \mu(k) }
 \label{Psoludoubletilde}
\end{eqnarray}
The two constraints of Eq. \ref{genett}
read in terms of the tilted non-conserved quantity of Eq. \ref{pexctilted}
\begin{eqnarray}
1 && =  \sum_{\tau=1}^{+\infty} \sum_{y( .)}  {\tilde  {\tilde P}}^{exc}[\tau ; y( .) ] 
=  \sum_{\tau=1}^{+\infty} e^{ -\tau  \mu(k) } \left( \sum_{y( .)}  { \tilde P}^{exc}[ \tau ; y(.)  ]  \right)
\nonumber \\
\frac{1}{n} && =  \sum_{\tau=1}^{+\infty}  \sum_{y(.) } \tau  {\tilde  {\tilde P}}^{exc}[\tau ; y(.) ] 
=  \sum_{\tau=1}^{+\infty} \tau e^{ -\tau  \mu(k) } \left( \sum_{y( .)}  { \tilde P}^{exc}[ \tau ; y(.)  ]  \right)
\label{constraintstt}
\end{eqnarray}
So the first condition determines the scaled cumulant generation function $\mu(k)$,
while the second condition determines the optimal density $n$ of excursions.

By consistency with the previous subsection, ${\tilde  {\tilde P}}^{exc}[\tau ; y( .) ]  $
represents the probability of excursions in the conditioned process generated by the matrix ${\tilde  {\tilde W}}$ 
of Eq. \ref{Wsoludoubletildedoob}.
It is now interesting to analyze a specific example in the next section.


\section{  Application to the Sisyphus Random Walk on the half line }

\label{sec_chain1d}

The Sisyphus Random Walk \cite{sisyphus} on the semi-infinite lattice $x=0,1,2,..$ 
corresponds to the matrices (Eq. \ref{markovchainreset})
\begin{eqnarray}
W_{x,y} && = \delta_{x,y+1}
\nonumber \\
W^{reset}_{x,y} && =  W_{x,y} (1-R_y)      +  \delta_{x,0} R_y  
\label{wxy1d}
\end{eqnarray}
so the dynamics of Eq. \ref{markovchain} reads
\begin{eqnarray}
P_x(t+1) =   (1-R_{x-1}) P_{x-1}(t)     +  \delta_{x,0} \left( \sum_{y=0}^{+\infty}  R_y   P_y(t) \right)
\label{markovchainreset1d}
\end{eqnarray}

In order to apply the general formalism described in the previous section,
one should first check that the hypothesis concerning the existence of a steady-state (Eq. \ref{markovchainst})
is satisfied.


\subsection{ Condition on the reset probabilities $R_.$ for the existence of a non-equilibrium steady-state  }

The stationary state $P_x^*$ of the dynamics of Eq. \ref{markovchainreset1d}
\begin{eqnarray}
P_x^* =   (1-R_{x-1}) P_{x-1}^*     +  \delta_{x,0} \sum_{y=0}^{+\infty} R_y   P_y^*
\label{markovchainresetst1d}
\end{eqnarray}
follows the simple recurrence for $x \geq 1$
\begin{eqnarray}
P_x^* =   (1-R_{x-1}) P_{x-1}^*  =  (1-R_{x-1}) (1-R_{x-2}) P_{x-2}^*  = ... =  \left[ \prod_{y=0}^{x-1} (1-R_y) \right] P_0^*
\label{markovchainresetrec1d}
\end{eqnarray}
while for $x=0$ Eq. \ref{markovchainreset} becomes
\begin{eqnarray}
P_0^* =  \sum_{x=0}^{+\infty} R_x   P_x^* = R_0 P_0^* + \sum_{x=1}^{+\infty} R_x   \left[ \prod_{y=0}^{x-1} (1-R_y) \right] P_0^*
\label{markovchainreset01d}
\end{eqnarray}
so $P_0^* $ disappears if it is not vanishing, and the remaining equation is just the normalization 
concerning the probability of the first reset
\begin{eqnarray}
1 =   R_0  +\sum_{x=1}^{+\infty} R_x   \left[ \prod_{y=0}^{x-1} (1-R_y) \right] 
= R_0  +  R_1    (1-R_0) + R_2    (1-R_1) (1-R_0)
+ ...
\label{markovchainresetnormaR}
\end{eqnarray}
The steady-state $P_0^*$ at the origin $x=0$ is thus determined by the normalization
\begin{eqnarray}
1=  \sum_{x=0}^{+\infty}   P_x^* =  P_0^*  \left( 1+ \sum_{x=1}^{+\infty}    \left[ \prod_{y=0}^{x-1} (1-R_y) \right] \right)
\label{markovchainresetnormaorigin}
\end{eqnarray}
The condition $P_0^*>0$ to produce a non-equilibrium steady-state localized around the origin
corresponds to the requirement of convergence for the series involving the reset probabilities $R_y$
\begin{eqnarray}
  \sum_{x=1}^{+\infty}    \left[ \prod_{y=0}^{x-1} (1-R_y) \right] < +\infty
\label{cvsteady1d}
\end{eqnarray}
In the uniform case $R_y=R$, this series converges for any value $0<R \leq 1$,
i.e. the only case of divergence corresponds to $R=0$, where the model without any reset is of course transient (Eq. \ref{wxy1d}).
In many applications, $R_y$ depends on the position $y$ either deterministically or randomly \cite{miles,daly_fire,daly_rainbis,sisyphus,optimal,pinsky,path} 
and it is thus interesting to discuss the criterion of Eq. \ref{cvsteady1d}
for these cases :

(i) Random cases :  if the reset probabilities $R_y \in [0,1]$ are independent random variables,
the specific structure of Eq. \ref{cvsteady1d} corresponds to 
the well-known class of Kesten random variables that appear in many disordered systems
\cite{Kesten,Der_Pom,Bou,Der_Hil,Cal,strong_review,c_microcano,c_watermelon,c_mblcayley}.
The condition of convergence for Eq. \ref{cvsteady1d}
is $ \overline{ \ln (1-R_y ) } <0$ and will be thus satisfied, 
since the random variable $R_y$ represents a positive reset probability $0 \leq R_y \leq 1$.

(ii) Deterministic cases : the series of Eq. \ref{cvsteady1d} will generically converge, except when
$R_y$ decays towards zero too rapidly for $y \to +\infty$,
for instance for the choice $R_y=\frac{1}{y+1}$ that leads to the logarithmic divergence of Eq. \ref{cvsteady1d}.

In the following, we assume the convergence of Eq. \ref{cvsteady1d}
in order to have a non-equilibrium steady-state that can be analyzed
via the large deviations at Level 2.5.


\subsection{  Large deviations at level 2.5 for the joint probability of the empirical density and the empirical currents }

The empirical 2-point density of Eq. \ref{rho2pt} contains two types of contributions,
namely the non-local reset currents from any point $x \geq 0$ 
towards the origin
\begin{eqnarray}
J_x && \equiv  \rho^{(2)}_{0,x}   
\label{Jreset1d}
\end{eqnarray}
and the local currents arriving at any point $x \geq 1$ from its left neighbor $(x-1)$
\begin{eqnarray}
j_x && \equiv  \rho^{(2)}_{x,x-1}   
\label{jlocal1d}
\end{eqnarray}
As a consequence, the consistency constraints of Eq. \ref{rho1pt} involving the 1-point density become
for the origin
\begin{eqnarray}
 \rho_0 =  J_0 + \sum_{x=1}^{+\infty}   J_{x} = J_0 + j_1
\label{rho1pt1dorigin}
\end{eqnarray}
and for the other points $x \geq 1$
\begin{eqnarray}
 \rho_x =   j_x = J_x + j_{x+1}
\label{rho1pt1d}
\end{eqnarray}

The large deviations at level 2.5 of Eqs \ref{proba2.5chain} \ref{constraints2.5chain} \ref{rate2.5chain} 
yield that the joint probability to see the empirical density $\rho_.$ and the empirical currents $j_.$ and $J_.$ read
\begin{eqnarray}
P_T ( \rho_. ,j_.,J_. )
 && \opsimeq_{T \to +\infty}  
\delta \left( \sum_{x=0}^{+\infty} \rho_{x} - 1 \right) 
\delta \left(  \sum_{x=0}^{+\infty}   J_{x} -\rho_0  \right)
\delta \left(  J_0 + j_1 -\rho_0 \right)
 \prod_{x=1}^{+\infty}  \left[  \delta \left(   j_x - \rho_x   \right)
\delta \left(  J_x + j_{x+1} -  \rho_x  \right) \right]
\nonumber \\
&& e^{ - \displaystyle T 
\left[   \sum_{x=1}^{+\infty}  j_x
 \ln \left( \frac{ j_x }{   (1-R_{x-1})  \rho_{x-1}}  \right)
+   \sum_{x=0}^{+\infty}  J_x  \ln \left( \frac{ J_x }{  R_x  \rho_x }  \right) \right] } 
\label{proba2.5chain1d}
\end{eqnarray}
The fourth constraint on the first line shows that all the local currents $j_x$ for $x \geq 1$
can be eliminated in terms of the density
\begin{eqnarray}
 j_x =\rho_x 
\label{elimj}
\end{eqnarray}
so that Eq. \ref{proba2.5chain1d} becomes for the joint distribution of the density $\rho_.$
and the non-local reset currents $J_.$
\begin{eqnarray}
P_T ( \rho_. ,J_. )
 && \opsimeq_{T \to +\infty}  
\delta \left( \sum_{x=0}^{+\infty} \rho_{x} - 1 \right) 
\delta \left(  \sum_{x=0}^{+\infty}   J_{x} -\rho_0  \right)
 \prod_{x=0}^{+\infty} 
\delta \left(  J_x + \rho_{x+1} -  \rho_x  \right) 
\nonumber \\
&& e^{ - \displaystyle T 
\left[   \sum_{x=1}^{+\infty}  \rho_x
 \ln \left( \frac{ \rho_x }{   (1-R_{x-1})  \rho_{x-1}}  \right)
+   \sum_{x=0}^{+\infty}  J_x  \ln \left( \frac{ J_x }{  R_x  \rho_x }  \right) \right] } 
 \label{proba2.5chain1dsansj}
\end{eqnarray}
One can now further use the constraints to obtain the large deviations for the empirical density alone 
or for the non-local reset currents alone, as described in the following two subsections.


\subsection { Large deviations at Level 2 for the empirical density $\rho_.$ alone }

The last constraint on the first line of Eq. \ref{proba2.5chain1dsansj}
can be used to eliminate the non-local reset currents $J_x$ 
in terms of the empirical density
\begin{eqnarray}
 J_x =   \rho_x -  \rho_{x+1}
\label{bigJ1d}
\end{eqnarray}
Then the second constraint on the first line of Eq. \ref{proba2.5chain1dsansj}
is automatically satisfied
\begin{eqnarray}
\sum_{x=0}^{+\infty} J_x =  \sum_{x=0}^{+\infty} ( \rho_x -  \rho_{x+1} ) = \rho_0
\label{bigJ1dsum}
\end{eqnarray}
so that Eq. \ref{proba2.5chain1dsansj}
yields for the large deviations for the empirical density $\rho_.$ alone
\begin{eqnarray}
P_T ( \rho_.  )
  \opsimeq_{T \to +\infty}  
\delta \left( \sum_{x=0}^{+\infty} \rho_{x} - 1 \right) 
 e^{ - \displaystyle T 
\left[   \sum_{x=1}^{+\infty}  \rho_x
 \ln \left( \frac{ \rho_x }{   (1-R_{x-1})  \rho_{x-1}}  \right)
+   \sum_{x=0}^{+\infty} ( \rho_x -  \rho_{x+1} )  \ln \left( \frac{ ( \rho_x -  \rho_{x+1} ) }{  R_x  \rho_x }  \right) \right] } 
 \label{proba2.5chain1dsansjrho}
\end{eqnarray}
One may further simplify the rate function in the exponential to obtain
\begin{eqnarray}
 P_T ( \rho_. )
  \opsimeq_{T \to +\infty}  
\delta \left( \sum_{x=0}^{+\infty} \rho_{x} - 1 \right) 
 e^{ - \displaystyle T 
\left[ -\rho_0 \ln (\rho_0) -  \sum_{x=1}^{+\infty}  \rho_x \ln \left(  1-R_{x-1}    \right)
+
 \sum_{x=0}^{+\infty} \left(\rho_x -  \rho_{x+1}  \right) 
 \ln \left( \frac{ \rho_x -  \rho_{x+1}  }{  R_x   }  \right) \right] } 
\label{proba2.5chaintreeonlyrhosimpli1d}
\end{eqnarray}

The fact that the Large deviations at Level 2 for the empirical density $\rho_.$ alone
can be written in closed form in the present model is not generic : indeed in most non-equilibrium models,
it is not possible to use the constraints of the large deviations at Level 2.5 to eliminate
completely the flows in terms of the density alone.


\subsection { Large deviations for the empirical reset currents $J_.$ alone }

If one wishes instead to eliminate the empirical density in terms of the empirical reset currents $J_.$ 
via
\begin{eqnarray}
 \rho_x && =   \sum_{y=x}^{+\infty} J_y
\label{rhobigJ1d}
\end{eqnarray}
Eq. \ref{proba2.5chaintreeonlyrhosimpli1d}
 yields for the large deviations for the empirical reset currents $J_.$ alone
 \small
\begin{eqnarray}
 P_T ( J_. )
&&  \opsimeq_{T \to +\infty} 
\delta \left( \sum_{x=0}^{+\infty} \sum_{y=x}^{+\infty} J_y - 1 \right) 
 \delta \left( \sum_{x=0}^{+\infty}  J_x - \rho_0 \right) 
 e^{ - \displaystyle T 
\left[ -\rho_0 \ln (\rho_0) -  \sum_{x=1}^{+\infty}  \left(\sum_{y=x}^{+\infty} J_y \right)  \ln \left(  1-R_{x-1}    \right)
+
 \sum_{x=0}^{+\infty} J_x 
 \ln \left( \frac{ J_x  }{  R_x   }  \right) \right] } 
\nonumber \\
&&  \opsimeq_{T \to +\infty}  
\delta \left( \sum_{y=0}^{+\infty} (y+1) J_y- 1 \right) 
\delta \left( \sum_{x=0}^{+\infty}  J_x- \rho_0 \right) 
e^{ - \displaystyle T 
\left[ J_0 \ln \left( \frac{  J_0   }{  R_0  \rho_0 }  \right)
+
 \sum_{x=1}^{+\infty}  J_x   \ln \left( \frac{ J_x   }{  R_x    
\left(  \prod_{y=0}^{x-1}  (1-R_{y})    \right)
\rho_0 } \right) \right] } 
\nonumber \\ && 
\label{proba2.5chaintreeonlybigJsimpli1d}
\end{eqnarray}
\normalsize
This expression allows to make the link with the large deviations for excursions between two consecutive  resets
as explained in next subsection.


\subsection{ Large deviations for the empirical density of excursions between two consecutive resets }

Since the internal trajectory $y( s ) =s$ of an excursion is deterministic,
the probability of excursion of Eq. \ref{pinter} only involves its duration $\tau$
\begin{eqnarray}
P^{exc}[ \tau  ] =R_{\tau-1} (1- R_{\tau-2}) ... (1-R_1) (1-R_0)
 =  R_{\tau-1} \left[    \prod_{s=0}^{\tau-2}   (1-R_{s} )\right]
\label{pinter1d}
\end{eqnarray}
As a consequence, 
the probability to see 
 the empirical density $n[\tau ]  $ of excursions between resets
and the total density $n$ follows the large deviation form of Eqs \ref{probaempifin} \ref{rateexc} \ref{constraints}
\begin{eqnarray}
P_T ( n[.], n )  \opsimeq_{T \to +\infty}  
 \delta \left[ \sum_{\tau=1}^{+\infty} \tau  n[\tau ]  -1 \right]
\delta \left[ \sum_{\tau=1}^{+\infty}  n[\tau  ] - n \right]
 e^{ - \displaystyle T \sum_{\tau=1}^{+\infty}   n[\tau ]
  \ln \left( \frac{n[\tau   ]  }{ P^{exc}[ \tau  ]   n  }  \right)   } 
\label{probaexctree1d}
\end{eqnarray}
and coincides with Eq \ref{proba2.5chaintreeonlybigJsimpli1d}
with the dictionary 
\begin{eqnarray}
\rho_0 && = n
\nonumber \\
J_x && = n[\tau=x+1]
\label{corres}
\end{eqnarray}


\subsection{ Large deviations of general time-additive observables }

In this section, we analyze the large deviations of
 time-additive observables of the form $(A_T+B_T )$ (Eqs \ref{additiveA} \ref{additiveB}
\ref{additiveArho} \ref{additiveBrho2})
via their generating function (Eq \ref{genek})
\begin{eqnarray}
Z_T(k) \equiv    < e^{\displaystyle T k \left( A_T + B_T \right) } >  
&& =  < e^{ \displaystyle  k  \sum_{t=1}^T \left[ \alpha_{x(t)}  +  \beta_{x(t),x(t-1)}  \right] } >
= < e^{\displaystyle T k \left( \sum_x \alpha_x \rho_x + \sum_x \sum_y \beta_{x,y} \rho^{(2)}_{x,y}\right) } >  
\label{genekchainsi}
\end{eqnarray}
In the present model where the empirical 2-point density $ \rho^{(2)}_{x,y}$ involves the two contributions
of Eqs \ref{Jreset1d}
and \ref{jlocal1d},
the generating function of Eq. \ref{genekchainsi}
reads more explicitly
in terms of the local currents $j_x= \rho^{(2)}_{x,x-1} $ and of the non-local reset currents $J_x=\rho^{(2)}_{0,x}$
\begin{eqnarray}
Z_T(k) 
= < e^{\displaystyle T k \left( \sum_{x=0}^{+\infty}  \alpha_x \rho_x + \sum_{x=1}^{+\infty}  \beta_{x,x-1} j_x
 + \sum_{x=0}^{+\infty}  \beta_{0,x} J_x\right) } >  
\label{genekchainsij}
\end{eqnarray}

As explained in detail above, the constraints existing in the present model
between the density $\rho_x$, the local currents $j_x$
and the non-local reset currents $J_x$ could be used to eliminate some variables in terms of others
via various choices.
This means that there will be some redundancy between the three functions $(\alpha_x,\beta_{x,x-1},\beta_{0,x})$.
However, in order to understand more clearly how the general formalism of the previous section works,
it is more pedagogical to keep these redundant notations and to see what combinations of the three functions
$(\alpha_x,\beta_{x,x-1},\beta_{0,x})$ naturally emerge in the solutions.


\subsubsection{ Analysis via the tilted matrix }

As explained in detail in section \ref{sec_additivechain}, 
the two first points of views both lead to the analysis of the tilted matrix of Eq. \ref{Wktilt}
for the model of Eq. \ref{wxy1d}
\begin{eqnarray}
 {\tilde W}^{[k]}_{x ,y} 
 = \left[ \delta_{x,y+1} (1-R_y)      +  \delta_{x,0} R_y  \right] e^{ k \left(\alpha_{x} + \beta_{x,y}  \right) }
 = \delta_{x,y+1} (1-R_y)  e^{ k \left(\alpha_{x} + \beta_{x,y}  \right) }    
 +  \delta_{x,0} R_y e^{ k \left(\alpha_0 + \beta_{0,y}  \right) }
\label{Wktiltsi}
\end{eqnarray}
The eigenvalues Eqs \ref{Wktiltright} and \ref{Wktiltleft}
for the right eigenvector ${\tilde r}^{[k]}_x $
and the left eigenvector ${\tilde l}^{[k]}_y  $
read
\begin{eqnarray}
e^{ \mu(k) }  {\tilde r}^{[k]}_x = \sum_y  {\tilde W}^{[k]}_{x ,y}  {\tilde r}^{[k]}_y
=  (1-R_{x-1})  e^{ k \left(\alpha_{x} + \beta_{x,x-1}  \right) }           {\tilde r}^{[k]}_{x-1}
+   \delta_{x,0} \sum_y   R_y e^{ k \left(\alpha_0 + \beta_{0,y}  \right) }            {\tilde r}^{[k]}_y
 \label{Wktiltrightsi}
\end{eqnarray}
and 
\begin{eqnarray}
e^{ \mu(k) }  {\tilde l}^{[k]}_y = \sum_x  {\tilde l}^{[k]}_x {\tilde W}^{[k]}_{x ,y}  
=   {\tilde l}^{[k]}_{y+1}  (1-R_y)  e^{ k \left(\alpha_{y+1} + \beta_{y+1,y}  \right) } 
+    {\tilde l}^{[k]}_0   R_y e^{ k \left(\alpha_0 + \beta_{0,y}  \right) }
 \label{Wktiltleftsi}
\end{eqnarray}
For $x>0$, Eq. \ref{Wktiltrightsi} is a simple recurrence with the solution
\begin{eqnarray}
  {\tilde r}^{[k]}_x
&& =  (1-R_{x-1})  e^{- \mu(k) +  k \left(\alpha_{x} + \beta_{x,x-1}  \right) }   {\tilde r}^{[k]}_{x-1}
=  \left[ \prod_{z=0}^{x-1} (1-R_z) e^{ - \mu(k) +  k \left(\alpha_{z+1} + \beta_{z+1,z}  \right) }\right] {\tilde r}^{[k]}_0
\nonumber \\
&& 
=  
e^{ \displaystyle - x \mu(k) + k \sum_{z=1}^{x}\left(\alpha_{z} + \beta_{z,z-1}  \right) } 
\left[ \prod_{z=0}^{x-1} (1-R_z) \right]
{\tilde r}^{[k]}_0
 \label{Wktiltrightrec}
\end{eqnarray}
while Eq. \ref{Wktiltright}
for the origin yields
\begin{eqnarray}
e^{ \mu(k) }  {\tilde r}^{[k]}_0 && =
 \sum_{x=0}^{+\infty}    R_x e^{ k \left(\alpha_0 + \beta_{0,x}  \right) }            {\tilde r}^{[k]}_x
\nonumber \\
&&  = R_0 e^{ k \left(\alpha_0 + \beta_{0,0}  \right) }            {\tilde r}^{[k]}_0
+  \sum_{x=1}^{+\infty}   
  e^{ \displaystyle - x \mu(k) +k \left(\alpha_0 + \beta_{0,x}  \right)
  + k \sum_{z=1}^{x}\left(\alpha_{z} + \beta_{z,z-1}  \right) } 
\left[ R_x \prod_{z=0}^{x-1} (1-R_z) \right]
{\tilde r}^{[k]}_0
 \label{Wktiltrightzero}
\end{eqnarray}
So the amplitude ${\tilde r}^{[k]}_0 $ of the right eigenvector at the origin
disappears and the remaining equation determines the eigenvalue $e^{ \mu(k) } $ 
in terms of the resetting probabilities $R_y$ and in terms of the functions $\alpha_.$ and $\beta_{.,.}$
that parametrize the time-additive observable under study
\begin{eqnarray}
1
  = R_0 e^{ - \mu(k) + k \left(\alpha_0 + \beta_{0,0}  \right) }      
+  \sum_{x=1}^{+\infty}   
  e^{ \displaystyle - (x+1) \mu(k) +k \left(\alpha_0 + \beta_{0,x}  \right)
  + k \sum_{z=1}^{x}\left(\alpha_{z} + \beta_{z,z-1}  \right) } 
 R_x \prod_{z=0}^{x-1} (1-R_z) 
 \label{Wktiltrightmu}
\end{eqnarray}
The solution of Eq. \ref{Wktiltleftsi} for the left eigenvector reads
\begin{eqnarray}
  {\tilde l}^{[k]}_y = 
  {\tilde l}^{[k]}_0  \left[ R_y e^{ - \mu(k) + k \left(\alpha_0 + \beta_{0,y}  \right) }
+ \sum_{x=y+1}^{+\infty}  e^{\displaystyle - (x+1-y) \mu(k) + k \left(\alpha_0 + \beta_{0,x} \right)
+ k \sum_{z=y+1}^{x}\left(\alpha_{z} + \beta_{z,z-1}  \right) }
 R_x \prod_{z=y}^{x-1} (1-R_z) 
\right]
 \label{Wktiltleftsolu}
\end{eqnarray}
where the consistency for $y=0$ reproduces the equation \ref{Wktiltrightmu} for the eigenvalue $\mu(k)$.
The matrix generating the conditioned process obtained via the generalization of Doob's h-transform (Eq. \ref{Wsoludoubletildedoob}) involves the same non-vanishing matrix elements as the initial matrix $W^{reset}_{x,y}$ of Eq. \ref{wxy1d}
\begin{eqnarray}
 {\tilde  {\tilde W}}_{x ,y} && = e^{ - \mu(k) } {\tilde l}^{[k]}_x {\tilde W}^{[k]}_{x ,y} \frac{1}{{\tilde l}^{[k]}_y } 
 =  \delta_{x,y+1}  {\tilde  {\tilde W}}_{x ,x-1} 
  +   \delta_{x,0}     {\tilde  {\tilde W}}_{0 ,y} 
  \label{Wsoludoubletildedoobch}
\end{eqnarray}
where the probability to jump from $(x-1)$ to $x$
has changed from its initial value $(1-R_y)$ of Eq. \ref{wxy1d}
to its new value
\begin{eqnarray}
 {\tilde  {\tilde W}}_{x ,x-1}  = 
(1-R_{x-1})  e^{ - \mu(k) + k \left(\alpha_{x} + \beta_{x,x-1}  \right) }    \frac{{\tilde l}^{[k]}_x}{{\tilde l}^{[k]}_{x-1 } }
   \label{Wxx1eff}
\end{eqnarray}
while the probability to have a reset from $y$ to the origin
has changed from its initial value $R_y$ of Eq. \ref{wxy1d}
to its new value
\begin{eqnarray}
 {\tilde  {\tilde W}}_{0 ,y}  =    R_y e^{ - \mu(k) +k \left(\alpha_0 + \beta_{0,y}  \right) }
 \frac{{\tilde l}^{[k]}_0}{{\tilde l}^{[k]}_y }
   \label{W0xeff}
\end{eqnarray}
Both involve explicitly the eigenvalue $e^{\mu(k)}$ and the left eigenvector ${\tilde l}^{[k]}_. $.

The stationary density (Eq. \ref{righttdk}) of this conditioned process reads
\begin{eqnarray}
 {\tilde  {\tilde \rho}}_{x }  =  {\tilde l}^{[k]}_x  {\tilde r}^{[k]}_x 
 =
  {\tilde l}^{[k]}_0  {\tilde r}^{[k]}_0
  \left[    \sum_{y=x}^{+\infty}  e^{\displaystyle - (y+1) \mu(k) + k \left(\alpha_0 + \beta_{0,y} \right)
+ k \sum_{z=1}^{y}\left(\alpha_{z} + \beta_{z,z-1}  \right) }
 R_y \prod_{z=0}^{y-1} (1-R_z) 
\right]
   \label{rhott}
\end{eqnarray}
where the normalization determines the value of the density at the origin
${\tilde  {\tilde \rho}}_{0 }= {\tilde l}^{[k]}_0  {\tilde r}^{[k]}_0$ 
\begin{eqnarray}
1= \sum_{x=0}^{+\infty}  {\tilde  {\tilde \rho}}_{x } 
&&  =
  {\tilde l}^{[k]}_0  {\tilde r}^{[k]}_0
  \left[  \sum_{x=0}^{+\infty}    \sum_{y=x}^{+\infty}  e^{\displaystyle - (y+1) \mu(k) + k \left(\alpha_0 + \beta_{0,y} \right)
+ k \sum_{z=1}^{y}\left(\alpha_{z} + \beta_{z,z-1}  \right) }
 R_y \prod_{z=0}^{y-1} (1-R_z) 
\right]
\nonumber \\
&& =
  {\tilde l}^{[k]}_0  {\tilde r}^{[k]}_0
  \left[     \sum_{y=0}^{+\infty}  (y+1) e^{\displaystyle - (y+1) \mu(k) + k \left(\alpha_0 + \beta_{0,y} \right)
+ k \sum_{z=1}^{y}\left(\alpha_{z} + \beta_{z,z-1}  \right) }
 R_y \prod_{z=0}^{y-1} (1-R_z) 
\right]
  \label{normarhott}
\end{eqnarray}
The stationary local currents and reset currents of this conditioned process are then given by
\begin{eqnarray}
{\tilde  {\tilde j}}_x && =  {\tilde  {\tilde W}}_{x ,x-1} {\tilde  {\tilde \rho}}_{x-1 } 
=(1-R_{x-1})  e^{ - \mu(k) + k \left(\alpha_{x} + \beta_{x,x-1}  \right) }    {\tilde l}^{[k]}_x {\tilde r}^{[k]}_{x-1 } 
\nonumber \\
{\tilde  {\tilde J}}_y && =  {\tilde  {\tilde W}}_{0 ,y} {\tilde  {\tilde \rho}}_{y } 
=   R_y e^{ - \mu(k) +k \left(\alpha_0 + \beta_{0,y}  \right) }
 {\tilde l}^{[k]}_0  {\tilde r}^{[k]}_y 
\label{jlocalJdoob}
\end{eqnarray}


\subsubsection{ Analysis via the tilted excursions }

Here one needs to consider the tilted quantity of Eq. \ref{pexctilted}
for the present model where the internal trajectory $y( s ) =s$ of an excursion is deterministic
 (Eq. \ref{pinter1d})
 \small
\begin{eqnarray}
{ \tilde P}^{exc}_k[ \tau  ] = P^{exc}[ \tau ]
e^{\displaystyle k \left[ \sum_{s=1}^{\tau-1} ( \alpha_{s} + \beta_{s,s-1} )+\alpha_0+ \beta_{0,\tau-1 }\right] } 
 = 
 \left[  R_{\tau-1}  \prod_{s=0}^{\tau-2}   (1-R_{s} )\right]
e^{\displaystyle k \left[ \sum_{s=1}^{\tau-1} ( \alpha_{s} + \beta_{s,s-1} )+\alpha_0+ \beta_{0,\tau-1 }\right] } 
\label{pexctiltedsi}
\end{eqnarray}
\normalsize
The conditioned excursion probability of Eq. \ref{Psoludoubletilde} reads
\begin{eqnarray}
 {\tilde  {\tilde P}}^{exc}[\tau] && = { \tilde P}^{exc}[ \tau ]  e^{ -\tau  \mu(k) }
 =  P^{exc}[ \tau ]
e^{\displaystyle  -\tau  \mu(k) + k \left[ \sum_{s=1}^{\tau-1} ( \alpha_{s} + \beta_{s,s-1} )+\alpha_0+ \beta_{0,\tau-1 }\right] } 
 \nonumber \\ &&
 = 
 \left[  R_{\tau-1}  \prod_{s=0}^{\tau-2}   (1-R_{s} )\right]
e^{\displaystyle  -\tau  \mu(k) +k \left[ \sum_{s=1}^{\tau-1} ( \alpha_{s} + \beta_{s,s-1} )+\alpha_0+ \beta_{0,\tau-1 }\right] } 
 \label{Psoludoubletildesi}
\end{eqnarray}
Its normalization (Eq. \ref{constraintstt}) determines
the scaled cumulant generating function $\mu(k)$  
\small
\begin{eqnarray}
1  && =  \sum_{\tau=1}^{+\infty}  {\tilde  {\tilde P}}^{exc}[\tau] 
 \nonumber \\ &&
 = R_0   e^{  -  \mu(k) k \left[\alpha_0+ \beta_{0,\tau-1 }\right] } 
+ \sum_{\tau=2}^{+\infty} 
e^{\displaystyle -\tau  \mu(k) + k \left[ \sum_{s=1}^{\tau-1} ( \alpha_{s} + \beta_{s,s-1} )+\alpha_0+ \beta_{0,\tau-1 }\right] } 
 \left[  R_{\tau-1}  \prod_{s=0}^{\tau-2}   (1-R_{s} )\right]
\label{constraintsttsi}
\end{eqnarray}
\normalsize
which is equivalent to Eq. \ref{Wktiltrightmu} via the correspondence $x=\tau-1$,
while its first moment (Eq. \ref{constraintstt})
determines the inverse of the density $n$ of excursions of the conditioned process
\begin{eqnarray}
\frac{1}{n} && =  
 \sum_{\tau=1}^{+\infty} \tau {\tilde  {\tilde P}}^{exc}[\tau] 
 \nonumber \\ &&
 = R_0   e^{  -  \mu(k) k \left[\alpha_0+ \beta_{0,\tau-1 }\right] } 
+ \sum_{\tau=2}^{+\infty} \tau
e^{\displaystyle -\tau  \mu(k) + k \left[ \sum_{s=1}^{\tau-1} ( \alpha_{s} + \beta_{s,s-1} )+\alpha_0+ \beta_{0,\tau-1 }\right] } 
 \left[  R_{\tau-1}  \prod_{s=0}^{\tau-2}   (1-R_{s} )\right]
\label{constraintsttn}
\end{eqnarray}
This equation is equivalent to Eq. \ref{normarhott} for the density at the origin
${\tilde  {\tilde \rho}}_{0 }= {\tilde l}^{[k]}_0  {\tilde r}^{[k]}_0$ 
in agreement with the dictionary ${\tilde  {\tilde \rho}}_{0 } = n $ and $\tau=y+1$ of Eq. \ref{corres}.


\section{ Markov Jump Process in continuous time and discrete space with resets }

\label{sec_jump}

\subsection{ Models and notations }

In this section, we consider the continuous-time dynamics in discrete space defined by the Master Equation
\begin{eqnarray}
\frac{\partial P_x(t)}{\partial t} =    \sum_{y }   w^{reset}_{x,y}  P_y(t) 
\label{mastereq}
\end{eqnarray}
where the off-diagonal matrix elements $x \ne y$ represent the transitions rates from $y$ to $x $
\begin{eqnarray}
w^{reset}_{x,y}  && = w_{x,y} +  \delta_{x,0}   r_y \ \ \  \ \ \ {\rm for } \ \ x \ne y
\label{woffdiag}
\end{eqnarray}
The physical meaning is that
the resetting procedure takes place with the resetting rate $r_{y}$ from $y $ towards the origin $0$,
while the Markov Jump Process that would exist without resetting
is described by the off-diagonal matrix elements $w_{x,y} $
and by the corresponding diagonal elements that are fixed by the conservation of probability to be
\begin{eqnarray}
w_{y,y}  && =  - \sum_{x \ne y} w_{x,y} \equiv - w^{out}_y
\label{wdiag}
\end{eqnarray}
where the notation $w^{out}_y $ represents the total escape rate out of the position $y$.
The diagonal matrix elements of the matrix $ w^{reset}$ of Eq. \ref{woffdiag}
are also fixed by the conservation of probability in terms of the off-diagonal elements
\begin{eqnarray}
w^{reset}_{y,y}  && = - \sum_{x \ne y} w^{reset}_{x,y} = - \sum_{x \ne y} w_{x,y}-   r_y  + r_0 \delta_{y,0}
= - (w^{out}_y +   r_y )  + r_0 \delta_{y,0}
\label{wdiagreset}
\end{eqnarray}
Note that the reset rate $r_0$ from the origin to itself disappears from the diagonal matrix element
\begin{eqnarray}
w^{reset}_{0,0}  && =   - (w^{out}_0 +   r_0) + r_0 = - w^{out}_0 
\label{wdiagzero}
\end{eqnarray}
 since it does not produce a change of position.
 However the rate $r_0$ is important for the book-keeping of the number of resets.
For instance, if one chooses $r_0=0$ in order to suppress the possibility of resets from the origin to itself,
the whole statistics of resets will be coupled to the time spent at the origin.
As a consequence, it is useful to keep the possibility of $r_0>0$ to remain more general,
as we have also considered the possibility of positive reset probability
$R_0>0$ from the origin to itself in the Markov Chain framework described in the previous sections.
As a consequence, we will consider that the diagonal term of Eq. \ref{wdiagzero}
contains the two contributions 'out' and 'in'
\begin{eqnarray}
w^{reset}_{0,0}  && =   - w^{reset(out)}_{0} + w^{reset(in)}_{0,0}
\nonumber \\
 w^{reset(out)}_{0} && \equiv w^{out}_0 +   r_0
 \nonumber \\
 w^{reset(in)}_{0,0} && \equiv    r_0
\label{wdiagzeroinout}
\end{eqnarray}
In summary, the Master Equation \ref{mastereq} can be rewritten more explicitly as
\begin{eqnarray}
\frac{\partial P_x(t)}{\partial t} =  -   \left[ w^{out}_{x} +r_x \right] P_x(t)
+  \sum_{y \ne x}  \left[ w_{x,y} + \delta_{x,0} r_y  \right] P_y(t)      +  \delta_{x,0}  r_0  P_0(t)
\label{mastereqexpli}
\end{eqnarray}
Again as in Eq. \ref{markovchainst}, we will assume that the steady-state solution $P^*_x$ of Eq. \ref{mastereq}
\begin{eqnarray}
0 =    \sum_{y }   w^{reset}_{x,y}  P_y^* 
\label{mastereqst}
\end{eqnarray}
exists in order to apply the large deviations at Level 2.5 for non-equilibrium steady-states.


\subsection{ Large deviations at level 2.5 for the empirical density and the empirical flows  }

 For this continuous-time jump process, the empirical density reads
\begin{eqnarray}
 \rho_{x} && \equiv \frac{1}{T} \int_0^T dt \  \delta_{x(t),x}  
 \label{rho1pj}
\end{eqnarray}
and satisfies the normalization
\begin{eqnarray}
\sum_x \rho_{x} && = 1
\label{rho1ptnormaj}
\end{eqnarray}
while the jump density from $y$ to $x \ne y$
\begin{eqnarray}
q_{x,y} \equiv  \frac{1}{T} \sum_{t : x(t^-) \ne x(t^+)} \delta_{x(t^+),x} \delta_{x(t^-),y} 
\label{jumpempiricaldensity}
\end{eqnarray}
satisfies the following stationarity constraint (for any $x$, the total incoming flow should be equal to the total outgoing flow)
\begin{eqnarray}
\sum_{y \ne x} q_{x,y} = \sum_{y \ne x} q_{y,x}
\label{contrainteq}
\end{eqnarray}
As explained around Eq. \ref{wdiagzero}, we will also need to introduce 
the density of resets from the origin to itself
\begin{eqnarray}
 q_{0,0} = \frac{\textrm{ Number of resets from 0 to 0 during $[0,T]$} }{T}
\label{q00}
\end{eqnarray}
that will also lead to Eq \ref{contrainteq} for $x=0$ since $q_{0,0}$ will appear on both sides and thus disappear.

The joint probability distribution of the empirical density $\rho_.$ and flows $q_{.,.}$
satisfy the following
large deviation form at level 2.5 
\cite{fortelle_thesis,fortelle_jump,maes_canonical,maes_onandbeyond,wynants_thesis,chetrite_formal,BFG1,BFG2,chetrite_HDR,c_ring,c_interactions,c_open,barato_periodic,chetrite_periodic}
\begin{eqnarray}
P_{T}[ \rho_. ; q_{.,.} ] \oppropto_{T \to +\infty} C[ \rho_. ; q_{.,.} ] e^{- T I[ \rho_. ; q_{.,.} ] }
\label{level2.5master}
\end{eqnarray}
with the constraints discussed in Eqs \ref{rho1ptnormaj} and \ref{contrainteq}
\begin{eqnarray}
&& C [ \rho_. ; q_{.,.} ]
  = \delta \left( \sum_x \rho_{x} - 1 \right) 
 \prod_x \left[ \sum_{y \ne x} q_{x,y} - \sum_{y \ne x} q_{y,x}  \right]
\label{constraints2.5master}
\end{eqnarray}
while the rate function involves the off-diagonal rates $w^{reset}_{x,y} = w_{x,y} +  \delta_{x,0}   r_y $
of Eq. \ref{woffdiag}
as well as the special contribution from the resets from the origin to itself with the rate $r_0$ (Eq \ref{q00})
\begin{eqnarray}
I[ \rho_. ; q_{.,.} ]=  \sum_{y } \sum_{x \ne y} 
\left[ q_{x,y}  \ln \left( \frac{ q_{x,y}  }{  w^{reset}_{x,y}  \rho_y }  \right) 
 - q_{x,y}  + w^{reset}_{x,y}  \rho_y  \right]
 + \left[ q_{0,0}  \ln \left( \frac{ q_{0,0}  }{  r_0  \rho_0 }  \right) 
 - q_{0,0}  + r_0  \rho_0 \right]
\label{rate2.5master}
\end{eqnarray}
Again one needs to discuss the possible empirical flows $q_{x,y} $ due to the 
local rates $w_{x,y} $ and to the resetting procedure in order to obtain more explicit expressions,
as explained on the specific example of section 
\ref{sec_jump1d}.


\subsection{ Large deviations for the empirical density of excursions between two consecutive resets }

\subsubsection{ Probabilities of excursions between two consecutive resets }

For the dynamics of Eq. \ref{mastereq},
the probability to have an excursion of duration $\tau$ and of internal trajectory $y( 0 \leq s \leq \tau) $
with the initial position $y(0^+)=0$ fixed by the reset
and the final position $y(\tau)$ being the last position before the reset jump to the origin $y(\tau^+) =0$ reads
\begin{eqnarray}
P^{exc}[ \tau ; y( .)  ] 
 = r_{y(\tau)} e^{ - \int_0^{\tau} ds \left[ w^{out}_{y(s)} + r_{y(s)} \right] }
\prod_{s : y(s^-) \ne y(s^+)} w_{y(s^+),y(s^-)}
\label{pinterjump}
\end{eqnarray}

To be more precise, the internal trajectory $y( 0 \leq s \leq \tau) $ will 
contain a certain number $M$ of jumps $m=1,..,M$ occurring at times $s_0=0<s_1<...<s_M<\tau=s_{M+1}$
between the successive positions $(y(0 \leq s <s_1)=0 ; y(s_1 \leq s <s_2)=z_1 ; ... y(s_M \leq s <\tau)=z_M )$ with $z_{m+1} \ne z_m$ that are visited between these jumps. 
The probability of this trajectory reads
\begin{eqnarray}
 P^{exc}[ \tau ;  0; s_1 ; z_1 ,s_2;... z_{M-1} ; s_M ;z_M )  ] 
 = r_{z_M} e^{ - \displaystyle \sum_{m=0}^M (s_{m+1}- s_m)  \left[ w^{out}_{z_m} + r_{z_m} \right] }
 \prod_{m=1}^M w_{z_m,z_{m-1}} 
\label{ptraject}
\end{eqnarray}
while the special case $M=0$ where the reset occurs while the particle is still at the origin reads (see the discussion around Eq. \ref{wdiagzero})
\begin{eqnarray}
 P^{exc}[ \tau ; 0 ]  = r_{0} e^{ - \tau  \left[ w^{out}_{0} + r_{0} \right] }
\label{ptraject0}
\end{eqnarray}

The normalization over all possible excursions reads, first in compact form and then in more explicit form
\begin{eqnarray}
&& 1   =
\int_{0}^{+\infty} d \tau \sum_{y( .)}P^{exc}[ \tau ; y( .)  ] 
\label{normatraj}
 \\
&& = \int_{0}^{+\infty} d \tau  \left[ P^{exc}[ \tau ;0 ] 
+  \sum_{M=1}^{+\infty}  
\int_0^{\tau} ds_M \int_0^{s_M} ds_{M-1} ... \int_0^{s_2} ds_{1} 
\sum_{z_M \ne z_{M-1}}  ...
\sum_{z_2 \ne z_1}  \sum_{z_1 \ne 0} 
  P^{exc}[ \tau ;  0; s_1 ; z_1 ;... ; s_M ;z_M )  ] 
  \right]
\nonumber
\end{eqnarray}


\subsubsection{ Large deviations for the excursions density  }

The probability to see 
 the empirical density $n[\tau ; y( 0 \leq s \leq \tau)]  $ of excursions between resets
and the total density $n$ follows the large deviation form analog to Eq. \ref{probaempifin}
\begin{eqnarray}
P_T ( n[.;..], n ) && \opsimeq_{T \to +\infty}  
\delta \left[ \int_{0}^{+\infty} d \tau \sum_{y( .)} n[\tau ; y( . ) ] 
- n \right]
 \delta \left[ \int_{0}^{+\infty} d \tau \sum_{y( .)} \tau n[\tau ; y( .)]   -1 \right]
 \nonumber \\ &&
  e^{ -  \displaystyle T  \int_{0}^{+\infty} d \tau \sum_{y( .)} n[\tau ; y( .) ]  
  \ln \left( \frac{n[\tau ; y( .)    ]  }{ P^{exc}[ \tau ; y(.)  ]   n  }  \right)   } 
\label{probaempiexcjump}
\end{eqnarray}
where the sums over trajectories can be written in more explicit form as in Eq. \ref{normatraj}.

\subsubsection{ Simplifications for uniform resetting rate $r_y=r$  }

When the resetting rate $r_y=r$ does not depend on the position $y$,
one obtains the factorization analogous to Eq. \ref{pinterfactor}
\begin{eqnarray}
  P^{exc}_{uniform} [ \tau ;(y( . )  ]  =p^{exp}(\tau)   p^{config}_{\tau} [ (y( . )  ]
\label{ptrajectsifactorj}
\end{eqnarray}
The exponential distribution 
\begin{eqnarray}
p^{exp}(\tau) =  r e^{-r \tau} 
\label{pexp}
\end{eqnarray}
describes the probability of the duration $\tau$ of the excursion
independently of the internal trajectory (and replaces the geometric distribution of Eq. \ref{pgeo} of the discrete-time framework),
while the probability $ p^{config}_{\tau}[y(.)] $ of the internal trajectory once its duration $\tau$ is given
characterizes the process without resetting.
In particular, the probability of the total density $n$ of excursions simply follows the Poisson distribution
with its corresponding large deviation form
\begin{eqnarray}
P^{Poisson}_{T}(n) = \frac{(rT)^{nT} }{(n T)!} e^{- r T}
\oppropto_{T \to +\infty} e^{- \displaystyle  T \left[ n \ln \left( \frac{ n }{ r  } \right) -n +r   \right] }
\label{poisson}
\end{eqnarray}
On the contrary, whenever the reset rates $r_y$ depends on the position $y$, 
the duration $\tau$ and the internal trajectory $y( .)  $ are coupled via Eq. \ref{pinterjump}.


\subsection{ Large deviations for general time-additive observables  }

The empirical density
of Eq. \ref{rho1pj}
allows to 
reconstruct any time-additive observable that involves some function $\alpha_x$ of the position $x(t)$  
\begin{eqnarray}
 A_T  && \equiv \frac{1}{T} \int_0^T dt \ \alpha_{x(t)}  = \sum_x \alpha_x \rho_x
\label{additiveAjump}
\end{eqnarray}
while 
the jump density of Eq. \ref{jumpempiricaldensity}
allows to reconstruct any time-additive observable that involves some function $\beta_{x,y}$ 
of the jumps $(x(t^+)\ne x(t^-))$ and of the resets from the origin to itself
\begin{eqnarray}
 B_T && \equiv  \frac{1}{T} \sum_{t: x(t^-) \ne x(t^+) } \beta_{x(t^+),x(t^-)}  
 + \beta_{0,0} \left(  \frac{\textrm{ Number of resets from 0 to 0 during $[0,T]$ } }{T} \right)   
 \nonumber \\
 && =  \sum_y \sum_{ x \ne y } \beta_{x,y} q_{x,y} +  \beta_{0,0} q_{0,0}
\label{additiveBjumpq}
\end{eqnarray}

The large deviations of the sum $(A_T+B_T )$
can be analyzed via the scaled cumulant generating function $\mu(k)$
appearing in the asymptotic behavior
\begin{eqnarray}
Z_T(k) \equiv    < e^{ T k \left( A_T + B_T \right) } >  
=<   e^{ \displaystyle T k  \left(  \sum_x  \rho_x   \alpha_x 
+ \sum_y \sum_{ x \ne y } q_{x,y}   \beta_{x,y}  +  q_{0,0}  \beta_{0,0}  \right) }  >
\opsimeq_{T \to +\infty} e^{T \mu(k) }
\label{genekj}
\end{eqnarray}


\subsubsection{ Analysis via the tilted dynamics }

The probability of a trajectory for the process of Eq. \ref{mastereqexpli}
can be written only in terms of its empirical observables 
\begin{eqnarray}
P^{Traj}_T    \opsimeq_{T \to +\infty}  e^{ \displaystyle T \left[ - \sum_x \rho_x \left[ w^{out}_{x} +r_x \right] 
+  \sum_y \sum_{x \ne y} q_{x,y} \ln  \left[ w_{x,y} + \delta_{x,0} r_y  \right] 
+ q_{0,0} \ln (r_0) \right] }
\label{ptraj}
\end{eqnarray}

As a consequence, the generating function of Eq. \ref{genekj}  
can be analyzed by tilting each term of the initial Master Eq. \ref{mastereqexpli}
to produce the non-conserved dynamics
\begin{eqnarray}
\frac{\partial {\tilde P}_x(t)}{\partial t} && =  -   \left[ w^{out}_{x} +r_x -k \alpha_x\right] {\tilde P}_x(t)
+  \sum_{y \ne x}  \left[ w_{x,y} + \delta_{x,0} r_y  \right] e^{k\beta_{x,y}}{\tilde P}_y(t)  
    +  \delta_{x,0}  r_0  
e^{k\beta_{0,0}} {\tilde P}_0(t)
 \nonumber \\ &&
 \equiv   \sum_{y }   {\tilde w}^{[k]}_{x,y}  {\tilde P}_y(t) 
\label{mastertilte}
\end{eqnarray}
where the tilted matrix involves the off-diagonal elements $x \ne y$
\begin{eqnarray}
{\tilde w}^{[k]}_{x,y} \equiv    \left[  w_{x,y} +  \delta_{x,0}   r_y \right] e^{k\beta_{x,y}}         \ \ \  \ \ \ {\rm for } \ \ x \ne y
 \label{wtiltedoff}
\end{eqnarray}
and the diagonal elements
\begin{eqnarray}
{\tilde w}^{[k]}_{x,x} \equiv    - w^{out}_x -   r_x  + k \alpha_x   +  \delta_{x,0}  r_0  e^{k\beta_{0,0}} 
 \label{wtiltediag}
\end{eqnarray}

The scaled cumulant generating function $\mu(k)$ of Eq. \ref{genek}
will then correspond to the highest eigenvalue $\mu(k)$
of the tilted matrix ${\tilde w}_{.,.} $ that will dominate the propagator
\begin{eqnarray}
{\tilde P}_{x , y}(T)  \opsimeq_{T \to +\infty}  e^{T \mu(k) } {\tilde r}^{[k]}_x {\tilde l}^{[k]}_y 
\label{wjktiltprop}
\end{eqnarray}
where $ {\tilde r}^{[k]}_x$ is the corresponding positive right eigenvector
\begin{eqnarray}
 \mu(k)   {\tilde r}^{[k]}_x = \sum_y  {\tilde w}^{[k]}_{x ,y}  {\tilde r}^{[k]}_y
 =   \left[ - w^{out}_x -   r_x  + k \alpha_x     \right]  {\tilde r}^{[k]}_x
 +  \sum_{y \ne x}     w_{x,y} e^{k\beta_{x,y}}   {\tilde r}^{[k]}_y
 +    \delta_{x,0} \sum_{y }     r_y  e^{k\beta_{0,y}}   {\tilde r}^{[k]}_y
 \label{wktiltright}
\end{eqnarray}
and where ${\tilde l}^{[k]}_y $ is the corresponding positive left eigenvector
\begin{eqnarray}
 \mu(k)   {\tilde l}^{[k]}_y = \sum_x  {\tilde l}^{[k]}_x {\tilde w}^{[k]}_{x ,y}  
 =   {\tilde l}^{[k]}_y   \left[  - w^{out}_y -   r_y  + k \alpha_y   \right]
 +  \sum_{ x \ne y}  {\tilde l}^{[k]}_x    w_{x,y} e^{k\beta_{x,y}} 
 +   {\tilde l}^{[k]}_0     r_y e^{k\beta_{0,y}} 
 \label{wktiltleft}
\end{eqnarray}
with the normalization
\begin{eqnarray}
 \sum_x  {\tilde l}^{[k]}_x   {\tilde r}^{[k]}_x=1
 \label{wktiltnorma}
\end{eqnarray}
The generator of the conditioned process obtained via the generalization of Doob's h-transform reads
\begin{eqnarray}
 {\tilde  {\tilde w}}_{x ,y} =  {\tilde l}^{[k]}_x {\tilde w}^{[k]}_{x ,y} \frac{1}{{\tilde l}^{[k]}_y } - \mu(k) \delta_{x,y}
  \label{Wdoubletildedoobj}
\end{eqnarray}


\subsubsection{ Analysis via the tilted excursions }

As in the discrete time context (Eq. \ref{additiveAexc}), 
one can use the empirical density of excursions
to write
any time-additive observable of the position (Eq. \ref{additiveAjump})
\begin{eqnarray}
 A_T  
= \int_{0}^{+\infty} d\tau  \sum_{y( .)} n[\tau ; y( .) ] 
\left[ \int_{0}^{\tau}  ds \alpha_{y(s)} \right]  
\label{additiveAjumpexc}
\end{eqnarray}
and any time-additive observable of the jumps (Eq. \ref{additiveBjumpq})
\begin{eqnarray}
 B_T  = \int_{0}^{+\infty} d\tau  \sum_{y( .)} n[\tau ; y( .) ] 
\left[ \sum_{ 0<s<\tau : y(s^-) \ne y(s^+) } \beta_{y(s^+),y(s^-)}
+ \beta_{0,y(\tau^-)}
\right]  
\label{additiveBjumpexc}
\end{eqnarray}
where the last term contains the contribution of $\beta_{0,0}$ when 
the excursion has never left the origin $y(\tau^-)=0$.

So the generating function of Eq. \ref{genek} can be evaluated from
the joint probability of Eq. \ref{probaempiexcjump} as 
\small
\begin{eqnarray}
&& Z_T(k)  = \int dn \int {\cal D} n[.;..] 
P_T ( n[.;..], n )
 e^{ \displaystyle T k \int_{0}^{+\infty} d\tau \sum_{y(.)} n[\tau ; y(.) ] 
\left[ 
\int_{0}^{\tau}  ds \alpha_{y(s)}
+ \sum_{ 0<s<\tau : y(s^-) \ne y(s^+) } \beta_{y(s^+),y(s^-)}
+ \beta_{0,y(\tau^-)}
\right]  }  
\nonumber
 \\
&& \opsimeq_{T \to +\infty}  
\int dn \int {\cal D} n[.;..] 
\delta \left[ \int_{0}^{+\infty} d \tau \sum_{y( .)} n[\tau ; y( . ) ] 
- n \right]
 \delta \left[ \int_{0}^{+\infty} d \tau \sum_{y( .)} \tau n[\tau ; y( .)]   -1 \right]
 \nonumber \\ &&
  e^{   \displaystyle T  \int_{0}^{+\infty} d \tau \sum_{y( .)} n[\tau ; y( .) ]  
\left[ -   \ln \left( \frac{n[\tau ; y( .)    ]  }{ P^{exc}[ \tau ; y(.)  ]   n  }  \right)  
+k\int_{0}^{\tau}  ds \alpha_{y(s)}
+ k\sum_{ 0<s<\tau : y(s^-) \ne y(s^+) } \beta_{y(s^+),y(s^-)}
+ k\beta_{0,y(\tau^-) } \right]
   } 
   \label{genekexcj}
\end{eqnarray}
\normalsize
It is thus convenient to introduce 
tilted non-conserved quantity
\begin{eqnarray}
{ \tilde P}^{exc}_k[ \tau ; y(.)  ] \equiv P^{exc}[ \tau ; y(.)  ] 
e^{\displaystyle k \left[  \int_{0}^{\tau}  ds \alpha_{y(s)}
+ \sum_{ 0<s<\tau : y(s^-) \ne y(s^+) } \beta_{y(s^+),y(s^-)}
+ \beta_{0,y(\tau^-) } \right] } 
\label{pexctiltedj}
\end{eqnarray}
and to make the change of variable from $n[\tau ; y( . ) ]  $ to the probability distribution that would make $n[\tau ; y( . ) ]  $ and $n$ typical
\begin{eqnarray}
{\tilde  {\tilde P}}^{exc} [\tau ; y( .)  ] \equiv \frac{n[\tau ; y( .)  ]}{ n} 
\label{Pdoubletildej}
\end{eqnarray}
to rewrite Eq. \ref{genekexcj} as
\begin{eqnarray}
 Z_T(k)  && \opsimeq_{T \to +\infty}  
 \int {\cal D} {\tilde  {\tilde P}}^{exc} [. ; ... ] 
 \int dn
\delta \left[ \int_{0}^{+\infty} d \tau \sum_{y( .)}  {\tilde  {\tilde P}}^{exc}[\tau ; y( . ) ] - 1 \right]
 \delta \left[ n \int_{0}^{+\infty} d \tau \tau \sum_{y( .)}  {\tilde  {\tilde P}}^{exc}[\tau ; y( .)]   -1 \right]
 \nonumber \\ &&
  e^{  -  \displaystyle T  n \int_{0}^{+\infty} d \tau \sum_{y( .)} {\tilde  {\tilde P}}^{exc}[\tau ; y( .) ]  
   \ln \left( \frac{ {\tilde  {\tilde P}}^{exc}[\tau ; y( .)    ]  }{ {\tilde P}^{exc}[ \tau ; y(.)  ]     }  \right)  
   } 
   \label{genekexcjtt}
\end{eqnarray}
Then the analysis is exactly as for \ref{genett} with the same output :
the conditioned distribution of excursions is given by (Eq \ref{Psoludoubletilde})
\begin{eqnarray}
 {\tilde  {\tilde P}}^{exc}[\tau ; y( .)  ] = { \tilde P}^{exc}[ \tau ; y(.)  ]  e^{ -\tau  \mu(k) }
 \label{Psoludoubletildej}
\end{eqnarray}
where $\mu(k)$ is fixed by its normalization (Eq \ref{constraintstt})
\begin{eqnarray}
1 && = \int_{0}^{+\infty} d \tau \sum_{y( .)}  {\tilde  {\tilde P}}^{exc}[\tau ; y( .) ] 
=  \int_{0}^{+\infty} d \tau e^{ -\tau  \mu(k) } \left( \sum_{y( .)}  { \tilde P}^{exc}[ \tau ; y(.)  ]  \right)
\label{constraintstt1}
\end{eqnarray}
while the density $n$ of the conditioned process is fixed by the first moment of the duration $\tau$ (Eq \ref{constraintstt})
\begin{eqnarray}
\frac{1}{n} && =  \int_{0}^{+\infty} d \tau  \sum_{y(.) } \tau  {\tilde  {\tilde P}}^{exc}[\tau ; y(.) ] 
=  \int_{0}^{+\infty} d \tau \tau e^{ -\tau  \mu(k) } \left( \sum_{y( .)}  { \tilde P}^{exc}[ \tau ; y(.)  ]  \right)
\label{constraintstt2}
\end{eqnarray}


\section{ Sisyphus Markov Jump process in continuous time on the half line }

\label{sec_jump1d}

The Sisyphus Markov Jump process defined on the half-line $x=0,1,2,..$
corresponds to the off-diagonal matrix elements $x \ne y$ (Eq. \ref{woffdiag})
\begin{eqnarray}
w_{x,y} && = w \delta_{x,y+1}
\nonumber \\
w^{reset}_{x,y} && =   w \delta_{x,y+1}     +  \delta_{x,0} r_y  
\label{wxy1dj}
\end{eqnarray}
so the dynamics of Eq. \ref{mastereqexpli} reads
\begin{eqnarray}
\frac{\partial P_{x}(t)}{\partial t} && =- \left[ w+r_x \right]P_{x}(t)
+  w   P_{x-1}(t)    \ \ {\rm for } \ \ x \geq 1
\nonumber \\
\frac{\partial P_0(t)}{\partial t}  && =  - \left[ w+r_0 \right] P_0(t) +  \sum_{x=0}^{+\infty} r_x   P_{x}(t) 
= - w  P_0(t) +  \sum_{x=1}^{+\infty} r_x   P_{x}(t) 
\label{treejumpdyn}
\end{eqnarray}

In order to apply the general formalism described in the previous section,
one should first check that the hypothesis concerning the existence of a steady-state (Eq. \ref{mastereqst})
is satisfied.


\subsection{ Condition on the reset rates $r_.$ for the existence of a non-equilibrium steady-state  }

The stationary solution $P^*_x $ of Eq. \ref{treejumpdyn} satisfies a simple recurrence for $x > 0$
\begin{eqnarray}
P_x^* =  \frac{ w }{ w +r_x }  P_{x-1}^*  = ...
= \left[ \prod_{y=1}^{x}  \frac{ w }{ w  +r_{y} }   \right] P_{0}^*
\label{markovintermi}
\end{eqnarray}
while Eq. \ref{treejumpdyn} reads for $x=0$
\begin{eqnarray}
w P_0^*  =  \sum_{x=1}^{+\infty}    r_x    P_{x}^*
=  \sum_{x=1}^{+\infty}    r_x    \left[ \prod_{y=1}^{x}  \frac{ w }{ w  +r_{y} }   \right] P_{0}^* 
\label{markovintermizero}
\end{eqnarray}
so $P_0^*$ disappears if it is not vanishing, and the remaining equation is just an identity.
So $P_0^*$ is determined by the normalization
\begin{eqnarray}
1 = 
   \sum_{x=0}^{+\infty}      P_x^*
= P_0^* \left( 1+  \sum_{x=1}^{+\infty}     \left[ \prod_{y=1}^{x}  \frac{ w }{ w  +r_{y} }   \right]   \right)
\label{markovinterminorma}
\end{eqnarray}
The condition $P_0^*>0$ to produce a non-equilibrium steady-state localized around the origin
corresponds to the requirement of convergence for the series involving the reset rates $r_y$
\begin{eqnarray}
  \sum_{x=1}^{+\infty}   \left[ \prod_{y=1}^{x}  \frac{ w }{ w  +r_{y} }   \right]   < +\infty
\label{cvsteady1dj}
\end{eqnarray}
This criterion has the same form as Eq. \ref{cvsteady1d} if one introduces the notation $\hat R_y \equiv \frac{r_y}{w+r_y}$
so that the discussion following Eq. \ref{cvsteady1d}
concerning the uniform case $r_y=r$, the random cases or the deterministic cases
can be directly translated and will not be repeated here.

In the following, we assume the convergence of Eq. \ref{cvsteady1dj}
in order to have a non-equilibrium steady-state that can be analyzed
via the large deviations at Level 2.5.


\subsection{ Large deviations at level 2.5 for the empirical density and the empirical currents }

The empirical jump density of Eq. \ref{jumpempiricaldensity}
contains two types of contributions,
namely the non-local reset currents from any point $x$ towards the origin
 (even $x=0$ as discussed around Eq. \ref{q00})
\begin{eqnarray}
J_{x} && \equiv  q_{0,x}  
\label{Jreset1dj}
\end{eqnarray}
and the local currents arriving at any point $x \geq 1$ from its left neighbor $(x-1)$
\begin{eqnarray}
j_{x} && \equiv  q_{x,x-1}  
\label{jlocal1dj}
\end{eqnarray}
For any position $x \geq 1$ the constraint of Eq. \ref{contrainteq} reads
\begin{eqnarray}
  j_{x} = J_{x} + j_{x+1}
\label{contrainteqg}
\end{eqnarray}
For the origin $x=0$, the constraint of Eq. \ref{contrainteq}
\begin{eqnarray}
\sum_{x=1}^{+\infty}  J_{x} = j_1
\label{contrainteq0}
\end{eqnarray}
can be recovered by summing Eq. \ref{contrainteqg} over $x=1,2,..$ 
so that it is sufficient to impose Eq. \ref{contrainteqg}.
As a consequence, the large deviation at level 2.5 for the empirical density and the empirical currents of Eqs \ref{level2.5master}
\ref{rate2.5master} \ref{constraints2.5master} become
\begin{eqnarray}
P_{T}[ \rho_. ,j_., J_.] && \oppropto_{T \to +\infty} 
\delta \left( \sum_{x=0}^{+\infty} \rho_{x} - 1 \right) 
\prod_{x=1}^{+\infty}  \delta \left( 
 J_x +  j_{x+1} -  j_x 
\right) 
\nonumber \\
&&  e^{- \displaystyle T
\left(  \sum_{x=1}^{+\infty} 
\left[ j_{x}   \ln \left( \frac{ j_{x}   }{  w  \rho_{x-1}  }  \right) 
 - j_{x}   + w  \rho_{x-1} \right] 
 + \sum_{x=0}^{+\infty} 
\left[ J_x   \ln \left( \frac{ J_x   }{  r_x  \rho_x  }  \right) 
 -  J_x   + r_x  \rho_x  \right] \right) }
\label{level2.5mastertree}
\end{eqnarray}
One can use the constraints to eliminate either the non-local reset currents $J_.$ or
the local currents $j_.$ as described in the following two subsections,
but it is not possible to obtain in closed form the Large deviations at Level 2 for the density $\rho_.$ alone,
in contrast to the case of the Sisyphus Random Walk (Eq \ref{proba2.5chaintreeonlyrhosimpli1d}).


\subsection{ Large deviations for the empirical density $\rho_.$ and the local currents $j_.$ }

The last constraints on the first line of Eq. \ref{level2.5mastertree}
can be used to eliminate the non-local reset currents $J_x$ for $x \geq 1$
in terms of the local currents
\begin{eqnarray}
 J_x =   j_x - j_{x+1}
\label{bigJj}
\end{eqnarray}
while the reset current $J_0$ from the origin to itself remains.
So Eq. \ref{level2.5mastertree}
yields the following large deviations for the empirical density $\rho_.$ and the local currents $j_.$
\begin{eqnarray}
&& P_{T}[ \rho_. , j_.,J_0] 
  \oppropto_{T \to +\infty} 
\delta \left( \sum_{x=0}^{+\infty} \rho_{x} - 1 \right)  e^{- \displaystyle T
\left[ J_0   \ln \left( \frac{ J_0   }{  r_0  \rho_0  }  \right) 
 -  J_0   + r_0  \rho_0  \right] }
 \label{level2.5mastertreeloc}
 \\
&&  e^{- \displaystyle T
\left(  \sum_{x=1}^{+\infty} 
\left[ j_{x}   \ln \left( \frac{ j_{x}   }{  w  \rho_{x-1}  }  \right) 
 - j_{x}   + w  \rho_{x-1} \right] 
 + \sum_{x=1}^{+\infty} 
\left[ ( j_x - j_{x+1})   \ln \left( \frac{ ( j_x - j_{x+1})   }{  r_x  \rho_x  }  \right) 
 -  ( j_x - j_{x+1})   + r_x  \rho_x  \right]   
 \right) }
\nonumber
\end{eqnarray}


\subsection{ Large deviations for the empirical density $\rho_.$ and the non-local reset currents $J_.$ }

If one wishes instead to eliminate the local currents in terms of the reset currents $J_.$ 
via
\begin{eqnarray}
 j_x && =   \sum_{y=x}^{+\infty} J_y
\label{jbigJ}
\end{eqnarray}
one obtains the following large deviations for the empirical density $\rho_.$ and the reset currents $J_.$
\begin{eqnarray}
&& P_{T}[ \rho_. , J_.] 
 \oppropto_{T \to +\infty} 
\delta \left( \sum_{x=0}^{+\infty} \rho_{x} - 1 \right) 
\nonumber \\
&&  e^{- \displaystyle T
\left(  \sum_{x=1}^{+\infty} 
\left[ \left( \sum_{y=x}^{+\infty} J_y \right)   \ln \left( \frac{ \left( \sum_{y=x}^{+\infty} J_y \right)  }{  w  \rho_{x-1}  }  \right) 
 - \left( \sum_{y=x}^{+\infty} J_y \right)  + w  \rho_{x-1} \right] 
 + \sum_{x=0}^{+\infty} 
\left[ J_x   \ln \left( \frac{ J_x   }{  r_x  \rho_x  }  \right) 
 -  J_x   + r_x  \rho_x  \right] \right) }
\label{level2.5mastertreenonloc}
\end{eqnarray}


\subsection{ Large deviations for excursions between two consecutive resets }

For the dynamics of Eq. \ref{mastereq}, the simplifications in Eq. \ref{ptraject} 
are that the rate $w$ is uniform
and the internal trajectory $y( 0 \leq s \leq \tau) $ can only involve the positions $z_m=m$.
As a consequence, it will be clearer to denote the trajectory by the times $0 \leq s_1 \leq s_2 ... \leq s_Y \leq Y$,
where $s_{y}$ denotes the time where the particle jumps from $(y-1)$ to $y$, 
while $Y$ represents the last position before the reset towards $y=0$ at time $\tau$.
So the probability of Eq. \ref{ptraject} reads (with $s_0=0$ and $s_{Y+1}=\tau$)
\begin{eqnarray}
 P^{exc}[ \tau ; 0; s_{y=1,..,Y}   ] 
 && = r_{Y} w^Y \ \ e^{ \displaystyle   - \sum_{y=0}^Y (s_{y+1}- s_y) (w+r_{y} )   }
 \nonumber \\
 && = w^Y e^{   - w \tau } \ \ r_{Y} e^{    - s_1 r_0 - (s_2-s_1) r_1 - ... (s_Y- s_{Y-1} )r_{Y-1} - (\tau-s_Y) r_{Y} }
 \label{ptraject1d}
\end{eqnarray}
while the special case $Y=0$ where the reset occurs while the particle is still at the origin reads 
\begin{eqnarray}
 P^{exc}[ \tau ; 0 ]  = r_{0} e^{ - \tau  ( w + r_{0} ) }
\label{ptraject01d}
\end{eqnarray}

The normalization over all possible excursions of Eq. \ref{normatraj} becomes
\begin{eqnarray}
 1   
 = \int_{0}^{+\infty} d \tau  \left[ P^{exc}[ \tau ;0 ] 
+  \sum_{Y=1}^{+\infty}  \int_{0 \leq s_1 ...\leq s_Y \leq \tau} 
\! \! \! \! \! \! \! \! \! \! \! \! \! \! \! \! \! \! \! \! \! \! \! \! \! \! \! \! \! \! \! 
ds_1  ... ds_Y
  P^{exc}[ \tau ;  0; s_{y=1,..,Y} )  ] 
  \right]
\label{normatrajsi}
\end{eqnarray}

In order to make the link with the steady-state $P^*(x)$ of Eq. \ref{markovintermi},
it is useful to introduce to introduce the probability $P^{exc}_{end}(Y) $
of the end-position $Y$ of an excursion after integration over the duration $\tau$
\begin{eqnarray}
 P^{exc}_{end}(Y)  
  \equiv \int_0^{+\infty} d \tau  P^{exc}[ \tau ; 0; s_{y=1,..,Y}   ] 
=   \frac{ r_{Y} } { w+r_0 } \prod_{y=1}^Y \frac{ w}{ w+r_y} =  \frac{ r_{Y} } { w+r_0 } \left(\frac{P^*(Y)}{P^*(0)} \right)
\label{ptrajectend}
\end{eqnarray}
where the normalization is satisfied using Eq. \ref{markovintermizero}
\begin{eqnarray}
\sum_{Y=0}^{+\infty} P^{exc}_{end}(Y)  
=  \frac{ r_{0} } { w+r_0 }
  + \frac{ 1} { (w+r_0 ) P^*(0)} \sum_{Y=1}^{+\infty} r_Y P^*(Y) =  \frac{ r_{0} } { w+r_0 }
  + \frac{ w} { w+r_0} =1
\label{ptrajectendn}
\end{eqnarray}

The empirical density $n[\tau ; 0; s_{y=1,..,Y}]  $ of excursions between resets
and the total density $n$ follows the large deviations of Eq. \ref{probaempiexcjump}
in the more explicit form
\begin{eqnarray}
&& P_T ( n[.;..], n )  \opsimeq_{T \to +\infty}  
\delta \left[ \int_{0}^{+\infty} d \tau  \left[ n[ \tau ;0 ] 
+  \sum_{Y=1}^{+\infty}  \int_{0 \leq s_1 ...\leq s_Y \leq \tau} 
\! \! \! \! \! \! \! \! \! \! \! \! \! \! \! \! \! \! \! \! \! \! \! \! \! \! \! \! \! \! \! 
ds_1  ... ds_Y
  n[ \tau ;  0; s_{y=1,..,Y} )  ] 
  \right] - n \right]
\nonumber \\ &&
 \delta \left[ \int_{0}^{+\infty} d \tau  \tau \left[ n[ \tau ;0 ] 
+  \sum_{Y=1}^{+\infty}  \int_{0 \leq s_1 ...\leq s_Y \leq \tau} 
\! \! \! \! \! \! \! \! \! \! \! \! \! \! \! \! \! \! \! \! \! \! \! \! \! \! \! \! \! \! \! 
ds_1  ... ds_Y
  n[ \tau ;  0; s_{y=1,..,Y} )  ] 
  \right]   -1 \right]
 \nonumber \\ &&
  e^{ - T \displaystyle 
  \int_{0}^{+\infty} d \tau  \left[ n[ \tau ;0 ] 
  \ln \left( \frac{n[\tau ; 0 ]  }{ P^{exc}[ \tau ; 0  ]   n  }  \right) 
+  \sum_{Y=1}^{+\infty}  \int_{0 \leq s_1 ...\leq s_Y \leq \tau} 
\! \! \! \! \! \! \! \! \! \! \! \! \! \! \! \! \! \! \! \! \! \! \! \! \! \! \! \! \! \! \! 
ds_1  ... ds_Y
  n[ \tau ;  0; s_{y=1,..,Y} )  ] 
   \ln \left( \frac{ n[ \tau ;  0; s_{y=1,..,Y} )  ]  }{ P^{exc}[ \tau ;  0; s_{y=1,..,Y} )  ]   n  }  \right)
  \right] 
    } 
\label{probaempiexcjumps}
\end{eqnarray}


\subsection{ Large deviations of general time-additive observables }

In this section, we consider time-additive observables of the form $(A_T+B_T )$
(Eqs \ref{additiveAjump} \ref{additiveBjumpq}).
Using Eqs \ref{Jreset1dj} and \ref{jlocal1dj},
 their generating function of Eq. \ref{genekj} reads more explicitly
 in terms of the density $\rho_x$, the local current $j_x$ and the non-local reset currents $J_x$
\begin{eqnarray}
Z_T(k) \equiv    < e^{ T k \left( A_T + B_T \right) } >  
=<   e^{ \displaystyle T k  \left(  \sum_{ x =0 }^{+\infty}  \alpha_x   \rho_x  
+  \sum_{ x =1 }^{+\infty}     \beta_{x,x-1} j_x  +  \sum_{ x =0 }^{+\infty}   \beta_{0,x} J_x  \right) }  >
\label{genekj1d}
\end{eqnarray}

\subsubsection{ Analysis via the tilted matrix }

For the present model (Eq. \ref{wxy1dj}), the tilted
matrix involves the off-diagonal elements $x \ne y$ (Eq. \ref{wtiltedoff})
\begin{eqnarray}
{\tilde w}^{[k]}_{x,y} \equiv    \left[  w \delta_{x,y+1} +  \delta_{x,0}   r_y \right] e^{k\beta_{x,y}}         \ \ \  \ \ \ {\rm for } \ \ x \ne y
 \label{wtiltedoffs}
\end{eqnarray}
and the diagonal elements (Eq. \ref{wtiltediag})
\begin{eqnarray}
{\tilde w}^{[k]}_{x,x} \equiv    - w -   r_x  + k \alpha_x   +  \delta_{x,0}  r_0  e^{k\beta_{0,0}} 
 \label{wtiltediags}
\end{eqnarray}
So the eigenvalue Eqs \ref{wktiltright}
and \ref{wktiltleft}
read
\begin{eqnarray}
\left[   w +   r_x  - k \alpha_x   + \mu(k) \right]  {\tilde r}^{[k]}_x 
 =     w  e^{k\beta_{x,x-1}}   {\tilde r}^{[k]}_{x-1}
+ \delta_{x,0} \left[    \sum_{y = 0}^{+\infty}  r_y  e^{k\beta_{0,y}}   {\tilde r}^{[k]}_y \right]
 \label{wktiltrightj}
\end{eqnarray}
and 
\begin{eqnarray}
\left[  w +  r_y  - k \alpha_y + \mu(k)  \right] {\tilde l}^{[k]}_y = 
   {\tilde l}^{[k]}_{y+1}      w   e^{k\beta_{y+1,y}}   
  +   {\tilde l}^{[k]}_0    r_y  e^{k\beta_{0,y}}   
 \label{wktiltleftj}
\end{eqnarray}
For $x \geq 1$, Eq. \ref{wktiltrightj} corresponds to the simple recurrence
\begin{eqnarray}
  {\tilde r}^{[k]}_x 
 =  \left[   \frac{  w  e^{k\beta_{x,x-1}}  }{  w +   r_x  - k \alpha_x   + \mu(k)} \right]  {\tilde r}^{[k]}_{x-1}
=  \left[ \prod_{z=1}^x  \frac{  w  e^{k\beta_{z,z-1}}  }{  w +   r_z  - k \alpha_z   + \mu(k)} \right]  {\tilde r}^{[k]}_{0}
 \label{wktiltrightjrec}
\end{eqnarray}
while Eq. \ref{wktiltrightj} for $x=0$ yields
\begin{eqnarray}
\left[   w +   r_0  - k \alpha_0   + \mu(k) \right]  {\tilde r}^{[k]}_0 
 =      \sum_{x = 0}^{+\infty}  r_x  e^{k\beta_{0,x}}   {\tilde r}^{[k]}_x 
=   r_0  e^{k\beta_{0,0}}   {\tilde r}^{[k]}_0
+\sum_{x = 1}^{+\infty}  r_x  e^{k\beta_{0,x}} \left[ \prod_{z=1}^x  \frac{  w  e^{k\beta_{z,z-1}}  }{  w +   r_z  - k \alpha_z   + \mu(k)} \right]  {\tilde r}^{[k]}_{0}
 \label{wktiltrightjsi}
\end{eqnarray}
So $ {\tilde r}^{[k]}_0 $ disappears
and the remaining equation determines the scaled cumulant generating function $\mu(k)$ 
\begin{eqnarray}
1
=   \frac{  r_0  e^{k\beta_{0,0}} } {w +   r_0  - k \alpha_0   + \mu(k) }
+\sum_{x = 1}^{+\infty} \frac{  r_x  e^{k\beta_{0,x}} } {w +   r_0  - k \alpha_0   + \mu(k) }
\left[ \prod_{z=1}^x  \frac{  w  e^{k\beta_{z,z-1}}  }{  w +   r_z  - k \alpha_z   + \mu(k)} \right]  
 \label{tilteqmuk}
\end{eqnarray}
The solution of Eq. \ref{wktiltleftj} for the left eigenvector reads
\begin{eqnarray}
 {\tilde l}^{[k]}_y = 
   {\tilde l}^{[k]}_0 
   \left[  \frac{   r_y  e^{k\beta_{0,y}}   }{ w +  r_y  - k \alpha_y + \mu(k) }
  + \sum_{x=y+1}^{+\infty} \frac{   r_x  e^{k\beta_{0,x}}   }{ w +  r_x  - k \alpha_x + \mu(k) }
\prod_{z=y}^{x-1}  \frac{   w   e^{k\beta_{z+1,z}}   }{ w +  r_z  - k \alpha_z + \mu(k) }
  \right]
 \label{soluleftj}
\end{eqnarray}
where the consistency for $y=0$ reproduces the equation \ref{tilteqmuk} for the eigenvalue $\mu(k)$
after some minimal rewriting.

The matrix generating the conditioned process obtained via the generalization of Doob's h-transform (Eq. \ref{Wdoubletildedoobj}) involves the same non-vanishing matrix elements as the initial matrix $w^{reset}_{x,y}$ of Eq. \ref{wxy1dj} : the off-diagonal elements $x \ne y$ read (Eq. \ref{wtiltedoffs})
\begin{eqnarray}
 {\tilde  {\tilde w}}_{x ,y} =  {\tilde l}^{[k]}_x {\tilde w}^{[k]}_{x ,y} \frac{1}{{\tilde l}^{[k]}_y } 
 =    \delta_{x,y+1} {\tilde  {\tilde w}}_{x ,x-1}  
  +  \delta_{x,0}  {\tilde  {\tilde w}}_{0 ,y}
  \label{Wttoff}
\end{eqnarray}
where the rate to jump from $(x-1)$ to $x$
has changed from its initial value $w$ of Eq. \ref{wxy1dj}
to its new value
\begin{eqnarray}
 {\tilde  {\tilde w}}_{x ,x-1}  = 
w e^{k\beta_{x,x-1}} \frac{ {\tilde l}^{[k]}_x}{{\tilde l}^{[k]}_{x-1}}
   \label{Wxx1effj}
\end{eqnarray}
while the rate for a reset from $y$ to the origin
has changed from its initial value $r_y$ of Eq. \ref{wxy1d}
to its new value
\begin{eqnarray}
 {\tilde  {\tilde w}}_{0 ,y}  =   r_y e^{k\beta_{0,y}} \frac{ {\tilde l}^{[k]}_0}{{\tilde l}^{[k]}_y} 
    \label{W0xeffj}
\end{eqnarray}
Both involve explicitly the left eigenvector ${\tilde l}^{[k]}_. $.
The diagonal elements (Eq. \ref{wtiltediags}) 
\begin{eqnarray}
 {\tilde  {\tilde w}}_{x ,x} = - w -   r_x  + k \alpha_x   +  \delta_{x,0}  r_0  e^{k\beta_{0,0}}   - \mu(k)
  \label{Wttdiag}
\end{eqnarray}
involve the eigenvalue  $\mu(k)$.
The stationary density of this conditioned process reads
\begin{eqnarray}
 {\tilde  {\tilde \rho}}_{x }  =  {\tilde l}^{[k]}_x  {\tilde r}^{[k]}_x 
 && =   {\tilde l}^{[k]}_0 {\tilde r}^{[k]}_{0}
   \left[ \sum_{y=x}^{+\infty}  \frac{   r_y  e^{k\beta_{0,y}}   }{ w +  r_x  - k \alpha_x + \mu(k) }
   \prod_{z=1}^y  \frac{  w  e^{k\beta_{z,z-1}}  }{  w +   r_z  - k \alpha_z   + \mu(k)}
\right]
   \label{rhottj}
\end{eqnarray}
where the normalization determines the value of the density at the origin
${\tilde  {\tilde \rho}}_{0 }= {\tilde l}^{[k]}_0  {\tilde r}^{[k]}_0$
\begin{eqnarray}
1= \sum_{x=0}^{+\infty}  {\tilde  {\tilde \rho}}_{x } 
 =
  {\tilde l}^{[k]}_0  {\tilde r}^{[k]}_0
  \left[  \sum_{x=0}^{+\infty}
   \sum_{y=x}^{+\infty}  \frac{   r_y  e^{k\beta_{0,y}}   }{ w +  r_x  - k \alpha_x + \mu(k) }
   \prod_{z=1}^y  \frac{  w  e^{k\beta_{z,z-1}}  }{  w +   r_z  - k \alpha_z   + \mu(k)}
\right]
  \label{normarhottj}
\end{eqnarray}
Finally, the stationary local currents and reset currents of this conditioned process are given by
\begin{eqnarray}
{\tilde  {\tilde j}}_x && =  {\tilde  {\tilde w}}_{x ,x-1} {\tilde  {\tilde \rho}}_{x-1 } 
=w e^{k\beta_{x,x-1}}  {\tilde l}^{[k]}_x {\tilde r}^{[k]}_{x-1}
\nonumber \\
{\tilde  {\tilde J}}_y && =  {\tilde  {\tilde w}}_{0 ,y} {\tilde  {\tilde \rho}}_{y }  
=  r_y e^{k\beta_{0,y}}  {\tilde l}^{[k]}_0 {\tilde r}^{[k]}_y 
\label{jlocalJdoobj}
\end{eqnarray}


\subsubsection{ Analysis via the tilted excursions }

For the present model where the probabilities of excursions are given by Eqs \ref{ptraject1d} and \ref{ptraject01d}
the non-conserved tilted quantities of Eq. \ref{pexctiltedj}
read for $Y \geq 1$
\begin{eqnarray}
{ \tilde P}^{exc}_k[ \tau ; 0; s_{y=1,..,Y}   ] && \equiv P^{exc}[ \tau ; 0; s_{y=1,..,Y}   ] 
e^{\displaystyle k \left[  \sum_{y=0}^Y (s_{y+1}- s_y) \alpha_{y}
+ \sum_{y=1}^Y \beta_{y,y-1}
+ \beta_{0,Y } \right] } 
\nonumber \\
&& =
 \left( r_{Y}  e^{ k \beta_{0,Y } }  \right)  \left[ \prod_{y=1}^Y \left( w e^{k \beta_{y,y-1} }  \right)  \right]
 \left[ \prod_{y=0}^Y e^{ - (s_{y+1}- s_y) (w+r_{y} - k \alpha_y)        } \right]
\label{pexctiltedjy}
\end{eqnarray}
and for $Y=0$
\begin{eqnarray}
{ \tilde P}^{exc}_k[ \tau ; 0  ] \equiv P^{exc}[ \tau ; 0 ] 
e^{ k  \tau \alpha_0+ k \beta_{0,0 }} 
= \left( r_{0} e^{ k \beta_{0,0 }} \right)
e^{ - \tau  ( w + r_{0} -k \alpha_0) }
\label{pexctiltedjz}
\end{eqnarray}
The conditioned excursion probability of Eq. \ref{Psoludoubletildej} reads
\begin{eqnarray}
 {\tilde  {\tilde P}}^{exc}[\tau ; 0; s_{y=1,..,Y}  ] = { \tilde P}^{exc}[ \tau ; 0; s_{y=1,..,Y}   ]  e^{ -\tau  \mu(k) }
 \label{Psoludoubletildejsi}
\end{eqnarray}
where the scaled cumulant generating function $\mu(k)$
is fixed by its normalization (Eq. \ref{constraintstt1})
\begin{eqnarray}
1 && 
=  \int_{0}^{+\infty} d \tau e^{ -\tau  \mu(k) } \left[ 
{ \tilde P}^{exc}[ \tau ;0 ] 
+  \sum_{Y=1}^{+\infty}  \int_{0 \leq s_1 ...\leq s_Y \leq \tau} 
\! \! \! \! \! \! \! \! \! \! \! \! \! \! \! \! \! \! \! \! \! \! \! \! \! \! \! \! \! \! \! 
ds_1  ... ds_Y
  { \tilde P}^{exc}[ \tau ;  0; s_{y=1,..,Y} )  ] 
 \right]
 \nonumber \\
&& = \frac{ r_{0} e^{ k \beta_{0,0 }} }
{ w + r_{0} -k \alpha_0 + \mu(k) }
+ \sum_{Y=1}^{+\infty} \frac{r_{Y}  e^{ k \beta_{0,Y } } }{ w + r_{0} -k \alpha_0 + \mu(k)} 
\prod_{y=1}^Y \frac{w e^{k \beta_{y,y-1} }}{w + r_{y} -k \alpha_y + \mu(k)}
\label{constraintstt1si}
\end{eqnarray}
that coincides with Eq. \ref{tilteqmuk} as it should for consistency between the two approaches.


\section{ Diffusion Processes in a force field in dimension $d$ with resets }

\label{sec_diff}

\subsection{ Models and notations }

In this section, we turn to the continuous time/continuous space framework.
We focus on the dynamics described by the Fokker-Planck equation
in the force field $\vec F(\vec x)$ in dimension $d$, with space-independent diffusion coefficient $D$
and where the reset from $\vec y$ towards the origin $\vec 0 $  is governed by the rate $r(\vec y)$
\begin{eqnarray}
\frac{ \partial P_t(\vec x)   }{\partial t}   =  -   \vec \nabla .  \left[ P_t(\vec x )   \vec F(\vec x ) 
-D  \vec \nabla   P_t(\vec x)  \right]
-   r( \vec x)  P_t(\vec x)+ \delta^{(d)} (\vec x) \int d^d \vec y  \ r(\vec y)  P_t(\vec y)
\label{fokkerplanckreset}
\end{eqnarray}
Again as in Eqs \ref{markovchainst} and \ref{mastereqst},
we will assume that the steady-state solution $P^*(\vec x)$ of Eq. \ref{fokkerplanckreset}
\begin{eqnarray}
0   =  -   \vec \nabla .  \left[ P^*(\vec x )   \vec F(\vec x ) 
-D  \vec \nabla   P^*(\vec x)  \right]
-   r( \vec x)  P^*(\vec x)+ \delta^{(d)} (\vec x) \int d^d \vec y  \ r(\vec y)  P^*(\vec y)
\label{fokkerplanckresetst}
\end{eqnarray}
exists in order to apply the large deviations at Level 2.5 for non-equilibrium steady-states.


\subsection{ Large deviations at level 2.5 for the empirical density and the empirical currents  }

\subsubsection{ Empirical density and the empirical currents with their constraints}

The empirical density
\begin{eqnarray}
 \rho(\vec x) && \equiv \frac{1}{T} \int_0^T dt \  \delta^{(d)} ( \vec x(t)- \vec x)  
\label{rhodiff}
\end{eqnarray}
satisfies the normalization
\begin{eqnarray}
\int d^d \vec x \ \rho (\vec x) && = 1
\label{rho1ptnormadiff}
\end{eqnarray}
The non-local reset currents $J(\vec x) \geq 0 $ measure the non-local jumps from $\vec x \ne \vec 0$ towards the origin $\vec 0$
\begin{eqnarray}
J(\vec x)  \equiv  \frac{1}{T} \sum_{t : \vec x(t^+) = \vec 0 \ne  \vec x(t^-) } \delta_{\vec x(t^-), \vec x } 
\label{diffjreset}
\end{eqnarray}
while the local current field $\vec j(\vec x)$ characterizes the diffusion process in the force field
\begin{eqnarray} 
\vec j(\vec x) \equiv   \frac{1}{T} \int_0^T dt \ \frac{d \vec x(t)}{dt}   \delta^{(d)}( \vec x(t)- \vec x)  
\label{diffjlocal}
\end{eqnarray}
Here the stationarity constraint for all positions $\vec x \ne \vec 0$
yields that the non-local reset currents $J(\vec x)  $ 
are related to the divergence of the local current field $\vec j(\vec x)$
\begin{eqnarray}
J(\vec x) = - \vec \nabla . \vec j(\vec x)
\label{i2.5diffusion}
\end{eqnarray}
that replaces the standard divergence-free condition for diffusion processes without resetting.


\subsubsection{ Large deviations at level 2.5 for the density, the local currents and the non-local reset currents }

The joint distribution of the empirical density $\rho(.)$, the empirical local currents $\vec j(\vec x)$
and the empirical non-local reset currents $J(\vec x)$ satisfy the large deviation form 
\begin{eqnarray}
&& P_T[ \rho(.), \vec j(.),J(.)]   \opsimeq_{T \to +\infty}  \delta \left(\int d^d \vec x \rho(\vec x) -1  \right)
\left[ \prod_{\vec x \ne \vec 0}  \delta \left( J(\vec x) + \vec \nabla . \vec j(\vec x) \right) \right] 
\nonumber \\
&& e^{- \displaystyle T \left( \frac{1}{4 D} 
\int \frac{d^d \vec x}{ \rho(\vec x) } \left[ \vec j(\vec x) - \rho(\vec x) \vec F(\vec x)+D \vec \nabla \rho(\vec x) \right]^2
+ 
 \int d^d \vec x
\left[  J( \vec x)   \ln \left( \frac{  J(\vec x)  }{     r(\vec x)   \rho(\vec x) }  \right) 
 -  J(\vec x)  +    r(\vec x)  \rho(\vec x) \right]  \right) }
\label{ld2.5diff}
\end{eqnarray}
with the constraints discussed in Eqs \ref{rho1ptnormadiff} and \ref{i2.5diffusion},
while the rate function contains two contributions :
the first contribution involving the local current field $\vec j(\vec x)  $
corresponds to the usual rate function for 
continuous-time/continuous-space diffusion processes 
\cite{wynants_thesis,maes_diffusion,chetrite_formal,engel,chetrite_HDR,c_lyapunov},
while the second contribution involving the non-local reset current $J( \vec x)$ from $\vec x$ to the origin $\vec 0$
corresponds to the usual rate function for jump processes, that we have already seen in Eq. \ref{rate2.5master}.


\subsubsection{ Large deviations for the empirical density $\rho(.)$ and the empirical local currents $\vec j(.) $ }

The constraints of Eq. \ref{i2.5diffusion} can be used to eliminate all the non-local reset currents to obtain
the large deviations for the empirical density $\rho(.)$ and the empirical local currents $\vec j(.) $
\begin{eqnarray}
&& P_T[ \rho(.), \vec j(.)]   \opsimeq_{T \to +\infty}  \delta \left(\int d^d \vec x \rho(\vec x) -1  \right)
\label{ld2.5diffloc}
 \\
&& e^{- \displaystyle T \left[\frac{1}{4 D} 
\int \frac{d^d \vec x}{ \rho(\vec x) } \left( \vec j(\vec x) - \rho(\vec x) \vec F(\vec x)+D \vec \nabla \rho(\vec x) \right)^2
+ 
 \int d^d \vec x
\left[  \left(  - \vec \nabla . \vec j(\vec x) \right)   \ln \left( \frac{ \left(  - \vec \nabla . \vec j(\vec x) \right)  }{     r(\vec x)   \rho(\vec x) }  \right) 
 + \vec \nabla . \vec j(\vec x)  +    r(\vec x)  \rho(\vec x) \right]  \right] }
\nonumber
\end{eqnarray}


\subsubsection{ Large deviations for the empirical density $\rho(.)$ and the empirical reset currents $J(.) $ in dimension $d=1$ }

In arbitrary dimension $d >1$, the constraints of Eq. \ref{i2.5diffusion} cannot be explicitly inverted to obtain 
the local current field in terms of the non-local reset currents.
However in dimension $d=1$, the constraints of Eq. \ref{i2.5diffusion} become for all $x \ne 0$
\begin{eqnarray}
J(x)= - \frac{ d   j(x) }{d x} 
\label{i2.5diffusion1d}
\end{eqnarray}
i.e. the positive non-local reset current $J(x) \geq 0 $ determines the gradient of the local current $j(x)$,
that can be thus reconstructed as follows.
In the region $x>0$, the local current is positive $j(x)>0$ and given by
\begin{eqnarray}
 j(x) = \int_{x}^{+\infty} dy J(y)
\label{i2.5diffusionplus}
\end{eqnarray}
while in the region $x<0$, the local current is negative $j(x)<0$ and given by
\begin{eqnarray}
 j(x) = - \int_{-\infty}^{x} dy J(y)
\label{i2.5diffusionmoins}
\end{eqnarray}
In particular, the discontinuity of the local current at the origin
will correspond to the integral of all the reset currents that are re-injected at the origin
\begin{eqnarray}
 j(0^+)-  j(0^-)=\int_{0^+}^{+\infty} dy J(y) +\int_{-\infty}^{0^-} dy J(y) =  \int_{-\infty}^{+\infty} dy J(y) >0
\label{i2.5diffusiondicontinuity}
\end{eqnarray}

Using Eqs \ref{i2.5diffusionplus} and \ref{i2.5diffusionmoins}, the local currents can be eliminated from 
Eq. \ref{ld2.5diff} to obtain the following large deviations for the empirical density $\rho(.)$ and the empirical reset currents $J(.) $ 
\begin{eqnarray}
&& P^{(d=1)}_T[ \rho_.,,J_.]   \opsimeq_{T \to +\infty}  \delta \left(\int_{-\infty}^{+\infty} dx \rho(x) -1  \right)
e^{- \displaystyle T 
 \int_{-\infty}^{+\infty} dx
\left[  J(x)   \ln \left( \frac{  J(x)  }{     r(x)   \rho(x) }  \right) 
 -  J(x)  +    r(x)  \rho(x) \right]  }
\label{ld2.5diffnonloc}
 \\
&& e^{- \displaystyle \frac{T}{4 D}  \left[
\int_{-\infty}^{0} \frac{dx}{ \rho(x) } \left( -\int_{-\infty}^{x} dy J(y)
 - \rho(x) F(x)+D  \rho'(x)    \right)^2
+  \int_{0}^{+\infty} \frac{dx}{ \rho(x) } \left( \int_{x}^{+\infty} dy J(y) 
- \rho(x) F(x)+D  \rho'(x)    \right)^2
   \right] }
\nonumber
\end{eqnarray}


\subsection{ Large deviations for excursions between two consecutive resets }

\subsubsection{ Probabilities of excursions between two consecutive resets }

The probability to have an excursion of duration $\tau$ and of internal trajectory $\vec y( 0 \leq s \leq \tau) $
with the initial condition $\vec y(s=0)=\vec 0$ fixed by the reset
and the end-position $\vec y(s=\tau)$ being the last position before the next reset leading to 
$\vec y(\tau^+) =\vec 0$
reads
\begin{eqnarray}
P^{exc}[ \tau ; \vec y( .)  ] 
 = r( \vec y(\tau)) \ e^{ - \displaystyle \int_0^{\tau} ds  r(\vec y(s)) 
- \frac{1}{4D} \int_0^{\tau} ds \left( \frac{d \vec y(s)}{ds} - \vec F(\vec y(s)) \right)^2 
- \frac{1}{2} \int_0^{\tau} ds \ \vec \nabla . \vec F( \vec y(s))  }
\label{pinterdiffusion}
\end{eqnarray}
The normalization over all possible excursions involves the integration over the duration $\tau$ and the path-integral over the internal trajectory $y(0 \leq s \leq \tau)  $
\begin{eqnarray}
 1  = && 
\int_{0}^{+\infty} d \tau \int_{y(0)=0} {\cal D} y(.) \ P^{exc}[ \tau ; y(.)  ] 
\label{normaBrown}
\end{eqnarray}


\subsubsection{ Large deviations for the excursions density  }

The probability to see 
 the empirical density $n[\tau ; y( .) ]  $ of excursions between resets
and the total density $n$ follows the large deviation form 
analogous to Eq. \ref{probaempiexcjump}
\begin{eqnarray}
P_T ( n[.;..], n ) && \opsimeq_{T \to +\infty}  
 \delta \left[ \int_{0}^{+\infty} d \tau \int_{y(0)=0} {\cal D} y(.)  \ n[\tau ; y(.) ] 
- n \right]
  \delta \left[ \int_{0}^{+\infty} d \tau \int_{y(0)=0} {\cal D} y(.) \  \tau \ n[\tau ; y(.) ]   -1 \right]
\nonumber \\
&&  e^{ - \displaystyle  T \int_{0}^{+\infty} d \tau \int_{y(0)=0} {\cal D} y(.) \  n[\tau ; y(.) ]  
  \ln \left( \frac{n[\tau ; y(.)    ]  }{ P^{exc}[ \tau ; y(.)  ]   n  }  \right)  } 
\label{probaempifinjump}
\end{eqnarray}


\subsubsection{ Simplifications for uniform resetting rate $r_y=r$  }

When the resetting rate $r_y=r$ does not depend on the position $y$,
the excursion probability of Eq. \ref{pinterdiffusion}
becomes factorized as in Eq. \ref{ptrajectsifactorj}
\begin{eqnarray}
  P^{exc}_{uniform} [ \tau ;(y( . )  ]  =p^{exp}(\tau)   p^{config}_{\tau} [ (y( . )  ]
\label{ptrajectsifactord}
\end{eqnarray}
into the exponential distribution of Eq. \ref{pexp}
that describes the probability of the duration $\tau$ of the excursion
independently of the internal trajectory
and the probability $ p^{config}_{\tau}[y(.)] $ of the internal trajectory once its duration $\tau$ is given
that characterizes the process without resetting 
\begin{eqnarray}
p^{config}_{\tau} [ (y( . )  ]
 =  \ e^{  \displaystyle 
- \frac{1}{4D} \int_0^{\tau} ds \left( \frac{d \vec y(s)}{ds} - \vec F(\vec y(s)) \right)^2 
- \frac{1}{2} \int_0^{\tau} ds \ \vec \nabla . \vec F( \vec y(s))  }
\label{pinterdiffusionc}
\end{eqnarray}
In particular,  
the probability of the total density $n$ of excursions simply follows the Poisson distribution
given in Eq. \ref{poisson} with its corresponding large deviation form.
On the contrary, whenever the reset rates $r(\vec y)$ depends on the position $\vec y$, 
the duration $\tau$ and the internal trajectory $\vec y( .)  $ are coupled via Eq. \ref{pinterdiffusion}.


\subsection{ Large deviations for general time-additive observables  }

The empirical density of Eq. \ref{rhodiff}
allows to 
reconstruct any time-additive observable that involves some function $\alpha(\vec x)$ of the position $\vec x(t)$
\begin{eqnarray}
 A_T  && \equiv   \frac{1}{T}\int_0^T dt \ \alpha( \vec x(t))  = \int d^d \vec x \  \alpha (\vec x )  \rho( \vec x )
\label{additiveAdiff}
\end{eqnarray}
The non-local reset currents $J(\vec x)  $ of Eq. \ref{diffjreset}
 allows to reconstruct any time-additive observable that involves some function $ \beta (\vec x )$ 
of the reset jumps from $\vec x$ towards the origin $\vec 0$
\begin{eqnarray}
 B_T  && \equiv   \frac{1}{T}\sum_{t: \vec x(t^+) = \vec 0 \ne  \vec x(t^-) } \beta( \vec x(t^-) ) = 
\int d^d \vec x \  \beta (\vec x )  J( \vec x )
\label{additiveBdiff}
\end{eqnarray}
Finally the local current field $\vec j(\vec x)$ of Eq. \ref{diffjlocal}
allows to reconstruct any time-additive observable involving the increments $ \frac{d \vec x(t)}{dt}$ 
of the diffusion process parametrized by some vector field $ \vec\gamma (\vec x )$ 
\begin{eqnarray}
  \Gamma_T  && \equiv   \frac{1}{T}\int_0^T dt \ \frac{d \vec x(t)}{dt}
. \vec\gamma( \vec x(t))  = \int d^d \vec x \   \vec j ( \vec x ) . \vec \gamma (\vec x ) 
\label{additivegamma}
\end{eqnarray}
The large deviations of the sum $(A_T+B_T +\Gamma_T)$
can be analyzed via the generating function
\begin{eqnarray}
Z_T(k) \equiv    < e^{ T k \left( A_T + B_T +\Gamma_T \right) } >  
= < e^{ \displaystyle T k  \int d^d \vec x \left[ \alpha (\vec x )  \rho( \vec x )+ \beta (\vec x )  J( \vec x ) +  \vec \gamma (\vec x ) . \vec j ( \vec x ) \right] } >  
\label{genekdiff}
\end{eqnarray}


\subsubsection{ Analysis via the tilted dynamics }

The probability of a trajectory for the process of Eq. \ref{fokkerplanckreset}
can be written in terms of its empirical observables apart from the Wiener measure on the first line
\begin{eqnarray}
P^{Traj}_T    \opsimeq_{T \to +\infty}  
&& e^{  \displaystyle - \frac{1}{4D} \int_0^{T} ds   \left( \frac{d \vec y(s)}{ds}  \right)^2 }
\nonumber \\
&& e^{ \displaystyle T  \int d^d \vec x 
\left[  - \rho( \vec x ) \left(  r (\vec x )  + \frac{\left(  \vec F(\vec x) \right)^2}{4D} 
+ \frac{\vec \nabla . \vec F( \vec x)}{2}   \right)
 +  J( \vec x )  \ln (r (\vec x )  ) 
 +   \vec j ( \vec x ) . \frac{   \vec F ( \vec x )}{2D}   \right] }
\label{ptrajempi}
\end{eqnarray}

As a consequence, the generating function of Eq. \ref{genekdiff}  reads
\begin{eqnarray}
&& Z_T(k)  = e^{  \displaystyle - \frac{1}{4D} \int_0^{T} ds   \left( \frac{d \vec y(s)}{ds}  \right)^2 }
\nonumber \\
&& e^{ \displaystyle T  \int d^d \vec x 
\left[  - \rho( \vec x ) \left(  r (\vec x )  + \frac{\left(  \vec F(\vec x) \right)^2}{4D} 
+ \frac{\vec \nabla . \vec F( \vec x)}{2}  -k \alpha (\vec x ) \right)
 +  J( \vec x )  \ln (r (\vec x ) e^{k \beta (\vec x ) }  ) 
 +   \vec j ( \vec x ) . \frac{   \vec F ( \vec x ) + 2D \vec \gamma (\vec x )}{2D}   \right] }
\label{genekdifftraj}
\end{eqnarray}
and can be analyzed by tilting each term of the Fokker-Planck Eq. \ref{fokkerplanckreset} with resets
to produce the non-conserved dynamics
\begin{eqnarray}
&& \frac{ \partial {\tilde P}_t(\vec x)   }{\partial t}  ={\tilde  {\cal F}}_k {\tilde P}_t(.) 
\label{fokkerplancktilt}
 \\
&&   =  -   \left( \vec \nabla -k \vec \gamma (\vec x ) \right) .  
\left[  \vec F(\vec x ) P_t(\vec x )  
-D   \left( \vec \nabla -k \vec \gamma (\vec x ) \right)   P_t(\vec x)  \right]
-  \left[ r( \vec x) -k \alpha (\vec x ) \right]   P_t(\vec x)
 + \delta^{(d)} (\vec x) \int d^d \vec y  \ r(\vec y) e^{k \beta (\vec y )} P_t(\vec y)
\nonumber
\end{eqnarray}

The scaled cumulant generating function $\mu(k)$
corresponds to the highest eigenvalue of the tilted operator $ {\tilde  {\cal F}}_k$,
with its right eigenvector $ {\tilde r}^{[k]}(\vec x) $
\begin{eqnarray}
&&  \mu(k) {\tilde r}^{[k]}(\vec x)  = {\tilde  {\cal F}}_k{\tilde r}^{[k]}(.)
 \nonumber \\
 &&  =  -   \left( \vec \nabla -k \vec \gamma (\vec x ) \right) .  
\left[    \vec F(\vec x )  -D   \left( \vec \nabla -k \vec \gamma (\vec x ) \right)      \right] {\tilde r}^{[k]}(\vec x)
-  \left[ r( \vec x) -k \alpha (\vec x ) \right]    {\tilde r}^{[k]}(\vec x) 
 + \delta^{(d)} (\vec x) \int d^d \vec y  \ r(\vec y) e^{k \beta (\vec y )}  {\tilde r}^{[k]}(\vec y) 
 \nonumber \\
 && = D \Delta {\tilde r}^{[k]}(\vec x)
 - \left( \vec F(\vec x )  +2 D  k \vec \gamma (\vec x )\right) . \vec \nabla {\tilde r}^{[k]}(\vec x)
\label{fokkerright}
  \\ &&
 + \left[ - \vec \nabla . \vec F(\vec x ) - D k  \vec \nabla .\vec \gamma (\vec x )
 + k \vec F(\vec x ) . \vec \gamma (\vec x ) + D k^2 \vec \gamma^2 (\vec x )
 - r( \vec x) +k \alpha (\vec x ) \right]    {\tilde r}^{[k]}(\vec x) 
 + \delta^{(d)} (\vec x) \int d^d \vec y  \ r(\vec y) e^{k \beta (\vec y )}  {\tilde r}^{[k]}(\vec y) 
\nonumber
\end{eqnarray}
and its left eigenvector 
$ {\tilde l}^{[k]}(\vec y) $
\begin{eqnarray}
 \mu(k){\tilde l}^{[k]}(\vec y) && =  {\tilde  {\cal F}}_k^{\dagger} {\tilde l}^{[k]}(.) 
\nonumber \\
&&   =    
\left[    \vec F(\vec y ) +D   \left(  \vec \nabla + k \vec \gamma (\vec y ) \right)     \right]
 \left(  \vec \nabla + k \vec \gamma (\vec y ) \right) 
 {\tilde l}^{[k]}(\vec y)
-  \left[ r( \vec y) -k \alpha (\vec y ) \right]    {\tilde l}^{[k]}(\vec y) 
 +   \ r(\vec y) e^{k \beta (\vec y )}  {\tilde l}^{[k]}(\vec 0) 
 \nonumber \\
 && 
 = 
  D \Delta {\tilde l}^{[k]}(\vec y)
 + \left( \vec F(\vec y )  +2 D  k \vec \gamma (\vec y )\right) . \vec \nabla {\tilde l}^{[k]}(\vec y)
\nonumber \\
 &&  + \left[   D k  \vec \nabla .\vec \gamma (\vec y )
 + k \vec F(\vec y ) . \vec \gamma (\vec y ) + D k^2 \vec \gamma^2 (\vec y )
 - r( \vec y) +k \alpha (\vec y ) \right]    {\tilde l}^{[k]}(\vec y)  
  +  {\tilde l}^{[k]}(\vec 0)   \ r(\vec y) e^{k \beta (\vec y )} 
\label{fokkerleft}
\end{eqnarray}
In order to compute explicitly the generator of the conditioned process obtained via the generalization of Doob's h-transform
\begin{eqnarray}
{\tilde {\tilde {\cal F}}}_k \equiv  {\tilde l}^{[k]}(.){\tilde  {\cal F}}_k \frac{1}{{\tilde l}^{[k]}(.)} -\mu(k)
\label{fokkerdoob}
\end{eqnarray}
 it is technically simpler to compute the action of its adjoint operator ${\tilde {\tilde {\cal F}}}_k^{\dagger} $ on some test function $\phi(.)$
\begin{eqnarray}
&& {\tilde {\tilde {\cal F}}}_k^{\dagger} \phi(.)
=  \frac{1}{{\tilde l}^{[k]}(.)} {\tilde  {\cal F}}^{\dagger} \left[ {\tilde l}^{[k]}(.) \phi(.) \right] -\mu(k) \phi(.)
\nonumber \\
 && 
 = 
 \frac{1}{{\tilde l}^{[k]}(y)}
  D \Delta \left[ {\tilde l}^{[k]}(\vec y) \phi(\vec y) \right]
 +  \frac{1}{{\tilde l}^{[k]}(y)}\left( \vec F(\vec y )  +2 D  k \vec \gamma (\vec y )\right) . \vec \nabla 
 \left[ {\tilde l}^{[k]}(\vec y) \phi(\vec y) \right]  +   \frac{ r(\vec y) e^{k \beta (\vec y )} }{ {\tilde l}^{[k]}(y)}  {\tilde l}^{[k]}(\vec 0)  \phi(\vec 0) 
\nonumber \\
 &&  + \left[   D k  \vec \nabla .\vec \gamma (\vec y )
 + k \vec F(\vec y ) . \vec \gamma (\vec y ) + D k^2 \vec \gamma^2 (\vec y )
 - r( \vec y) +k \alpha (\vec y ) -\mu(k) \right]     \phi(\vec y)
\label{fokkerdoobdagger}
\end{eqnarray}
One can then use the eigenvalue Eq. \ref{fokkerleft} for the left eigenvector ${\tilde l}^{[k]}(.) $
to rewrite the coefficient of $\phi(\vec y)$ in the last line of Eq. \ref{fokkerdoobdagger} as
\begin{eqnarray}
&& \left[   D k  \vec \nabla .\vec \gamma (\vec y )
 + k \vec F(\vec y ) . \vec \gamma (\vec y ) + D k^2 \vec \gamma^2 (\vec y )
 - r( \vec y) +k \alpha (\vec y ) -  \mu(k)\right]
\nonumber \\
&& =    
 - \frac{1}{{\tilde l}^{[k]}(y)} D \Delta {\tilde l}^{[k]}(\vec y)
 - \frac{1}{{\tilde l}^{[k]}(y)} \left( \vec F(\vec y )  +2 D  k \vec \gamma (\vec y )\right) . \vec \nabla {\tilde l}^{[k]}(\vec y)
 -     \frac{{\tilde l}^{[k]}(\vec 0)   \ r(\vec y) e^{k \beta (\vec y )}}{{\tilde l}^{[k]}(y)}
\label{fokkerleftaux}
\end{eqnarray}
so that Eq. \ref{fokkerdoobdagger}
reduces to the form
\begin{eqnarray}
 {\tilde {\tilde {\cal F}}}_k^{\dagger} \phi(.)
    =   D  \Delta \phi(\vec y)
  +  \left[ \vec F(\vec y )  +2 D  k \vec \gamma (\vec y ) 
+2D  \frac{ (\vec \nabla  {\tilde l}^{[k]}(\vec y) )}{{\tilde l}^{[k]}(y)}
\right] .  \vec \nabla \phi(\vec y)      
  +     \frac{{\tilde l}^{[k]}(\vec 0)   \ r(\vec y) e^{k \beta (\vec y )}}{{\tilde l}^{[k]}(y)}
\left[   \phi(\vec y)  -\phi(\vec 0) \right]
\label{fokkerdoobdaggercalcul}
\end{eqnarray}
where it is explicit that the constant ${\tilde {\tilde l}}^{[k]} (\vec y)=1 $ is eigenvector of $ {\tilde {\tilde {\cal F}}}_k^{\dagger}  $
associated to the eigenvalue zero.

So the operator $  {\tilde {\tilde {\cal F}}}_k = ({\tilde {\tilde {\cal F}}}_k^{\dagger} )^{\dagger} $ generates the following 
conservative dynamics analogous to the initial dynamics of Eq. \ref{fokkerplanckreset}
\begin{eqnarray}
 \frac{ \partial {\tilde {\tilde P}}_t(\vec x)   }{\partial t}  = {\tilde {\tilde {\cal F}}}_k {\tilde {\tilde P}}_t (.) 
&&   =  - \vec \nabla . \left(
\vec F^{eff} (\vec x ) 
 {\tilde {\tilde P}}_t (\vec x) 
- D  \vec \nabla {\tilde {\tilde P}}_t (\vec x)
\right)
  -    r^{eff} (\vec x) 
   {\tilde {\tilde P}}_t (\vec x)
 +   \delta^{(d)} (\vec x) 
\int d^d \vec y   r^{eff} (\vec y)   {\tilde {\tilde P}}_t (\vec y)
\label{fokkerplancktt}
\end{eqnarray}
where the effective force that replaces the initial value $ \vec F(\vec x )$ of Eq. \ref{fokkerplanckreset} 
\begin{eqnarray}
\vec F^{eff} (\vec x )   \equiv \vec F(\vec x )  +2 D  k \vec \gamma (\vec x ) 
+2D  \frac{ (\vec \nabla  {\tilde l}^{[k]}(\vec x) )}{{\tilde l}^{[k]}(x)}
\label{forceeff}
\end{eqnarray}
and where the effective resetting rate from $\vec x$ to the origin $\vec 0$ 
that replaces the initial value $ r(\vec x )$ of Eq. \ref{fokkerplanckreset}
\begin{eqnarray}
 r^{eff} (\vec x) \equiv  r(\vec x) e^{k \beta (\vec x )} \frac{{\tilde l}^{[k]}(\vec 0)    }{{\tilde l}^{[k]}(x)} 
\label{reseteff}
\end{eqnarray}
involve both explicitly the left eigenvector ${\tilde l}^{[k]}(.)  $.
The right eigenvector ${\tilde r}^{[k]}(.)  $ will appear in the corresponding stationary density 
\begin{eqnarray}
 {\tilde  {\tilde \rho}}(\vec x)  =  {\tilde l}^{[k]} (\vec x) {\tilde r}^{[k]} (\vec x)
   \label{rhottdi}
\end{eqnarray}
in the corresponding local currents
\begin{eqnarray}
 {\tilde  {\tilde {\vec j}}} (\vec x)&& =  
 \vec F^{eff} (\vec x ) {\tilde  {\tilde \rho}} (x) 
 - D  \vec \nabla {\tilde  {\tilde \rho}} (x)
 = \left[ \vec F(\vec x )  +2 D  k \vec \gamma (\vec x ) +2D  \frac{ (\vec \nabla  {\tilde l}^{[k]}(\vec x) )}{{\tilde l}^{[k]}(x)} \right] 
 {\tilde l}^{[k]} (\vec x) {\tilde r}^{[k]} (\vec x)
  - D  \vec \nabla ({\tilde l}^{[k]} (\vec x) {\tilde r}^{[k]} (\vec x) ) 
  \nonumber \\
  && = 
  \left[ \vec F(\vec x )  +2 D  k \vec \gamma (\vec x )  \right] 
 {\tilde l}^{[k]} (\vec x) {\tilde r}^{[k]} (\vec x) 
 +D \left[   (\vec \nabla  {\tilde l}^{[k]}(\vec x) )  {\tilde r}^{[k]} (\vec x)
  -  {\tilde l}^{[k]} (\vec x) \vec \nabla ( {\tilde r}^{[k]} (\vec x) ) \right]
\label{jlocalJdoobdi}
\end{eqnarray}
and in the corresponding non-local reset currents
\begin{eqnarray}
{\tilde  {\tilde J}} (\vec x) =      r^{eff} (\vec x)  {\tilde  {\tilde \rho}}(\vec x) 
=  r(\vec x) e^{k \beta (\vec x )} {\tilde l}^{[k]}(\vec 0)    {\tilde r}^{[k]}(x)
\label{jresetJdoobdi}
\end{eqnarray}


\subsubsection{ Analysis via the tilted excursions }

The empirical density $ n[\tau ; y( .) ] $ of excursions can be used
to write
any time-additive observable of the forms of Eqs \ref{additiveAdiff} \ref{additiveBdiff} and \ref{additivegamma}
\begin{eqnarray}
 A_T  
&& = \int_{0}^{+\infty} d\tau  \int {\cal D} y(.) n[\tau ; y( .) ] 
\left[ \int_{0}^{\tau}  ds \alpha(y(s)) \right]  
\nonumber \\
 B_T  && = \int_{0}^{+\infty} d\tau  \int {\cal D} y(.) n[\tau ; y( .) ]  \beta(y(\tau^-))
\nonumber \\
  \Gamma_T  &&
  =   \int_{0}^{+\infty} d\tau  \int {\cal D} y(.) n[\tau ; y( .) ]
  \left[ \int_{0}^{\tau}  ds 
  \ \frac{d \vec y(s)}{ds}
. \vec\gamma( \vec y(s))
     \right]  
\label{additivediffexc}
\end{eqnarray}
So the generating function of Eq. \ref{genekdiff} can be evaluated from
the joint probability of Eq. \ref{probaempifinjump} as 
\begin{eqnarray}
Z_T(k)  
&& = \int dn \int {\cal D} n[.;..] 
P_T ( n[.;..], n )
 e^{ \displaystyle T k  
\int_{0}^{+\infty} d\tau  \int {\cal D} y(.) n[\tau ; y( .) ] 
\left[ \int_{0}^{\tau}  ds \left( \alpha(y(s)) + \ \frac{d \vec y(s)}{ds}. \vec\gamma( \vec y(s)) \right)
+\beta(y(\tau^-))
\right] 
 }   
 \nonumber \\
 && \opsimeq_{T \to + \infty}  \int dn \int {\cal D} n[.;..] 
 \delta \left[ \int_{0}^{+\infty} d \tau \int {\cal D} y(.)  \ n[\tau ; y(.) ] - n \right]
  \delta \left[ \int_{0}^{+\infty} d \tau \int {\cal D} y(.) \  \tau \ n[\tau ; y(.) ]   -1 \right]
\nonumber \\
&& 
 e^{  \displaystyle  - T \int_{0}^{+\infty} d \tau \int {\cal D} y(.) \  n[\tau ; y(.) ]  
  \ln \left( \frac{n[\tau ; y(.)    ]  }{ { \tilde P}^{exc}[ \tau ; y(.)  ]   n  }  \right) 
 }   
\label{genekdiffexc}
\end{eqnarray}
in terms of the
tilted non-conserved quantity
\begin{eqnarray}
{ \tilde P}^{exc}_k[ \tau ; y(.)  ] \equiv P^{exc}[ \tau ; y(.)  ] \ 
e^{\displaystyle k 
 \int_{0}^{\tau}  ds \left( \alpha(y(s)) + \ \frac{d \vec y(s)}{ds}. \vec\gamma( \vec y(s)) \right)
+k \beta(y(\tau^-))
} 
\label{pexctiltediff}
\end{eqnarray}
Again the change of variable from $n[\tau ; y( . ) ]  $ to the probability of excursions that will make $n[\tau ; y( .)  ] $
and $n$ typical 
\begin{eqnarray}
{\tilde  {\tilde P}}^{exc} [\tau ; y( .)  ] \equiv \frac{n[\tau ; y( .)  ]}{ n} 
\label{Pdoubletildediff}
\end{eqnarray}
allows to analyze Eq. \ref{pexctiltediff} exactly as Eq. \ref{genekexcjtt}
 with the same output :
the conditioned distribution of excursions is given by (Eq \ref{Psoludoubletildej})
\begin{eqnarray}
 {\tilde  {\tilde P}}^{exc}[\tau ; y( .)  ] = { \tilde P}^{exc}[ \tau ; y(.)  ]  e^{ -\tau  \mu(k) }
 \label{Psoludoubletildediff}
\end{eqnarray}
where the scaled cumulant generating function $\mu(k)$ is fixed by its normalization (Eq \ref{constraintstt1})
\begin{eqnarray}
1 && = \int_{0}^{+\infty} d \tau \int{\cal D} y(.)  {\tilde  {\tilde P}}^{exc}[\tau ; y( .) ] 
=  \int_{0}^{+\infty} d \tau e^{ -\tau  \mu(k) } \left( \int{\cal D} y(.)  { \tilde P}^{exc}[ \tau ; y(.)  ]  \right)
\label{constraintstt1d}
\end{eqnarray}
while the inverse of the density $n$ of the conditioned process is given by the first moment of the duration $\tau$ (Eq \ref{constraintstt2})
\begin{eqnarray}
\frac{1}{n} && =  \int_{0}^{+\infty} d \tau \int {\cal D} y(.) \tau  {\tilde  {\tilde P}}^{exc}[\tau ; y(.) ] 
=  \int_{0}^{+\infty} d \tau \tau e^{ -\tau  \mu(k) } \left( \int{\cal D} y(.)  { \tilde P}^{exc}[ \tau ; y(.)  ]  \right)
\label{constraintstt2d}
\end{eqnarray}


\section{Sisyphus process in continuous time and continuous space  }

\label{sec_neuronal}

In this section, we consider the dynamics of Eq. \ref{fokkerplanckreset}
in dimension $d=1$, in the absence of diffusion $D=0$ and in the constant force field 
of value unity $F(x)=1$ : this Sisyphus process taking place on the half-line $x \in [0,+\infty[$
\begin{eqnarray}
\frac{ \partial P_t(x)   }{\partial t}  =  -   \frac{ \partial P_t(x)   }{\partial x}  
-   r(x)  P_t(x)+ \delta(x) \int_{0}^{+\infty} dy  r(y)  P_t(y)
\label{neuronal}
\end{eqnarray}
is the continuous time/continuous space analog of the one-dimension Sisyphus Random Walk \cite{sisyphus}
discussed in section \ref{sec_chain1d}.
It also corresponds to the diffusionless version of the neuronal integrate-and-fire model studied in Ref \cite{miles}.

In order to apply the general formalism described in the previous section,
one should first check that the hypothesis concerning the existence of a steady-state (Eq. \ref{fokkerplanckresetst})
is satisfied.


\subsection{ Condition on the reset rates $r(.)$ for the existence of a non-equilibrium steady-state  }

The stationary solution $P^*(x)$ of Eq. \ref{neuronal} 
\begin{eqnarray}
0 =  -   \frac{ d P^*(x)   }{d x}  
-   r(x)  P^*(x)+ \delta(x) \int_{0}^{+\infty} dy  r(y)  P^*(y)
\label{neuronalsteq}
\end{eqnarray}
reads in terms of the heaviside step function $\theta(x)$
\begin{eqnarray}
  P^*(x) = \theta(x) P^*(0) e^{- \int_0^x dy r(y) }
\label{neuronalstsolu}
\end{eqnarray}
where the steady-state $P^*(0) $ at the origin is fixed by the normalization 
\begin{eqnarray}
1=\int_0^{+\infty} dx  P^*(x) = P^*(0) \int_0^{+\infty} dx e^{- \int_0^x dy r(y) }
\label{neuronalstnorma}
\end{eqnarray}
The condition on the reset rates $r(.)$ to have a non-equilibrium steady-state
localized around the origin $ P^*(0)>0 $ is thus given by the convergence of the integral
\begin{eqnarray}
 \int_0^{+\infty} dx e^{- \int_0^x dy r(y) } < + \infty
\label{neuronalstcv}
\end{eqnarray}
that represents the continuous analog of Eq. \ref{cvsteady1d}.
In the uniform case $r(y)=r$, this integral converges for any value $r>0$,
i.e. the only case of divergence corresponds to $r=0$, where the model without any reset is of course transient (Eq. \ref{neuronal}).
In the random case,
the specific structure of Eq. \ref{neuronalstcv} corresponds to the continuous version
of Kesten random variables 
\cite{Kesten,Der_Pom,Bou,Der_Hil,Cal,strong_review,c_microcano,c_watermelon,c_mblcayley}
so that the condition of convergence $ \overline{ r_y } >0$ will be satisfied since the random variable $r_y$ 
represents a positive reset rate $r_y \geq 0$.

Finally for the case where the reset rate $r(y)$ is a deterministic positive function of the position $y$,
the integral of Eq. \ref{neuronalstcv} will generically converge, except when
$r(y)$ decays towards zero too rapidly for $y \to +\infty$,
for instance for the choice $r(y \to + \infty) \simeq 1/y$ that leads to the logarithmic divergence of Eq. \ref{neuronalstcv}.

In the following, we assume the convergence of Eq. \ref{neuronalstcv}
in order to have a non-equilibrium steady-state that can be analyzed
via the large deviations at Level 2.5.


\subsection{ Large deviations at Level 2.5 for the density $\rho(.)$ and the non-local reset currents $J(.)$}

In the limit of vanishing diffusion $D \to 0$, the first contribution of the rate function of Eq. \ref{ld2.5diff}
becomes the constraint
\begin{eqnarray}
j(x) =  \rho(x)
\label{jfrho}
\end{eqnarray}
between the local current $j(x)$ and the density $\rho(x)$,
as it should since the motion is deterministic and ballistic between resets
(Eq. \ref{jfrho} is the analog of Eq. \ref{elimj}).
As a consequence, the large deviations at Level 2.5 for the joint distribution of the density $\rho(.)$
and the non-local reset currents $J(.)$ read (Eq. \ref{ld2.5diff})
\begin{eqnarray}
 P_T[ \rho(.), J(.)]  
 &&  
 \opsimeq_{T \to +\infty}  \delta \left(\int_0^{+\infty} dx \rho( x) -1  \right)
  \prod_{x>0} \delta \left( J(x) + \rho'(x)  \right) 
 \nonumber \\ && 
e^{- \displaystyle T 
 \int_0^{+\infty} d x
\left[  J( x)   \ln \left( \frac{  J( x)  }{     r( x)   \rho( x) }  \right) 
 -  J(x)  +    r( x)  \rho( x) \right]   }
\label{ld2.5diffsi}
\end{eqnarray}


\subsection{ Large deviations at Level 2 for the density $\rho(.)$ alone }

One may use the last constraints of the first line of Eq. \ref{ld2.5diffsi} to eliminate all the non-local reset currents
\begin{eqnarray}
J(x) = -  \rho'(x)
\label{bigjrho}
\end{eqnarray}
and one obtains the large deviations at Level 2 for the density $\rho(.)$ alone
\begin{eqnarray}
&& P_T[ \rho(.)]   \opsimeq_{T \to +\infty}  \delta \left(\int_{0}^{+\infty} dx \rho(x) -1  \right)
 e^{- \displaystyle T 
 \int_{0}^{+\infty} dx
\left[ -  \rho'(x) \ln \left( \frac{  \left( -  \rho'(x)\right)  }{     r(x)   \rho(x) }  \right) 
 +  \rho'(x)+    r(x)  \rho(x) \right]   }
\nonumber \\
&& \opsimeq_{T \to +\infty}  \delta \left(\int_{0}^{+\infty} dx \rho(x) -1  \right)
 e^{- \displaystyle T 
 \left[ 
  \int_{0}^{+\infty} dx  \left( -  \rho'(x)\right) \ln \left( \frac{  \left( -  \rho'(x)\right)  }{     r(x)    }  \right) 
 +  \int_{0}^{+\infty} dx \frac{ d}{dx}\left[ \rho(x) \ln (\rho(x))  \right]  +  \int_{0}^{+\infty} dx   r(x)  \rho(x)
  \right]   }
\nonumber \\
&& \opsimeq_{T \to +\infty}  \delta \left(\int_{0}^{+\infty} dx \rho(x) -1  \right)
 e^{- \displaystyle T 
 \left[ 
  \int_{0}^{+\infty} dx  \left( - \rho'(x)\right) \ln \left( \frac{  \left( -  \rho'(x)\right)  }{     r(x)    }  \right) 
 - \rho(0) \ln (\rho(0))  +  \int_{0}^{+\infty} dx   r(x)  \rho(x)
  \right]   }
\label{ld2.5diffrho}
\end{eqnarray}


\subsection{ Large deviations for the non-local reset currents $J(.)$ alone }

One may instead eliminate the empirical density 
\begin{eqnarray}
\rho(x)= \int_{x}^{+\infty} dy J(y) 
\label{rhobigj}
\end{eqnarray}
to translate Eq. \ref{ld2.5diffrho} into the large deviations for the non-local reset currents $J(.)$ alone
\begin{eqnarray}
&& P_T[ \rho(.)]     \opsimeq_{T \to +\infty}  \delta \left(\int_{0}^{+\infty} dx x J(x) -1  \right)
\delta \left(\int_{0}^{+\infty} dx  J(x) - \rho(0)  \right)
\nonumber \\
&& 
 e^{- \displaystyle T 
 \left[ 
  \int_{0}^{+\infty} dx  J(x) \ln \left( \frac{  J(x)  }{     r(x)    }  \right) 
 - \rho(0) \ln (\rho(0))  +  \int_{0}^{+\infty} dx   J(x)  \int_{0}^{x} dy r(y) 
  \right]   }
\nonumber \\
&&   \opsimeq_{T \to +\infty}  \delta \left(\int_{0}^{+\infty} dx x J(x) -1  \right)
\delta \left(\int_{0}^{+\infty} dx  J(x) - \rho(0)  \right)
 e^{- \displaystyle T 
  \int_{0}^{+\infty} dx  J(x) \ln \left( \frac{  J(x)  }{     r(x)  e^{-  \int_{0}^{x} dy r(y) }  \rho(0) }  \right)   }
\label{ld2.5diffbigJ}
\end{eqnarray}
The compact form of the last line allows to make the link with the large deviations for the excursions between two consecutive resets as explained below.


\subsection{ Large deviations for the excursions between two consecutive resets}

Here the internal trajectory of an excursion between to resets is deterministic and ballistic
\begin{eqnarray}
y( 0 \leq s \leq \tau) = s
\label{ballistic}
\end{eqnarray}
so that the probability for an excursion of Eq. \ref{pinterdiffusion}
can be labelled by its duration $\tau$ only and reduces to
\begin{eqnarray}
P^{exc}(\tau) =P^{exc}[ \tau ; y( 0 \leq s \leq \tau) =s ] 
=  r(\tau) e^{- \displaystyle \int_0^{\tau} ds r( s)  } = - \frac{d}{d\tau} e^{- \displaystyle \int_0^{\tau} ds r( s)  }
\label{excball}
\end{eqnarray}

The large deviation form for the empirical density $n(\tau)$ of excursions of duration $\tau$
and the total density $n$ of excursions 
\begin{eqnarray}
P_T ( n(.), n ) && \opsimeq_{T \to +\infty} 
\delta \left[ \int_0^{+\infty} d\tau n(\tau)- n \right]
 \delta \left[ \int_0^{+\infty} d\tau \tau n(\tau) -1 \right]
\ 
 e^{ - \displaystyle T \int_0^{+\infty} d\tau
  n(\tau)
  \ln \left( \frac{n(\tau)  }{ P^{exc}( \tau ) n  }  \right)   } 
\nonumber \\
&&  \opsimeq_{T \to +\infty} 
\delta \left[ \int_0^{+\infty} d\tau n(\tau)- n \right]
 \delta \left[ \int_0^{+\infty} d\tau \tau n(\tau) -1 \right]
\ 
 e^{ - \displaystyle T \int_0^{+\infty} d\tau
  n(\tau)
  \ln \left( \frac{n(\tau)  }{  r(\tau) e^{- \int_0^{\tau} ds r( s)  } n  }  \right)   } 
\label{largedevnsi}
\end{eqnarray}
 thus coincides with Eq. \ref{ld2.5diffbigJ} with the following dictionary
analogous to Eqs \ref{rhozeron} and \ref{Jn} :
the empirical density at the origin $\rho(0)$ corresponds to the empirical density $n$ of excursions
\begin{eqnarray}
\rho(0)= n
\label{diffrhon}
\end{eqnarray}
while the non-local reset current $J(x)$ corresponds to the empirical density of excursions of duration $\tau=x$
\begin{eqnarray}
J(x) = n(\tau=x)
\label{Jndiff}
\end{eqnarray}


\subsection{ Large deviations for general time-additive observables  }

In this section, we focus on the generating function
of time-additive observables of the form $(A_T+B_T +\Gamma_T)$ (Eqs \ref{additiveAdiff} \ref{additiveBdiff} \ref{additivegamma})
involving a priori three function $\alpha(x),\beta(x),\gamma(x)$ 
\begin{eqnarray}
Z_T(k)  \equiv    < e^{ T k \left( A_T  +\Gamma_T + B_T \right) } >  
&& =  < e^{ \displaystyle k \left[ \int_0^T dt \left[ \alpha(  x(t)) + \dot x(t) \gamma(x(t)) \right] +\sum_{t:  x(t^+) =  0 \ne   x(t^-) } \beta(  x(t^-) ) \right] } >
\nonumber \\
&& = < e^ {  \displaystyle T k \int_0^{+\infty} dx \left[ \alpha(x) \rho(x) + j(x) \gamma(x) + J(x) \beta(x) \right] } >
\label{genekdiff1dred}
\end{eqnarray}
In the present model, the local current $j(x)$ coincides with the density $\rho(x)$ (Eq. \ref{jfrho}),
so the function $\gamma(x)$ is redundant with respect to $\alpha(x)$,
but as in the previous applications,  it is more pedagogical to keep them both to see more clearly how the general formalism of the previous section works.


\subsubsection{ Analysis via the tilted dynamics }

In the present one-dimensional model (Eq. \ref{neuronal})
where the diffusion coefficient vanishes $D=0$ and where the force is constant of value unity $F(x)=1$,
the eigenvalue equations for the right eigenvector $ {\tilde r}^{[k]}(x) $ (Eq. \ref{fokkerright})
and for the left eigenvector ${\tilde l}^{[k]}(x) $ (Eq. \ref{fokkerleft}) become
\begin{eqnarray} 
\mu(k) {\tilde r}^{[k]}( x) 
 &&
   =  -    \frac{d {\tilde r}^{[k]}( x)  }{dx}    
 + \left[  k \alpha ( x )+ k \gamma ( x ) - r(  x) \right]    {\tilde r}^{[k]}( x) 
 + \delta ( x) \int_0^{+\infty}  dy  \ r( y) e^{k \beta ( y )}  {\tilde r}^{[k]}( y) 
\label{fokkerrightsi}
\end{eqnarray}
and
\begin{eqnarray}
\mu(k){\tilde l}^{[k]}( y) 
&&   =    
   \frac{d  {\tilde l}^{[k]}( y)}{dy} 
 + \left[  k \alpha ( y )+ k \gamma ( y ) - r(  y) \right]  {\tilde l}^{[k]}( y) 
 +   \ r( y) e^{k \beta ( y )}  {\tilde l}^{[k]}( 0) 
\label{fokkerleftsi}
\end{eqnarray}
The solution of Eq. \ref{fokkerrightsi} for the right eigenvector reads
\begin{eqnarray}
   {\tilde r}^{[k]}( x) = \theta(x)  {\tilde r}^{[k]}( 0) e^{ \displaystyle -x \mu(k) 
   + \int_0^x dx' \left[ k  \alpha ( x' ) + k  \gamma ( x' )- r(x') \right]}
\label{soluright}
\end{eqnarray}
with the condition at the origin
\begin{eqnarray}
  {\tilde r}^{[k]}( 0) = \int_0^{+\infty}  dx  \ r( x) e^{k \beta ( x )}  {\tilde r}^{[k]}( x) 
  = \int_0^{+\infty}  dx  \ r( x) e^{k \beta ( x )} {\tilde r}^{[k]}( 0) e^{ \displaystyle -x \mu(k) 
  + \int_0^x dx' \left[ k  \alpha ( x' ) + k  \gamma ( x' )- r(x') \right]
  }
\label{solurightc}
\end{eqnarray}
so $ {\tilde r}^{[k]}( 0) $ disappears and the remaining equation determines the scaled cumulant generating function
$\mu(k)$ 
\begin{eqnarray}
 1
  = \int_0^{+\infty}  dx  \ r( x)   e^{ \displaystyle -x \mu(k) +k \beta(x) 
  + \int_0^x dx' \left[ k  \alpha ( x' ) + k  \gamma ( x' )- r(x') \right]
 }
\label{eqmukdiff}
\end{eqnarray}
The solution of Eq. \ref{fokkerleftsi} for the left eigenvector reads
\begin{eqnarray}
   {\tilde l}^{[k]}( y) 
  = {\tilde l}^{[k]}( 0) 
  \int_y^{+\infty}  dx  \ r( x)   e^{ \displaystyle +k \beta(x)
  +  \int_y^x dx'   \left[ - \mu(k) +  k  \alpha ( x' ) + k  \gamma ( x' ) - r(x') \right]}
\label{soluleftc}
\end{eqnarray}
where the consistency at $y=0$ reproduces Eq. \ref{eqmukdiff} for the eigenvalue $\mu(k)$.

The operator $  {\tilde {\tilde {\cal F}}}_k$ that generates 
the conditioned process (Eq. \ref{fokkerplancktt}) has the same form 
as the initial dynamics with a constant force $F(x)=1$ and no diffusion $D=0$
\begin{eqnarray}
 \frac{ \partial {\tilde {\tilde P}}_t(x)   }{\partial t}  = {\tilde {\tilde {\cal F}}}_k {\tilde {\tilde P}}_t (.) 
&&   =  - \frac{ \partial {\tilde {\tilde P}}_t (x)}  {\partial x}    -    r^{eff} ( x) 
   {\tilde {\tilde P}}_t (x)
 +   \delta ( x) 
\int_0^{+\infty} dy   r^{eff} ( y)   {\tilde {\tilde P}}_t (y)
\label{fokkerplanckttsi}
\end{eqnarray}
where the effective resetting rate from $ x$ to the origin $ 0$ 
has changed from its initial value $ r( x )$ of Eq. \ref{neuronal}
to its new value of Eq. \ref{reseteff} that involves the left eigenvector ${\tilde l}^{[k]}(.)  $
\begin{eqnarray}
 r^{eff} (x) \equiv  r(x) e^{k \beta (x )} \frac{{\tilde l}^{[k]}(0)    }{{\tilde l}^{[k]}(x)} 
\label{reseteffsi}
\end{eqnarray}
The corresponding stationary density reads
\begin{eqnarray}
 {\tilde  {\tilde \rho}}(x)  =  {\tilde l}^{[k]} (x) {\tilde r}^{[k]} (x)
&&   = {\tilde l}^{[k]}( 0) {\tilde r}^{[k]}( 0)
  \int_x^{+\infty}  dz  \ r( z)   e^{ \displaystyle - z \mu(k) +k \beta(z)
  +  \int_0^z dx'   \left[   k  \alpha ( x' ) + k  \gamma ( x' ) - r(x') \right]}
   \label{rhottdisi}
\end{eqnarray}
where the normalization determines the value of the density at the origin
${\tilde  {\tilde \rho}}(0 )= {\tilde l}^{[k]}_0  {\tilde r}^{[k]}_0$ 
\begin{eqnarray}
1=\int_0^{+\infty} dx {\tilde  {\tilde \rho}}(x)  
&&   = {\tilde l}^{[k]}( 0) {\tilde r}^{[k]}( 0)
\int_0^{+\infty} dx  \int_x^{+\infty}  dz  \ r( z)   e^{ \displaystyle - z \mu(k) +k \beta(z)
  +  \int_0^z dx'   \left[   k  \alpha ( x' ) + k  \gamma ( x' ) - r(x') \right]}
  \nonumber \\
  &&
  =  {\tilde l}^{[k]}( 0) {\tilde r}^{[k]}( 0)
  \int_0^{+\infty}  dz  \ z  \ r( z)   e^{ \displaystyle - z \mu(k) +k \beta(z)
  +  \int_0^z dx'   \left[   k  \alpha ( x' ) + k  \gamma ( x' ) - r(x') \right]}
   \label{rhottdisinorma}
\end{eqnarray}
The corresponding local currents coincide with the density (Eq. \ref{jlocalJdoobdi})
\begin{eqnarray}
 {\tilde  {\tilde { j}}} ( x)&& =  
 {\tilde  {\tilde \rho}} (x) 
\label{jlocalJdoobdisi}
\end{eqnarray}
while the corresponding non-local reset currents read (Eq. \ref{jresetJdoobdi})
\begin{eqnarray}
{\tilde  {\tilde J}} ( x)&& 
=  r( x) e^{k \beta ( x )} {\tilde l}^{[k]}( 0)    {\tilde r}^{[k]}(x)
\label{jresetJdoobdisi}
\end{eqnarray}


\subsubsection{ Analysis via the tilted excursions }

In the present model where the probability of excursions is given by Eq. \ref{excball},
 the tilted non-conserved quantity of Eq. \ref{pexctiltediff} reads
\begin{eqnarray}
{ \tilde P}^{exc}_k[ \tau  ] && = P^{exc}[ \tau   ] \ 
e^{\displaystyle +k \beta(\tau) +k 
 \int_{0}^{\tau}  ds \left[ \alpha(s) + \gamma( s) \right]
} 
=  r(\tau) e^{ \displaystyle k \beta(\tau)
+ \int_0^{\tau} ds \left[ - r( s) + k  \alpha(s) + k \gamma( s) \right] 
} 
\label{pexctiltediffsi}
\end{eqnarray}
The conditioned distribution of excursions is given by (Eq \ref{Psoludoubletildediff})
\begin{eqnarray}
 {\tilde  {\tilde P}}^{exc}[\tau   ] = { \tilde P}^{exc}[ \tau   ]  e^{ -\tau  \mu(k) }
 =  r(\tau) e^{ \displaystyle  -\tau  \mu(k) 
 +k \beta(\tau) +\int_0^{\tau} ds \left[ - r( s) + k  \alpha(s) + k \gamma( s) \right] 
} 
 \label{Psoludoubletildediffsi}
\end{eqnarray}
where $\mu(k)$ is fixed by its normalization (Eq \ref{constraintstt1d})
\begin{eqnarray}
1 && = \int_{0}^{+\infty} d \tau   {\tilde  {\tilde P}}^{exc}[\tau  ] 
=  \int_{0}^{+\infty} d \tau r(\tau) e^{ \displaystyle  -\tau  \mu(k) +k \beta(\tau)
 +\int_0^{\tau} ds \left[ - r( s) + k  \alpha(s) + k \gamma( s) \right] 
} 
\label{constraintstt1dsi}
\end{eqnarray}
that coincides with Eq. \ref{eqmukdiff}
as it should by consistency between the two points of view.
Finally the inverse of the density $n$ of the conditioned process is fixed by the first moment of the duration $\tau$ (Eq \ref{constraintstt2d})
\begin{eqnarray}
\frac{1}{n} && =  \int_{0}^{+\infty} d \tau \  \tau \  {\tilde  {\tilde P}}^{exc}[\tau ] 
=  \int_{0}^{+\infty} d \tau \ \tau \ 
 r(\tau) e^{ \displaystyle  -\tau  \mu(k) +k \beta(\tau) +\int_0^{\tau} ds \left[ - r( s) + k  \alpha(s) + k \gamma( s) \right] 
} 
\label{constraintstt2dsi}
\end{eqnarray}
This equation is equivalent to Eq. \ref{rhottdisinorma} for the density at the origin
${\tilde  {\tilde \rho}}_{0 }= {\tilde l}^{[k]}_0  {\tilde r}^{[k]}_0$ 
in agreement with the dictionary Eqs
\ref{diffrhon} and \ref{Jndiff}.


\section{ Conclusions  }

\label{sec_conclusion}

In order to characterize the non-equilibrium steady-states of
Markov processes with stochastic resetting towards the origin,
we have described the large deviations 
at level 2.5 for the joint probability of the empirical density and the empirical flows,
the large deviations for the empirical excursions between consecutive resets,
and the large deviations of general time-additive observables
within the three possible frameworks : 
discrete-time/discrete-space Markov chains, continuous-time/discrete-space Markov jump processes, and continuous-time/continuous-space diffusion processes. 
In each case, we have illustrated the general formalism via simple examples based on the Sisyphus Random Walk
and its variants.
Since the equations containing
the main results have already been pointed out in the subsections I-A,I-B,I-C of the Introduction,
we shall not repeat them here. We hope that these explicit formula
 will be useful to analyze the various applications mentioned in the Introduction,
where the resetting probabilities or the resetting rates depend
on the position either deterministically or randomly.

\appendix


\section{  Application to the Sisyphus Random Walk on the Cayley tree }

\label{sec_chaintree}

In this section, we focus on the generalization of the one-dimensional
 Sisyphus Random Walk discussed in section \ref{sec_chain1d}
 for the geometry of the Cayley tree of branching ratio $b$.
(The one-dimensional model of section \ref{sec_chain1d}
corresponds to the special case $b=1$ with no branching).

\subsection{ Model and notations }

We consider a tree of branching parameter $b$, 
starting at the root $(0)$, 
with $b$ sites labelled by $i_1=1,2,..,b$ at the first generation $g=1$,
$b^2$ sites labelled by $(i_1,i_2)$ at the second generation $g=2$, 
i.e. the $b^g$ sites of generation $g$ will be labelled by the $g$ indices $(i_1,i_2,..,i_g)$
containing the information on the whole line of ancestors.

In the absence of resetting, a particle on site $(i_1,..,i_g)$ at time $t$ can only jump towards 
one of its $b$ children, so one can use the simplified notation for the corresponding probabilities
  $W_{(i_1,i_2..,i_g,i_{g+1}),(i_1,i_2,...,i_g)} \equiv W_{i_1,i_2..,i_g,i_{g+1}}$
 with the normalization
\begin{eqnarray}
1 = \sum_{i_{g+1}=1}^{b} W_{i_1,...,i_g,i_{g+1}}
\label{normawoutoftree}
\end{eqnarray}
The simplest example corresponds to the uniform case, where each of the $b$ children is chosen with equal probability $1/b$ at each step
\begin{eqnarray}
W^{uniform}_{i_1,...,i_g,i_{g+1}} = \frac{1}{b}
\label{normawoutoftreeunif}
\end{eqnarray}
but other choices are also interesting, so we will continue the analysis for the general case.

In the presence of resetting towards the origin, the dynamics of Eq. \ref{markovchainreset} reads
for the root and for the sites of other generations $g \geq 1$ respectively
\begin{eqnarray}
P_0(t+1) && = R_0 P_0(t) + \sum_{g=1}^{+\infty} \sum_{i_1,..,i_g}   R_{i_1,..,i_g}   P_{i_1,..,i_g}(t) 
\nonumber \\
P_{i_1,..,i_{g}}(t+1) && = W_{i_1,...,i_g}   (1-R_{i_1,..,i_{g-1}}) P_{i_1,..,i_{g-1}}(t)    
\label{treechaindyn}
\end{eqnarray}

In order to apply the general formalism described in section \ref{sec_chain},
one should first check that the hypothesis concerning the existence of a steady-state (Eq. \ref{markovchainst})
is satisfied.


\subsection{ Condition on the reset probabilities $R_{.....}$ for the existence of a non-equilibrium steady-state  }

The equations for the stationary state $P^*_.$ of the dynamics of Eq. \ref{treechaindyn}
\begin{eqnarray}
P_0^* && =  R_0 P_0^* + \sum_{g=1}^{+\infty} \sum_{i_1,..,i_g}   R_{i_1,..,i_g}   P_{i_1,..,i_g}^* 
\nonumber \\
P_{i_1,..,i_g}^* && = W_{i_1,...,i_g}  (1-R_{i_1,..,i_{g-1}}) P_{i_1,..,i_{g-1}}^*   
\label{treechainstatio}
\end{eqnarray}
can be directly solved by recurrence for all the sites of generation $g \geq 1$ in terms of $P_0^*$
\begin{eqnarray}
P_{i_1,..,i_{g}}^* && = W_{i_1,...,i_g}   (1-R_{i_1,..,i_{g-1}}) ... W_{i_1} (1-R_0) P_0^*   
= \left[  \prod_{g'=1  }^{g} W_{i_1,...,i_{g'}}  (1-R_{i_1,..,i_{g'-1}}) \right]  P_0^*   
\label{treechainstationotrootsolu}
\end{eqnarray}
while $P_0^*$ is then determined by the normalization
\begin{eqnarray}
1= P_0^* + \sum_{g=1}^{+\infty} \sum_{i_1,..,i_g }P_{i_1,..,i_{g}}^* && 
=P_0^*   \left( 1+ \sum_{g=1}^{+\infty} \sum_{i_1,..,i_g } \left[  \prod_{g'=1  }^{g} W(i_1,...,i_{g'})   (1-R_{i_1,..,i_{g'-1}}) \right] \right)  
\label{treechainstationotroot}
\end{eqnarray}
The condition $P_0^*>0$ to produce a non-equilibrium steady-state localized around the origin
corresponds to the requirement of convergence for the following series 
\begin{eqnarray}
 \sum_{g=1}^{+\infty} \sum_{i_1,..,i_g } \left[  \prod_{g'=1  }^{g} W(i_1,...,i_{g'})   (1-R_{i_1,..,i_{g'-1}}) \right] < +\infty  
\label{treechainstationotrootcv}
\end{eqnarray}
In the following, we assume the convergence of Eq. \ref{treechainstationotrootcv}
in order to have a non-equilibrium steady-state that can be analyzed
via the large deviations at Level 2.5.


\subsection{  Empirical density and empirical currents with their constraints }

The empirical 2-point density of Eq. \ref{rho2pt} contains two types of contributions,
namely the non-local reset currents from any point $(i_1,..,i_g)$ of any generation $g \geq 0$
towards the origin
\begin{eqnarray}
J_{i_1,...,i_g} && \equiv  \rho^{(2)}_{0,(i_1,..,i_g)}  
\nonumber \\
J_{0} && \equiv  \rho^{(2)}_{0,0}  
\label{Jresettree}
\end{eqnarray}
and the local currents arriving at any point $(i_1,..,i_g)$ of any generation $g \geq 1$ from its ancestor
$(i_1,..,i_{g-1})$
\begin{eqnarray}
j_{i_1,...,i_g} && \equiv  \rho^{(2)}_{(i_1,..,i_g),(i_1,..,i_{g-1})}  
\nonumber \\
j_{i_1} && \equiv  \rho^{(2)}_{i_1,0}  
\label{jlocaltree}
\end{eqnarray}
As a consequence, the consistency constraints of Eq. \ref{rho1pt} involving the 1-point density become
for the origin
\begin{eqnarray}
 \rho_0 = J_0+ \sum_{g=1}^{+\infty} \sum_{i_1,..,i_g}  J_{i_1,...,i_g} = J_0 + \sum_{i_1} j_{i_1}
\label{rho1ptreeorigin}
\end{eqnarray}
and for the other points of generation $g \geq 1$
\begin{eqnarray}
 \rho_{i_1,..,i_g} =   j_{i_1,...,i_g} = J_{i_1,..,i_g} + \sum_{i_{g+1}} j_{i_1,..,i_g,i_{g+1}}
\label{rho1pttree}
\end{eqnarray}


\subsection{  Large deviations at level 2.5 for the joint probability of the empirical density and the empirical currents }

The large deviations at level 2.5 of Eqs \ref{proba2.5chain} \ref{constraints2.5chain} \ref{rate2.5chain} 
yield that the joint probability to see the empirical density $\rho_.$ and the empirical currents $j_.$ and $J_.$ read
\begin{eqnarray}
P_T ( \rho_. ,j_.,J_. )
 && \opsimeq_{T \to +\infty}  
\delta \left( \rho_0+\sum_{g=1}^{+\infty} \sum_{i_1,..,i_g }\rho_{i_1,..,i_{g}}
 - 1 \right) 
\delta \left( J_0+ \sum_{g=1}^{+\infty} \sum_{i_1,..,i_g}  J_{i_1,...,i_g} -\rho_0  \right)
\delta \left(  J_0 + \sum_{i_1} j_{i_1} -\rho_0 \right)
\nonumber \\
&&  \prod_{g=1}^{+\infty} \prod_{i_1,..,i_g}  \left[  \delta \left(   j_{i_1,...,i_g} - \rho_{i_1,..,i_g}   \right)
\delta \left(  J_{i_1,..,i_g} + \sum_{i_{g+1}} j_{i_1,..,i_g,i_{g+1}} -  \rho_{i_1,..,i_g}  \right) \right]
\nonumber \\
&& e^{ - \displaystyle T 
\left[  \sum_{i_1} j_{i_1} \ln \left( \frac{ j_{i_1} }{  W_{i_1} (1-R_0)  \rho_0 }  \right)
+ \sum_{g=2}^{+\infty} \sum_{i_1,..,i_g } j_{i_1,..,i_g}
 \ln \left( \frac{ j_{i_1,..,i_g} }{  W_{i_1,..,i_g} (1-R_{i_1,..,i_{g-1}})  \rho_{i_1,..,i_{g-1}}}  \right)
\right] } 
\nonumber \\
&& e^{ - \displaystyle T 
\left[  J_0  \ln \left( \frac{ J_0 }{  R_0  \rho_0 }  \right)
+ \sum_{g=1}^{+\infty} \sum_{i_1,..,i_g } J_{i_1,..,i_g}  \ln \left( \frac{ J_{i_1,..,i_g} }{  R_{i_1,..,i_g}  \rho_{i_1,..,i_g} }  \right) \right] } 
\label{proba2.5chaintree}
\end{eqnarray}
The first constraints on the second line shows that all the local currents $j_.$ can be eliminated in terms of the density
\begin{eqnarray}
 j_{i_1,...,i_g} =\rho_{i_1,..,i_g} 
\label{elimjtree}
\end{eqnarray}
so that Eq. \ref{proba2.5chaintree} becomes for the joint distribution of the density $\rho_.$
and the non-local reset currents $J_.$
\begin{eqnarray}
P_T ( \rho_. ,J_. )
 && \opsimeq_{T \to +\infty}  
\delta \left( \rho_0+\sum_{g=1}^{+\infty} \sum_{i_1,..,i_g }\rho_{i_1,..,i_{g}}
 - 1 \right) 
\delta \left( J_0+ \sum_{g=1}^{+\infty} \sum_{i_1,..,i_g}  J_{i_1,...,i_g} -\rho_0  \right)
\delta \left(  J_0 + \sum_{i_1} \rho_{i_1} -\rho_0 \right)
\nonumber \\
&&  \prod_{g=1}^{+\infty} \prod_{i_1,..,i_g}  
\delta \left(  J_{i_1,..,i_g} + \sum_{i_{g+1}} \rho_{i_1,..,i_g,i_{g+1}} -  \rho_{i_1,..,i_g}  \right) 
\nonumber \\
&& e^{ - \displaystyle T 
\left[  \sum_{i_1} \rho_{i_1} \ln \left( \frac{ \rho_{i_1} }{  W_{i_1} (1-R_0)  \rho_0 }  \right)
+ \sum_{g=2}^{+\infty} \sum_{i_1,..,i_g } \rho_{i_1,..,i_g}
 \ln \left( \frac{ \rho_{i_1,..,i_g} }{  W_{i_1,..,i_g} (1-R_{i_1,..,i_{g-1}})  \rho_{i_1,..,i_{g-1}}}  \right)
\right] } 
\nonumber \\
&& e^{ - \displaystyle T 
\left[  J_0  \ln \left( \frac{ J_0 }{  R_0  \rho_0 }  \right)
+ \sum_{g=1}^{+\infty} \sum_{i_1,..,i_g } J_{i_1,..,i_g}  \ln \left( \frac{ J_{i_1,..,i_g} }{  R_{i_1,..,i_g}  \rho_{i_1,..,i_g} }  \right) \right] } 
\label{proba2.5chaintreesansj}
\end{eqnarray}
One can now further use the constraints to obtain the large deviations for the empirical density alone 
or for the non-local reset currents alone, as described in the following two subsections.


\subsection { Large deviations at Level 2 for the empirical density $\rho_.$ alone }

The constraints on the second line of Eq. \ref{proba2.5chaintreesansj}
can be used to eliminate the non-local reset currents associated to generations $g \geq 1$
in terms of the empirical density
\begin{eqnarray}
 J_{i_1,..,i_g} =   \rho_{i_1,..,i_g} - \sum_{i_{g+1}} \rho_{i_1,..,i_g,i_{g+1}} 
\label{bigJ}
\end{eqnarray}
while the third constraint on the first line of Eq. \ref{proba2.5chaintreesansj} gives the similar equation for the 
origin
\begin{eqnarray}
 J_0 =   \rho_0 - \sum_{i_1} \rho_{i_1} 
\label{bigJ0}
\end{eqnarray}
Then the second constraint on the first line of Eq. \ref{proba2.5chaintreesansj}
is automatically satisfied,
so that Eq. \ref{proba2.5chaintreesansj}
yields for the large deviations for the empirical density $\rho_.$ alone
\begin{eqnarray}
&& P_T ( \rho_. )
  \opsimeq_{T \to +\infty}  
\delta \left( \rho_0+\sum_{g=1}^{+\infty} \sum_{i_1,..,i_g }\rho_{i_1,..,i_{g}}
 - 1 \right) 
\label{proba2.5chaintreeonlyrho}
\\
&& e^{ - \displaystyle T 
\left[  \sum_{i_1} \rho_{i_1} \ln \left( \frac{ \rho_{i_1} }{  W_{i_1} (1-R_0)  \rho_0 }  \right)
+ \sum_{g=2}^{+\infty} \sum_{i_1,..,i_g } \rho_{i_1,..,i_g}
 \ln \left( \frac{ \rho_{i_1,..,i_g} }{  W_{i_1,..,i_g} (1-R_{i_1,..,i_{g-1}})  \rho_{i_1,..,i_{g-1}}}  \right)
\right] } 
\nonumber \\
&& e^{ - \displaystyle T 
\left[  \left(  \rho_0 - \sum_{i_1} \rho_{i_1}  \right)  \ln \left( \frac{  \rho_0 - \sum_{i_1} \rho_{i_1}  }{  R_0  \rho_0 }  \right)
+ \sum_{g=1}^{+\infty} \sum_{i_1,..,i_g } \left(\rho_{i_1,..,i_g} - \sum_{i_{g+1}} \rho_{i_1,..,i_g,i_{g+1}}  \right)  \ln \left( \frac{ \rho_{i_1,..,i_g} - \sum_{i_{g+1}} \rho_{i_1,..,i_g,i_{g+1}}  }{  R_{i_1,..,i_g}  \rho_{i_1,..,i_g} }  \right) \right] } 
\nonumber 
\end{eqnarray}

One may further simplify the rate function of the two last lines to obtain
\begin{eqnarray}
&& P_T ( \rho_. )
  \opsimeq_{T \to +\infty}  
\delta \left( \rho_0+\sum_{g=1}^{+\infty} \sum_{i_1,..,i_g }\rho_{i_1,..,i_{g}} - 1 \right) 
\label{proba2.5chaintreeonlyrhosimpli}
 \\
&& e^{ - \displaystyle T 
\left[ -\rho_0 \ln (\rho_0) - \sum_{i_1} \rho_{i_1} \ln \left(  W_{i_1} (1-R_0)    \right)
-  \sum_{g=2}^{+\infty} \sum_{i_1,..,i_g } \rho_{i_1,..,i_g} \ln \left(  W_{i_1,..,i_g} (1-R_{i_1,..,i_{g-1}})    \right)
\right] } 
\nonumber \\
&& e^{ - \displaystyle T 
\left[ \left(  \rho_0 - \sum_{i_1} \rho_{i_1}  \right)  \ln \left( \frac{  \rho_0 - \sum_{i_1} \rho_{i_1}  }{  R_0   }  \right)
+
 \sum_{g=1}^{+\infty} \sum_{i_1,..,i_g } \left(\rho_{i_1,..,i_g} - \sum_{i_{g+1}} \rho_{i_1,..,i_g,i_{g+1}}  \right)  \ln \left( \frac{ \rho_{i_1,..,i_g} - \sum_{i_{g+1}} \rho_{i_1,..,i_g,i_{g+1}}  }{  R_{i_1,..,i_g}   }  \right) \right] } 
\nonumber 
\end{eqnarray}


\subsection { Large deviations for the empirical reset currents $J_.$ alone }

If one wishes instead to eliminate the empirical density in terms of the empirical reset currents $J_.$ 
via
\begin{eqnarray}
 \rho_{i_1,..,i_g} && =   J_{i_1,..,i_g} +\sum_{g'=g+1}^{+\infty} \sum_{i_{g+1},...,i_{g'}} J_{i_1,..,i_g,i_{g+1},...,i_{g'}} 
\nonumber \\
\rho_0 && = J_0 + \sum_{g=1}^{+\infty} \sum_{i_1,...,i_{g}} J_{i_1,..,i_g} 
\label{rhobigJ}
\end{eqnarray}
Eq. \ref{proba2.5chaintreeonlyrhosimpli}
 yields 
for the large deviations for the empirical reset currents $J_.$ alone
\begin{eqnarray}
&& P_T ( J_. )
  \opsimeq_{T \to +\infty}  
\delta \left( J_0+\sum_{g=1}^{+\infty} (g+1) \sum_{i_1,..,i_g }J_{i_1,..,i_{g}}- 1 \right) 
\delta \left( J_0+\sum_{g=1}^{+\infty}  \sum_{i_1,..,i_g }J_{i_1,..,i_{g}}- \rho_0 \right) 
\nonumber \\
&& e^{ - \displaystyle T 
\left[ -\rho_0 \ln (\rho_0) 
-  \sum_{g=1}^{+\infty} \sum_{i_1,..,i_g } J_{i_1,..,i_g} \ln \left(  \prod_{g'=1}^g W_{i_1,..,i_{g'}} (1-R_{i_1,..,i_{g'-1}})    \right)
\right] } 
\nonumber \\
&& e^{ - \displaystyle T 
\left[ J_0 \ln \left( \frac{  J_0   }{  R_0   }  \right)
+
 \sum_{g=1}^{+\infty} \sum_{i_1,..,i_g } J_{i_1,..,i_g}   \ln \left( \frac{ J_{i_1,..,i_g}   }{  R_{i_1,..,i_g}   }  \right) \right] } 
\label{proba2.5chaintreeonlybigJ}
\end{eqnarray}
One may further simplify the rate function of the two last lines to obtain the more compact form
\begin{eqnarray}
&& P_T ( J_. )
  \opsimeq_{T \to +\infty}  
\delta \left( J_0+\sum_{g=1}^{+\infty} (g+1) \sum_{i_1,..,i_g }J_{i_1,..,i_{g}}- 1 \right) 
\delta \left( J_0+\sum_{g=1}^{+\infty}  \sum_{i_1,..,i_g }J_{i_1,..,i_{g}}- \rho_0 \right) 
\nonumber \\
&& e^{ - \displaystyle T 
\left[ J_0 \ln \left( \frac{  J_0   }{  R_0  \rho_0 }  \right)
+
 \sum_{g=1}^{+\infty} \sum_{i_1,..,i_g } J_{i_1,..,i_g}   \ln \left( \frac{ J_{i_1,..,i_g}   }{  R_{i_1,..,i_g}    
\left(  \prod_{g'=1}^g W_{i_1,..,i_{g'}} (1-R_{i_1,..,i_{g'-1}})    \right)
\rho_0 } \right) \right] } 
\label{proba2.5chaintreeonlybigJsimpli}
\end{eqnarray}
in order to make the link with the large deviations for excursions between two consecutive  resets
as explained in next subsection.


\subsection{ Large deviations for the empirical density of excursions between two consecutive resets }

For the Sisyphus Random Walk on the Cayley tree, 
the simplification in Eq. \ref{pinter} is that the internal trajectory $y( 1 \leq s \leq \tau-1) $
of an excursion is ballistic along one branch of the tree, and can be thus labelled only by its end-point 
$y(\tau-1)=(i_1,..,i_{\tau-1})  $, so that its probability of Eq. \ref{pinter}
becomes
\begin{eqnarray}
P^{exc}[ \tau ; (i_1,..,i_{\tau-1})  ] 
&& = R_{i_1,..,i_{\tau-1}} W_{i_1,..,i_{\tau-1}} (1-R_{i_1,..,i_{\tau-2}}) ... W_{i_1,i_2}(1-R_{i_1})W_{i_1} (1-R_{0})
 \nonumber \\ &&
 =  R_{i_1,..,i_{\tau-1}} \left[    \prod_{g'=1}^{\tau-1} W_{i_1,..,i_{g'}}  (1-R_{i_1,..,i_{g'-1}} )\right]
\label{pintertree}
\end{eqnarray}

As a consequence, 
the probability to see 
 the empirical density $n[\tau ; (i_1,..,i_{\tau-1}) ]  $ of excursions between resets
and the total density $n$ follows the large deviation form of Eqs \ref{probaempifin} \ref{rateexc} \ref{constraints}
\begin{eqnarray}
P_T ( n[.;..], n ) && \opsimeq_{T \to +\infty}  
 \delta \left[ \sum_{\tau=1}^{+\infty} \tau \sum_{i_1,..,i_{\tau-1} } n[\tau ; i_1,..,i_{\tau-1} ]  -1 \right]
\delta \left[ \sum_{\tau=1}^{+\infty} \sum_{i_1,..,i_{\tau-1} } n[\tau ; i_1,..,i_{\tau-1} ] 
- n \right]
\nonumber \\
&& 
 e^{ - \displaystyle T \sum_{\tau=1}^{+\infty} \sum_{i_1,..,i_{\tau-1} }
  n[\tau ;  i_1,..,i_{\tau-1} ]
  \ln \left( \frac{n[\tau ;  i_1,..,i_{\tau-1}  ]  }{  R_{i_1,..,i_{\tau-1}} \left[    \prod_{g'=1}^{\tau-1} W_{i_1,..,i_{g'}}  (1-R_{i_1,..,i_{g'-1}} )\right]  n  }  \right)   } 
\label{probaexctree}
\end{eqnarray}

The correspondence with Eq \ref{proba2.5chaintreeonlybigJsimpli} is now obvious with the
following dictionary :
the empirical density at the origin $\rho_0$ corresponds to the empirical density $n$ of excursions
\begin{eqnarray}
\rho_0= n
\label{rhozeron}
\end{eqnarray}
while the non-local reset current $J_{i_1,..,i_g} $ corresponds to the empirical density of excursions of duration $\tau=g+1$ with end-point $(i_1,..,i_{\tau-1})=(i_1,..,i_g)$
\begin{eqnarray}
J_{i_1,..,i_g} = n[g+1 ;  i_1,..,i_{g} ]
\label{Jn}
\end{eqnarray}


\subsection{ Large deviations of general time-additive observables }

In this section, we analyze the large deviations of
 time-additive observables of the form $(A_T+B_T )$ (Eqs \ref{additiveA} \ref{additiveB}
\ref{additiveArho} \ref{additiveBrho2})
via their generating function (Eq \ref{genek})
that reads in terms of the empirical density $\rho_.$, the empirical local currents $j_.$
and the empirical non-local reset currents $J_.$
\begin{eqnarray}
&& Z_T(k) \equiv    < e^{\displaystyle T k \left( A_T + B_T \right) } >  
 \nonumber \\ && 
= < e^{\displaystyle T k \left( \alpha_0 \rho_0+\beta_{0,0} J_0
+ \sum_{g=1}^{+\infty} \sum_{i_1,..,i_g }\left[  \alpha_{i_1,..,i_{g}} \rho_{i_1,..,i_{g}}
+ \beta_{i_1,..,i_{g}} j_{i_1,..,i_{g}}
  + \beta_{0;{i_1,..,i_{g}}} J_{i_1,..,i_{g}} \right]
   \right) } >  
\label{genektree}
\end{eqnarray}

The tilted dynamics of Eq. \ref{Wktilt}
reads for the model of Eq. \ref{treechaindyn}
for the root and for the sites of other generations $g \geq 1$ respectively
\begin{eqnarray}
 {\tilde P}_0(t+1) && = R_0 e^{ k \left(\alpha_0 + \beta_{0; 0}  \right)} {\tilde P}_0(t) 
 + \sum_{g=1}^{+\infty} \sum_{i_1,..,i_g}   
 R_{i_1,..,i_g} e^{ k \left(\alpha_0 + \beta_{0; {i_1,..,i_{g}}}  \right) }
    {\tilde P}_{i_1,..,i_g}(t) 
\nonumber \\
 {\tilde P}_{i_1,..,i_g}(t+1) && = W_{i_1,...,i_g}   (1-R_{i_1,..,i_{g-1}})
 e^{ k \left(\alpha_{i_1,..,i_{g}} + \beta_{i_1,..,i_{g}}  \right) }
   {\tilde P}_{i_1,..,i_{g-1}}(t)    
\label{treechaindynt}
\end{eqnarray}

The eigenvalue Eq. \ref{Wktiltright} 
for the right eigenvector ${\tilde r}^{[k]}_{.} $
reads
\begin{eqnarray}
e^{ \mu(k) } {\tilde r}^{[k]}_0 && =  R_0 e^{ k \left(\alpha_0 + \beta_{0; 0}  \right)} {\tilde r}^{[k]}_0
 + \sum_{g=1}^{+\infty} \sum_{i_1,..,i_g}   
 R_{i_1,..,i_g} e^{ k \left(\alpha_0 + \beta_{0; {i_1,..,i_{g}}}  \right) }
    {\tilde r}^{[k]}_{i_1,..,i_g} 
\nonumber \\
e^{ \mu(k) } {\tilde r}^{[k]}_{i_1,..,i_g} && = W_{i_1,...,i_g}   (1-R_{i_1,..,i_{g-1}})
 e^{ k \left(\alpha_{i_1,..,i_{g}} + \beta_{i_1,..,i_{g}}  \right) }
   {\tilde r}^{[k]}_{i_1,..,i_{g-1}}    
\label{treechaindyntr}
\end{eqnarray}
Plugging the solution for $g \geq 1$
\begin{eqnarray}
 {\tilde r}^{[k]}_{i_1,..,i_{g}} &&    
= \left[  \prod_{g'=1  }^{g} e^{- \mu(k) +k \left(\alpha_{i_1,..,i_{g'}} + \beta_{i_1,..,i_{g'}}  \right) } W_{i_1,...,i_{g'}}  (1-R_{i_1,..,i_{g'-1}}) \right]  {\tilde r}^{[k]}_0   
\label{treechainstationotrootsolur}
\end{eqnarray}
into Eq. \ref{treechaindyntr} for the root
yields the equation that determines the scaled cumulant generating function $\mu(k)$
\begin{eqnarray}
1  && =  R_0 e^{-  \mu(k) + k \left(\alpha_0 + \beta_{0; 0}  \right)} 
\nonumber \\
&&  + \sum_{g=1}^{+\infty} \sum_{i_1,..,i_g}   
 R_{i_1,..,i_g} e^{ -(g+1)  \mu(k) + k \left(\alpha_0 + \beta_{0; {i_1,..,i_{g}}}  \right) }
     \left[  \prod_{g'=1  }^{g} e^{k \left(\alpha_{i_1,..,i_{g'}} + \beta_{i_1,..,i_{g'}}  \right) } W_{i_1,...,i_{g'}}  (1-R_{i_1,..,i_{g'-1}}) \right]     
\label{treemuk}
\end{eqnarray}



\begin{thebibliography}{99}



\bibitem{review_reset}
M. R. Evans, S. N. Majumdar, G. Schehr,
	J. Phys. A: Math. Theor. 53 193001 (2020).

\bibitem{review_search}
O. B\'enichou, C. Loverdo, M. Moreau, R. Voituriez
Reviews of Modern Physics 83, 81 (2011).

\bibitem{book_finance}
R. Cont and P. Tankov, Financial Modelling with Jump Processes, Chapman and Hall / CRC Press, 2004.

\bibitem{review_finance}
P. Tankov and E. Voltchkova, Banque et Marchés, vol. 99, 2009.

\bibitem{dumont}
G. Dumont, J. Henry, C. O. Tarniceriu,
Journal of Mathematical Biology 73, 1413 (2016).

\bibitem{miles}
C. E. Miles, J. P. Keener, J. Phys. A: Math. Theor. 50 425003 (2017)

\bibitem{daly_fire}
E. Daly and A. Porporato, PRE 74, 041112 (2006).

 
\bibitem{daly_rain}
E. Daly and A. Porporato, PRE 73, 026108 (2006).

\bibitem{daly_rainbis}
E. Daly and A. Porporato, PRE 75, 011119 (2007).

 
 
 \bibitem{quantum2.5doob}
F. Carollo, R. L. Jack, J. P. Garrahan, Phys. Rev. Lett. 122, 130605 (2019).


\bibitem{previousquantum2.5doob}
F. Carollo, J. P. Garrahan, I. Lesanovsky, C. Perez-Espigares, Phys. Rev. A 98, 010103 (2018).

 \bibitem{garrahan_reset}
D. C. Rose, H. Touchette, I. Lesanovsky, J. P. Garrahan, Phys. Rev. E 98, 022129 (2018)


\bibitem{manette}
S. C. Manrubia and D. H. Zanette, Phys. Rev. E 59, 4945 (1999).



\bibitem{derrida-lecture}
B. Derrida, J. Stat. Mech. P07023 (2007).

\bibitem{harris_Schu}
R J Harris and G M Sch\"utz,
J. Stat. Mech.  P07020 (2007).

\bibitem{searles}
E.M. Sevick, R. Prabhakar, S. R. Williams, D. J. Searles,
Ann. Rev. of Phys. Chem.  Vol 59, 603 (2008). 

\bibitem{harris}
H. Touchette and R.J. Harris, chapter "Large deviation approach to nonequilibrium systems"
of the book "Nonequilibrium Statistical Physics of Small Systems: Fluctuation Relations and Beyond", Wiley 2013.

\bibitem{mft}
L. Bertini, A. De Sole, D. Gabrielli, G. Jona-Lasinio, and C. Landim
Rev. Mod. Phys. 87, 593 (2015).

\bibitem{sollich_review}
R. L. Jack, P. Sollich, The European Physical Journal Special Topics  224, 2351 (2015).

\bibitem{lazarescu_companion}
A. Lazarescu, J. Phys. A: Math. Theor. 48 503001 (2015).

\bibitem{lazarescu_generic}
A. Lazarescu, J. Phys. A: Math. Theor. 50 254004 (2017).

\bibitem{jack_review}
R. L. Jack, Eur. Phy. J. B  93, 74 (2020)


\bibitem{fortelle_thesis}
A. de La Fortelle, PhD (2000)
"Contributions to the theory of large deviations and applications" INRIA Rocquencourt.


\bibitem{vivien_thesis}
V. Lecomte, PhD Thesis (2007)
"Thermodynamique des histoires et fluctuations hors d'\'equilibre"
Universit\'e Paris 7.

\bibitem{chetrite_thesis}
R. Ch\'etrite, PhD Thesis 2008 
"Grandes d\'eviations et relations de fluctuation dans certains mod\`eles de syst\`emes
hors d'\'equilibre"  ENS Lyon

\bibitem{wynants_thesis}
B. Wynants, arXiv:1011.4210, PhD Thesis (2010), "Structures of Nonequilibrium Fluctuations", Catholic University of Leuven.



\bibitem{chetrite_HDR}
R. Ch\'etrite, HDR Thesis (2018)
"P\'er\'egrinations sur les ph\'enom\`enes al\'eatoires dans la nature",
 Laboratoire J.A. Dieudonn\'e, Universit\' e de Nice.





\bibitem{oono}
Y. Oono,
Progress of Theoretical Physics Supplement 99, 165 (1989).

\bibitem{ellis}
R.S. Ellis, Physica D 133, 106 (1999).

\bibitem{review_touchette}
H. Touchette, Phys. Rep. 478, 1 (2009).



\bibitem{fortelle_chain}
G. Fayolle and A. de La Fortelle,
Problems of Information Transmission 38, 354 (2002).


\bibitem{c_largedevdisorder}
C. Monthus, Eur. Phys. J. B 92, 149 (2019) in the
topical issue " Recent Advances in the Theory of Disordered Systems"
edited by F. Igloi and H. Rieger.



\bibitem{fortelle_jump}
A. de La Fortelle, 
Problems of Information Transmission 37 , 120 (2001).



\bibitem{maes_canonical}
C. Maes and K. Netocny, Europhys. Lett. 82, 30003 (2008)

\bibitem{maes_onandbeyond}
C. Maes, K. Netocny and B. Wynants, Markov Proc. Rel. Fields. 14, 445 (2008).


\bibitem{chetrite_formal}
A. C. Barato and R. Ch\'etrite, J. Stat. Phys. 160, 1154 (2015).

\bibitem{BFG1}
L. Bertini, A. Faggionato and D. Gabrielli, 
Ann. Inst. Henri Poincare Prob. and Stat. 51, 867 (2015).

\bibitem{BFG2}
L. Bertini, A. Faggionato and D. Gabrielli, 
Stoch. Process. Appli. 125, 2786 (2015).


\bibitem{c_ring}
C. Monthus, J. Stat. Mech. (2019) 023206

\bibitem{c_interactions}
C. Monthus, J. Phys. A: Math. Theor. 52, 135003 (2019)


\bibitem{c_open}
C. Monthus, J. Phys. A: Math. Theor. 52, 025001 (2019)

\bibitem{barato_periodic}
A. C. Barato, R. Ch\'etrite, J. Stat. Mech. (2018) 053207

\bibitem{chetrite_periodic}
L. Chabane, R. Ch\'etrite, G. Verley, J. Stat. Mech. (2020) 033208.




\bibitem{maes_diffusion}
C. Maes, K. Netocny and B.  Wynants
Physica A 387, 2675 (2008).


\bibitem{engel}
J. Hoppenau, D. Nickelsen and A. Engel,
 New J. Phys. 18 083010 (2016).

\bibitem{c_lyapunov}
C. Monthus, arxiv 2010.14994




\bibitem{lecomte_chaotic}
V. Lecomte, C. Appert-Rolland and F. van Wijland,
Phys. Rev. Lett. 95 010601 (2005).

\bibitem{lecomte_thermo}
V. Lecomte, C. Appert-Rolland and F. van Wijland,
J. Stat. Phys. 127 51-106 (2007).

\bibitem{lecomte_formalism}
V. Lecomte, C. Appert-Rolland and F. van Wijland,
Comptes Rendus Physique 8, 609 (2007).

\bibitem{lecomte_glass}
J.P. Garrahan, R.L. Jack, V. Lecomte, E. Pitard, K. van Duijvendijk, F. van Wijland,
Phys. Rev. Lett. 98, 195702 (2007).

\bibitem{kristina1}
J.P. Garrahan, R.L. Jack, V. Lecomte, E. Pitard, K. van Duijvendijk and F. van Wijland, 
J. Phys. A 42, 075007 (2009).

\bibitem{kristina2}
K. van Duijvendijk, R.L. Jack and F. van Wijland, 
Phys. Rev. E 81, 011110 (2010).


\bibitem{jack_ensemble}
R. L. Jack, P. Sollich, Prog. Theor. Phys. Supp. 184, 304 (2010)

\bibitem{simon1}
D. Simon, J. Stat. Mech. (2009) P07017



\bibitem{simon2}
V. Popkov, G. M. Schuetz, D. Simon, J. Stat. Mech. P10007 (2010).

\bibitem{simon3}
D. Simon, J. Stat. Phys. 142,  931 (2011)

\bibitem{Gunter1}
V. Popkov, G. M. Schuetz, J. Stat. Phys 142,  627 (2011)


\bibitem{Gunter2}
V. Belitsky, G. M. Schuetz, J. Stat. Phys. 152, 93 (2013)


\bibitem{Gunter3}
O. Hirschberg, D. Mukamel, G. M. Schuetz, J. Stat. Mech. P11023 (2015).

\bibitem{Gunter4}
G. M. Schuetz, From Particle Systems to Partial Differential Equations II, Springer Proceedings in Mathematics and Statistics Volume 129, pp 371-393, P. Gonçalves and A.J. Soares (Eds.), (Springer, Cham, 2015).





\bibitem{chetrite_canonical}
R. Ch\'etrite and H. Touchette,
Phys. Rev. Lett. 111, 120601 (2013).

\bibitem{chetrite_conditioned}
R. Ch\'etrite and H. Touchette
 Ann. Henri Poincare 16, 2005 (2015).

\bibitem{chetrite_optimal}
R. Ch\'etrite, H. Touchette, J. Stat. Mech. P12001 (2015).



\bibitem{touchette_circle}
P. T. Nyawo, H. Touchette, Phys. Rev. E 94, 032101 (2016)

\bibitem{touchette_langevin}
H. Touchette, Physica A 504, 5 (2018).

\bibitem{touchette_occ}
F. Angeletti, H. Touchette, Journal of Mathematical Physics 57, 023303 (2016).

\bibitem{touchette_occupation}
P. T. Nyawo, H. Touchette, Europhys. Lett. 116, 50009 (2016); \\
P. T. Nyawo, H. Touchette, Phys. Rev. E 98, 052103 (2018).

\bibitem{derrida-conditioned}
B. Derrida and T. Sadhu, Journal of Statistical Physics 176, 773 (2019); \\
B. Derrida and T. Sadhu, 
Journal of Statistical Physics 177, 151 (2019).

\bibitem{derrida-ring}
K. Proesmans, B. Derrida, J. Stat. Mech. (2019) 023201.

\bibitem{bertin-conditioned}
N. Tizon-Escamilla, V. Lecomte and E. Bertin, 	J. Stat. Mech. (2019) 013201.



\bibitem{touchette-reflected}
J. du Buisson, H. Touchette, Phys. Rev. E 102, 012148 (2020)





\bibitem{touchette2015}
J.M. Meylahn, S. Sabhapandit, H. Touchette, Phys. Rev. E 92, 062148 (2015).


\bibitem{harris_reset}
 R. J. Harris, H. Touchette, J. Phys. A: Math. Theor. 50 10LT01 (2017)


\bibitem{pal} 
A. Pal, R. Chatterjee, S. Reuveni, A. Kundu,  J. Phys. A: Math. Theor. 52, 264002 (2019).


\bibitem{maj2019}
F. den Hollander, S.N. Majumdar, J.M. Meylahn, H. Touchette, J. Phys. A: Math. Theor. 52, 175001 (2019).

\bibitem{coghi_reset}
F. Coghi, R. J. Harris, J. Stat. Phys. 179,131 (2020)



\bibitem{sisyphus}
M. Montero and J. Villarroel, PRE 94, 032132 (2016).


\bibitem{optimal}
M.R. Evans, S.N. Majumdar, J. Phys. A: Math. Theor. 44, 435001 (2011).

\bibitem{pinsky}
R.G. Pinsky, Stochastic Processes and their Applications
Volume 130, Issue 5, May 2020, Pages 2954-2973.

\bibitem{path}
E. Roldan, S. Gupta, Phys. Rev. E 96 022130 (2017)



\bibitem{gaspard}
D. Andrieux and P. Gaspard, JSTAT P11007 (2008).

\bibitem{maes_semi}
C. Maes, K.l Netocny and B. Wynants,
J. Phys. A: Math. Theor. 42 (2009) 365002


\bibitem{zambotti}
M. Mariani and L. Zambotti, Adv. Appl. Prob. 48, 648 (2016).

\bibitem{faggionato}
A. Faggionato, arxiv:1709.05653



\bibitem{Kesten}
H. Kesten, Acta Math. 131, 208 (1973);
H. Kesten et al. , Compositio Math 30, 145 (1975).


\bibitem{Der_Pom}
B. Derrida and Y. Pomeau, Phys. Rev. Lett. 48 , 627 (1982).

\bibitem{Bou}
J. P. Bouchaud and A. Georges, Phys. Rep. 195, 127 (1990).

\bibitem{Der_Hil}
B. Derrida and H. Hilhorst, J. Phys. A 16, 2641 (1983).

\bibitem{Cal}
C. de Callan, J.M. Luck, Th. Nieuwenhuizen and D. Petritis, 
 J. Phys. A 18, 501 (1985).
 

 
 \bibitem{strong_review}
F. Igloi and C. Monthus, Phys. Rep. 412, 277 (2005).


\bibitem{c_microcano}
 C. Monthus,  Phys. Rev. B  69, 054431  (2004).
 
 \bibitem{c_watermelon}
C. Monthus,  J. Stat. Mech.  P06036 (2015).

\bibitem{c_mblcayley}
C. Monthus,  J. Stat. Mech. 123304 (2017).

\end{thebibliography}
\end{document}